%% file: main.tex
\PassOptionsToPackage{table,prologue}{xcolor}
\documentclass[acmsmall,screen,nonacm]{acmart}
\usepackage{subcaption} %

\usepackage{multirow}
\usepackage{multicol}

\usepackage{mathpartir} %

\usepackage{xspace}

\usepackage[inline]{enumitem}

\usepackage{hyperref}

\usepackage{stmaryrd} %

\usepackage{listings}
\usepackage{tikz}
\usetikzlibrary{trees}

\usepackage{wrapfig}

\usepackage[capitalize]{cleveref} %

\include{macros-specific} %

\include{macros-generic}

\include{macros-authors}

\let\svthefootnote\thefootnote
\newcommand\freefootnote[1]{%
  \let\thefootnote\relax%
  \footnotetext{#1}%
  \let\thefootnote\svthefootnote%
}

\allowdisplaybreaks

\begin{document}

\title[A Verified High-Performance Composable Object Library for Remote Direct Memory Access]
{A Verified High-Performance Composable Object Library for Remote Direct Memory Access (Extended Version)}

\author[G. Ambal]{Guillaume Ambal\text{*}}
\orcid{0000-0002-4667-7266}
\affiliation{\institution{Imperial College London}
\country{UK}}
\email{g.ambal@imperial.ac.uk}

\author[G. Hodgkins]{George Hodgkins\text{*}}
\orcid{0009-0005-7327-9561}
\affiliation{
  \institution{University of Colorado, Boulder}
  \country{USA}}
\email{George.Hodgkins@colorado.edu}

\author[M. Madler]{Mark Madler}
\orcid{0009-0004-3433-7338}
\affiliation{
  \institution{University of Colorado, Boulder}
  \country{USA}}
\email{Mark.Madler@colorado.edu}

\author[G. Chockler]{Gregory Chockler}
\orcid{0000-0001-6700-9235}
\affiliation{
  \institution{University of Surrey}
  \country{UK}}
\email{g.chockler@surrey.ac.uk}

\author[B. Dongol]{Brijesh Dongol}
\orcid{0000-0003-0446-3507}
\affiliation{
  \institution{University of Surrey}
  \country{UK}}
\email{b.dongol@surrey.ac.uk}

\author[J. Izraelevitz]{Joseph Izraelevitz}
\orcid{0009-0002-1267-5024}
\affiliation{
  \institution{University of Colorado, Boulder}
  \country{USA}}
\email{Joseph.Izraelevitz@colorado.edu }

\author[A. Raad]{Azalea Raad}
\orcid{0000-0002-2319-3242}
\affiliation{\institution{Imperial College London}
\country{UK}}
\email{azalea.raad@imperial.ac.uk}

\author[V. Vafeiadis]{Viktor Vafeiadis}
\orcid{0000-0001-8436-0334}
\affiliation{\institution{MPI-SWS}
\country{Germany}}
\email{viktor@mpi-sws.org }

\begin{abstract}
\input{abstract}

\end{abstract}

\maketitle
\freefootnote{\text{*}co-first authors.}

\input{intro}

\input{overview}

\input{metalanguage} %

\input{barrier}

\input{ringbuffer}

\input{example-LOCO}

\input{related}

\input{conc}

\bibliography{refs,paper}

\appendix
\newpage

\input{design}

\input{eval}

\input{backend}

\input{bugs}

\input{example}

\input{proofs}

\input{proof}

\end{document}

%% file: macros-specific.tex
\usepackage{pifont}
\newcommand{\newxspacedcommand}[2]{\newcommand{#1}{\xspace{#2}\xspace}}

\newcommand{\img}[1]{\ensuremath{\mathtt{img}(#1)}\xspace}

\newcommand{\assign}{:=}
\newcommand{\assignbrl}{\assign_{\brl}}

\newcommand{\code}[1]{\ensuremath{\mathtt{#1}}\xspace}
\newcommand{\pcode}[1]{\ensuremath{\mathtt{#1}}\xspace}

\newcommand{\textget}{\text{get}\xspace}

\newcommand{\textput}{\text{put}\xspace}
\newcommand{\textputs}{\text{puts}\xspace}

\newcommand{\Nodes}{{\sf Node}}
\newcommand{\Threads}{{\sf Tid}}

\newcommand{\Loc}{{\sf Loc}}
\newcommand{\Val}{{\sf Val}}
\newcommand{\Method}{{\sf Method}}

\newcommand{\Act}{\Labs}
\newcommand{\act}{\lab}
\newcommand{\ev}{\action} %

\newcommand{\node}{n\xspace}
\newcommand{\nodefun}[1]{{\tt \node}(#1)\xspace}
\newcommand{\threadt}{t\xspace}
\newcommand{\threadfun}[1]{{\tt \threadt}(#1)\xspace}

\newcommand{\Wid}{{\sf Wid}}
\newcommand{\wid}{\ensuremath{d}\xspace}

\newcommand{\assignwid}{\assign^\wid}

\newcommand{\loc}{\ensuremath{\mathtt{loc}}\xspace}

\newcommand{\neqn}[1]{\overline{#1}}

\newcommand{\set}[1]{\left\{{#1}\right\}}
\newcommand{\tup}[1]{\langle{#1}\rangle}

\newcommand{\aident}{\iota} %
\newcommand{\AId}{{\sf ActionId}}
\newcommand{\EId}{{\sf EventId}}

\newcommand{\action}{\ensuremath{\mathsf{e}}}
\newcommand{\saction}{\ensuremath{\mathsf{s}}}

\newcommand{\Events}{\mathsf{Event}}
\newcommand{\SEvents}{\mathsf{SEvent}}

\newcommand{\Labs}{\ensuremath{\mathsf{Lab}}\xspace}
\newcommand{\lab}{l}

\newcommand{\rtsowrite}{\ensuremath{\mathtt{Write}^{\rm\sc TSO}}\xspace}
\newcommand{\rtsoread}{\ensuremath{\mathtt{Read}^{\rm\sc TSO}}\xspace}
\newcommand{\rtsopoll}{\ensuremath{\mathtt{Poll}}\xspace}

\newcommand{\rlwrite}{\ensuremath{\mathtt{Write}}\xspace}
\newcommand{\rlread}{\ensuremath{\mathtt{Read}}\xspace}
\newcommand{\rlmf}{\ensuremath{\mathtt{Mfence}}\xspace}
\newcommand{\rlcas}{\ensuremath{\mathtt{CAS}}\xspace}
\newcommand{\rlwait}{\ensuremath{\mathtt{Wait}}\xspace}
\newcommand{\rlget}{\ensuremath{\mathtt{Get}}\xspace}
\newcommand{\rlput}{\ensuremath{\mathtt{Put}}\xspace}
\newcommand{\rlrf}{\ensuremath{\mathtt{Rfence}}\xspace}

\newcommand{\mswwrite}{\ensuremath{\mathtt{Write}^{\rm\sc MSW}}\xspace}
\newcommand{\mswread}{\ensuremath{\mathtt{TryRead}^{\rm\sc MSW}}\xspace}
\newcommand{\mswwait}{\ensuremath{\mathtt{Wait}^{\rm\sc MSW}}\xspace}
\newcommand{\mswget}{\ensuremath{\mathtt{Get}^{\rm\sc MSW}}\xspace}
\newcommand{\mswput}{\ensuremath{\mathtt{Put}^{\rm\sc MSW}}\xspace}

\newcommand{\brlwrite}{\ensuremath{\mathtt{Write}_{\brl}}\xspace}
\newcommand{\brlread}{\ensuremath{\mathtt{Read}_{\brl}}\xspace}
\newcommand{\brlwait}{\ensuremath{\mathtt{Wait}_{\brl}}\xspace}
\newcommand{\brlgf}{\ensuremath{\mathtt{GF}_{\brl}}\xspace}
\newcommand{\brlbr}{\ensuremath{\mathtt{Bcast}_{\brl}}\xspace}

\newcommand{\balbar}{\ensuremath{\mathtt{BAR}_\bal}\xspace}

\newcommand{\rblsub}{\ensuremath{\mathtt{Submit}^{\rm\sc RBL}}\xspace}
\newcommand{\rblrec}{\ensuremath{\mathtt{Receive}^{\rm\sc RBL}}\xspace}

\newcommand{\dumloc}{\ensuremath{\bot}\xspace}

\newcommand{\mytextsf}[1]{\textnormal{\textsf{#1}}}
\newcommand{\makerel}[2][.]{\ensuremath{{\color{#1}\mytextsf{#2}}}\xspace}
\newcommand{\makerelext}[2][]{\ensuremath{#2_{\mytextsf{e}}\xspace}}
\newcommand{\makerelint}[2][]{\ensuremath{#2_{\mytextsf{i}}}\xspace}

\definecolor{darkgrey}{rgb}{0.4,0.4,0.4}
\definecolor{darkturquoise}{rgb}{0.012, 0.502, 0.486}
\definecolor{rocolor}{HTML}{4B0082}
\colorlet{colorSC}{red!80!black}
\colorlet{colorPO}{darkgray!80!black}
\colorlet{colorPPO}{magenta}
\colorlet{colorOPPO}{magenta}
\colorlet{colorIPPO}{cyan}
\colorlet{colorRF}{green!60!black}
\colorlet{colorMO}{orange}
\colorlet{colorRO}{rocolor}
\colorlet{colorFR}{purple}
\colorlet{colorSO}{red}
\colorlet{colorHB}{blue}
\colorlet{colorPF}{brown!80!black}
\colorlet{colorOB}{teal}
\colorlet{colorIB}{blue}

\newcommand{\po}{\makerel[colorPO]{po}}
\newcommand{\ppo}{\makerel[colorPPO]{ppo}}
\newcommand{\ippo}{\makerel[colorIPPO]{ippo}}

\newcommand{\tagppo}{\makerel[colorPPO]{to}}
\newcommand{\rf}{\makerel[colorRF]{rf}}
\newcommand{\rfe}{\makerelext[colorRF]{\rf}}
\newcommand{\rfi}{\makerelint[colorRF]{\rf}}
\newcommand{\mo}{\makerel[colorMO]{mo}}
\newcommand{\ro}{\makerel[colorRO]{nfo}}
\newcommand{\fr}{\makerel[colorFR]{rb}}
\newcommand{\rblfr}{\makerel[colorFR]{fb}}

\newcommand{\fri}{\makerelint[colorFR]{\fr}}
\newcommand{\hb}{\makerel[colorHB]{hb}}
\newcommand{\pf}{\makerel[colorPF]{pf}}
\newcommand{\pfget}{\makerel[colorPF]{pfg}}
\newcommand{\pfput}{\makerel[colorPF]{pfp}}
\newcommand{\ib}{\makerel[colorIB]{ib}}

\newcommand{\iso}{\makerel[colorSO]{iso}}
\newcommand{\so}{\makerel[colorSO]{so}}

\newcommand{\arr}[2][]{\xrightarrow[#1]{#2}}

\newcommand{\Inst}{\ensuremath{\mathtt{Inst}}\xspace}

\newcommand{\vr}{\ensuremath{\mathtt{v}_\mathtt{R}}\xspace}
\newcommand{\vw}{\ensuremath{\mathtt{v}_\mathtt{W}}\xspace}

\newxspacedcommand{\Read}{\mathcal{R}}
\newxspacedcommand{\Write}{\mathcal{W}}

\newcommand{\defeq}{\triangleq}

\newcommand{\setpred}[2]{ \left\{ #1 \;\middle|\; #2 \right\} }

\newcommand{\rst}[1]{|_{#1}}
\newcommand{\imm}[1]{{#1}{\rst{\text{imm}}}}

\newcommand{\cmark}{\textcolor{green!60!black}{\ding{51}}\xspace}
\newcommand{\xmark}{\textcolor{red!80!black}{\ding{55}}\xspace}
\newcommand{\checkyes}{\cmark}
\newcommand{\checkno}{\xmark}

\newcommand{\cyes}{\checkyes}
\newcommand{\cno}{\checkno}
\newcommand{\cqp}{{\sc sn}\xspace}

\newcommand{\intab}[1]{\begin{tabular}{@{}c@{}}#1\end{tabular}}

\newcommand{\inarrTTT}[1]{\begin{array}[t]{@{}l@{}}#1\end{array}}

\newcommand{\inarr}[1]{\begin{array}{@{}l@{}}#1\end{array}}
\newcommand{\inarrII}[2]{\begin{array}{@{}l@{~~}|@{~~}l@{}}\inarr{#1}&\inarr{#2}\end{array}}

\newcommand{\pcie}{\text{PCIe}\xspace}
\newcommand{\rdmatso}{\textsc{rdma}\textsuperscript{\textsc{tso}}\xspace}

\newcommand{\rdmawait}{\textsc{rdma}\textsuperscript{\textsc{wait}}\xspace}

\newcommand{\msw}{\textsc{msw}\xspace}
\newcommand{\brl}{\textsc{sv}\xspace} %
\newcommand{\bal}{\textsc{bal}\xspace}
\newcommand{\rbl}{\textsc{rbl}\xspace}

\newcommand{\corerma}{\text{coreRMA}\xspace}

\newcommand{\framework}{{\sc Mowgli}\xspace}

\newcommand{\purple}[1]{{\color{purple}#1}}

\newcommand{\teal}[1]{{\color{teal}#1}}

\newcounter{rowcounter}
\newcounter{colcounter}

\newcommand{\rowc}{\Alph{rowcounter}\refstepcounter{rowcounter}}
\newcommand{\colc}{\thecolcounter\refstepcounter{colcounter}}
\definecolor{cellname}{rgb}{0.5,0.5,0.5}
\newcommand\colheader{\cellcolor{cellname}\color{white}\textbf{\colc}}
\newcommand\rowheader[1]{\cellcolor{cellname}\color{white}\textbf{\rowc}&#1}

\newcommand{\Tags}{{\sf Stamp}}

\newcommand{\cons}{\mathcal{C}}

\newcommand{\exec}{\mathcal{G}}

\newcommand{\tagt}{\ensuremath{a}\xspace}
\newcommand{\gettags}{\ensuremath{\mathtt{stmp}}\xspace}

\newcommand{\gettagsmsw}{\ensuremath{\mathtt{stmp}_\mathtt{MSW}}\xspace}
\newcommand{\gettagsrl}{\ensuremath{\mathtt{stmp}_\mathtt{RL}}\xspace}
\newcommand{\gettagsbrl}{\ensuremath{\mathtt{stmp}_\mathtt{SV}}\xspace}
\newcommand{\gettagsbal}{\ensuremath{\mathtt{stmp}_\mathtt{BAL}}\xspace}
\newcommand{\gettagsrbl}{\ensuremath{\mathtt{stmp}_\mathtt{RBL}}\xspace}

\newcommand{\tagcread}{\ensuremath{\mathtt{aCR}}\xspace}
\newcommand{\tagcwrite}{\ensuremath{\mathtt{aCW}}\xspace}
\newcommand{\tagcas}{\ensuremath{\mathtt{aCAS}}\xspace}
\newcommand{\tagmf}{\ensuremath{\mathtt{aMF}}\xspace}
\newcommand{\tagwait}{\ensuremath{\mathtt{aWT}}\xspace}

\newcommand{\tagnlr}[1][\node]{\ensuremath{\mathtt{aNLR}_{#1}}\xspace}
\newcommand{\tagnlw}[1][\node]{\ensuremath{\mathtt{aNLW}_{#1}}\xspace}
\newcommand{\tagnrr}[1][\node]{\ensuremath{\mathtt{aNRR}_{#1}}\xspace}
\newcommand{\tagnrw}[1][\node]{\ensuremath{\mathtt{aNRW}_{#1}}\xspace}
\newcommand{\tagnf}[1][\node]{\ensuremath{\mathtt{aRF}_{#1}}\xspace}
\newcommand{\taggf}[1][\node]{\ensuremath{\mathtt{aGF}_{#1}}\xspace}

\newcommand{\implbrl}{\ensuremath{I_\mathtt{SV}}\xspace}
\newcommand{\implbalb}{\ensuremath{\texttt{b}}\xspace}
\newcommand{\implbal}{\ensuremath{I^{\implbalb}_\mathtt{BAL}}\xspace}
\newcommand{\ordero}{\ensuremath{o}\xspace}
\newcommand{\mswsize}{\ensuremath{\texttt{size}}\xspace}
\newcommand{\mswhash}{\ensuremath{\texttt{hash}}\xspace}
\newcommand{\implmsw}{\ensuremath{I^{\mswsize}_\mathtt{MSW}}\xspace}
\newcommand{\rblthdr}{\ensuremath{\texttt{rthd}}\xspace}
\newcommand{\rblthdw}{\ensuremath{\texttt{wthd}}\xspace}
\newcommand{\implrbl}{\ensuremath{I^{\rblthdw,\rblthdr}_\mathtt{S,RBL}}\xspace}
\newcommand{\rblfailread}{\mathcal{F}\xspace}
\newcommand{\rblin}{\ensuremath{\texttt{in}}\xspace}
\newcommand{\rblout}{\ensuremath{\texttt{out}}\xspace}

\newcommand{\rtsomf}{\ensuremath{\mathtt{Mfence}^{\rm\sc TSO}}\xspace}
\newcommand{\rtsocas}{\ensuremath{\mathtt{CAS}^{\rm\sc TSO}}\xspace}
\newcommand{\rtsoget}{\ensuremath{\mathtt{Get}^{\rm\sc TSO}}\xspace}
\newcommand{\rtsoput}{\ensuremath{\mathtt{Put}^{\rm\sc TSO}}\xspace}
\newcommand{\rtsorf}{\ensuremath{\mathtt{Rfence}^{\rm\sc TSO}}\xspace}
\newcommand{\rtsoadd}{\ensuremath{\mathtt{SetAdd}}\xspace}
\newcommand{\rtsorm}{\ensuremath{\mathtt{SetRemove}}\xspace}
\newcommand{\rtsoempty}{\ensuremath{\mathtt{SetIsEmpty}}\xspace}

\newcommand{\gettagsrtso}{\ensuremath{\mathtt{stmp}_\mathtt{TSO}}\xspace}

\newcommand{\implwait}{\ensuremath{I_\mathtt{W}}\xspace}
\newcommand{\impliw}[1]{\llfloor #1 \rrfloor_{\implwait}}

\newcommand{\true}{\texttt{true}}
\newcommand{\false}{\texttt{false}}

\newcommand{\vect}[1]{\widetilde{#1}}
\newcommand{\outcome}{\ensuremath{\mathtt{outcome}}\xspace}
\newcommand{\abs}{\ensuremath{\mathtt{abs}}\xspace}
\newcommand{\absf}[5][f]{\ensuremath{\abs_{#3,#2}^{#1}(#4, #5)}}
\newcommand{\wideabs}{\ensuremath{\mathtt{wideabs}}\xspace}
\newcommand{\wideabsf}[5][f]{\ensuremath{\wideabs_{#3,#2}^{#1}(#4, #5)}}

\newcommand{\interp}[1]{\llbracket #1 \rrbracket}
\newcommand{\interpt}[1]{\llbracket #1 \rrbracket_{\threadt}}
\newcommand{\impl}[1]{\llfloor #1 \rrfloor}
\newcommand{\impli}[1]{\llfloor #1 \rrfloor_{I}}
\newcommand{\implti}[1]{\llfloor #1 \rrfloor_{\threadt,I}}

\newcommand{\progf}{\ensuremath{\mathsf{f}}}
\newcommand{\progp}{\ensuremath{\mathsf{p}}}
\newcommand{\Progp}{\ensuremath{\mathsf{SeqProg}}}
\newcommand{\progps}{\vect{\progp}}

\newcommand{\Id}{\texttt{Id}}

\newcommand{\cardinal}[1]{\ensuremath{\#(#1)}}

\newcommand{\SVar}{{\rm SVar} \xspace}

\newcommand{\LetC}[3]{\code{let}\, #1  =  #2 \, \code{in} \, #3}
\newcommand{\LetF}[2]{\code{let}\, #1 \  #2}
\newcommand{\ITE}[3]{\code{if} \,\, #1 \,\, \code{then} \,\,#2 \,\, \code{else} \,\,#3}

%% file: macros-generic.tex
\newcommand{\DONE}[1]{ }

\newcommand{\ie}{\text{i.e.}\xspace}
\newcommand{\Ie}{\text{I.e.}\xspace}
\newcommand{\eg}{\text{e.g.}\xspace}
\newcommand{\Eg}{\text{E.g.}\xspace}

\newcommand{\resp}{\text{resp.}\xspace}

\newcommand{\cf}{\textit{cf.}\xspace}

\Crefname{figure}{\text{Fig.}}{\text{Figs.}}
\crefname{figure}{\text{Fig.}}{\text{Figs.}}

\crefname{section}{\S{}}{Section}
\crefformat{section}{#2\S{}#1#3}
\Crefformat{section}{Section~#2#1#3}
\crefname{theorem}{Theorem}{Theorems}
\crefformat{theorem}{Theorem~#2#1#3}
\Crefformat{theorem}{Theorem~#2#1#3}
\crefname{corollary}{Corollary}{Corollaries}
\crefformat{corollary}{Corollary~#2#1#3}
\Crefformat{corollary}{Corollary~#2#1#3}

\crefformat{appendix}{#2\S{}#1#3}
\Crefname{appendix}{Appendix}{Appendix}
\Crefformat{appendix}{Appendix~#2#1#3}

\crefname{definition}{\text{Def.\xspace}}{\text{Definitions}}
\Crefname{definition}{\text{Definition}}{\text{Definitions}}

\crefname{lemma}{\text{Lem.\xspace}}{\text{Lemmas}}
\Crefname{lemma}{\text{Lemma}}{\text{Lemmas}}

\newcounter{mylabelcounter}

\makeatletter
\newcommand{\labelAxiom}[2]{%
\hfill{\textsc{(#1)}}\refstepcounter{mylabelcounter}
\immediate\write\@auxout{%
  \string\newlabel{#2}{{\unexpanded{\textsc{#1}}}{\thepage}{{\unexpanded{\textsc{#1}}}}{mylabelcounter.\number\value{mylabelcounter}}{}}
}%
}
\makeatother

\definecolor{highlightcolor}{HTML}{F0E0F0}

\newcommand{\suq}{\subseteq}
\newcommand{\inv}[1]{\ensuremath{#1^{-1}}}

\newlength{\Oldarrayrulewidth}
\newcommand{\Cline}[2]{%
  \noalign{\global\setlength{\Oldarrayrulewidth}{\arrayrulewidth}}%
  \noalign{\global\setlength{\arrayrulewidth}{#1}}\cline{#2}%
  \noalign{\global\setlength{\arrayrulewidth}{\Oldarrayrulewidth}}
}

\newcolumntype{?}{!{\vrule width 1pt}}

%% file: macros-authors.tex
\newcommand{\ignore}[1]{}

\newcommand{\orix}[1]{}

\newcommand{\gatxt}[1]{{\color{teal} [[GA: #1]]}}
\renewcommand{\gatxt}[1]{}
\newcommand{\gax}[1]{}

\newcommand{\azaleax}[1]{}

\newcommand{\haggaix}[1]{}

\definecolor{bdcolor}{HTML}{e83858}

%% file: abstract.tex
Remote Direct Memory Access (RDMA) is a memory technology that allows
remote devices to directly write to and read from each other's memory,
bypassing components such as the CPU and operating system. This
enables low-latency high-throughput networking, as required for many
modern data centres, HPC applications and AI/ML workloads. However,
baseline RDMA comprises a highly permissive weak memory model that is
difficult to use in practice and has only recently been formalised.

In this paper, we introduce the Library of Composable Objects (LOCO), a
formally verified library for building multi-node objects on RDMA,
filling the gap between shared memory and distributed system
programming. LOCO objects are well-encapsulated and
take advantage of the strong locality and the weak consistency
characteristics of RDMA. They have performance comparable to
custom RDMA systems (e.g.\ distributed maps), but with a far simpler
programming model amenable to formal proofs of correctness.

To support verification, we develop a novel modular declarative
verification framework, called \framework, that is flexible enough
to model multinode objects and is independent of a memory consistency model.
We instantiate \framework with the RDMA memory model, and use it to verify
correctness of LOCO libraries.

%% file: intro.tex
\section{Introduction}
\label{sec:intro}

The \emph{remote direct memory access} (RDMA) protocol allows a
machine to access the memory of a remote machine across a network
without communicating with the remote processor.
Instead, the memory access is performed directly by the \emph{network interface card} (NIC).
Like memory, RDMA exports a load/store
interface, allowing a machine to copy from or write to remote memory.
Because it bypasses the software networking stack on both ends
of the connection, RDMA achieves low-latency,
high-throughput communication, making it a key technology in many
production-grade data centres such as those at
Microsoft~\cite{zhu2015congestion}, Google~\cite{lu2018multi},
Alibaba~\cite{wang2023srnic}, and Meta~\cite{gangidi2024rdma}.

Despite its memory-like interface, RDMA is a hardware-accelerated
networking protocol, and has traditionally been programmed as such---not as shared memory.
This has resulted in a very weak memory model with out-of-order behaviours
visible even in a sequential setting \cite{OOPSLA-24}.
Consider, for example, the following program, where all memories are zero-initialised.
$$\inarr{
\neqn{z} := x ; \quad \text{\color{gray!70!teal}// RDMA put: write the value of local variable $x$ to remote location $z$}
\\
 x := 1 \phantom{;} \quad \text{\color{gray!70!teal}// update local variable $x$ to $1$}
}$$
Somewhat counterintuitively, this program can result in $z$ getting
the value $1$, with the following execution steps:
\begin{enumerate*}
\item the put instruction ($\neqn{z} := x$) is offloaded to the NIC;
\item the CPU executes $x := 1$ updating the value of $x$ in the local memory; and
\item the NIC executes the put instruction, fetching
  the \emph{new} value of $x$ from local memory before performing the
  remote write.
\end{enumerate*}

Since programming RDMA directly is challenging, prior work has developed custom
RDMA libraries.
Most existing libraries are monolithic: they encapsulate a useful
distributed protocol (such as consensus~\cite{aguilera-osdi-2020} or
distributed storage~\cite{wang-sigmod-2022, dragojevic-nsdi-2014}) as
a single, global entity---not one that can be reused by other RDMA libraries.
Some other libraries (e.g.~\cite{wang-osdi-2020, cai-vldb-2018})
provide a simple high-level memory abstraction
that hides all the complexities of a highly non-uniform, weakly consistent
network memory, but also loses a lot of the performance that can be achieved
by knowing the system layout ~\cite{liu-ppopp-2014, tang-hpca-2013, majo-pc-2017}.
Other intermediate layers, such as MPI~\cite{mpi} or NCCL~\cite{nccl}
are designed explicitly for networks and present a message passing interface
that is ideal for embarrassingly parallel or task-oriented workflows,
but ill-suited for irregular and data-dependent workloads,
such as data stores or stateful transactional systems,
for which shared-memory solutions excel~\cite{liu-sigmodrec-2021}.
Although these library implementations are impressive engineering artefacts
and have often been carefully tuned to achieve very good performance,
they are almost impossible to verify formally due to their lack of modularity.

In this paper, we argue for a new way for programming RDMA applications%
---and more generally systems with non-uniform weakly consistent memories---%
with \emph{flexible} libraries that can expose the non-uniform memory aspects and
that support formal verification.
Key to our approach is \emph{composability}---namely, the ability to put together
smaller/simpler objects to build larger ones---and this composability is
reflected both in the design and implementation of our library as well as
in the formal proofs about its correctness.

\paragraph{\textbf{LOCO}}

As a first contribution, we introduce the \emph{Library of Composable Objects (LOCO)}.
A LOCO object is a concurrent object as in \citet{DBLP:journals/toplas/HerlihyW90},
exposing a collection of methods,
but storing its state in a distributed fashion across all participating nodes.
Familiar examples include cross-node locks, barriers, queues, and maps.
LOCO objects provide encapsulation and can be composed together to build other LOCO objects.
For instance, we can use simpler objects, such as the underlying RDMA operations
and the local CPU instructions,
to build intermediate objects, such as barriers, which in turn can be used to
build larger objects, such as a concurrent map.

For concreteness, we implement and verify LOCO over \rdmatso
(which combines an RDMA networking fabric with Intel x86-TSO nodes),
making use of an existing formalisation by \citet{OOPSLA-24}.
\rdmatso is, however, too low-level for our purposes because it does
not provide a compositional way for waiting for RDMA operations to complete.
To this end, we extend \rdmatso and define \rdmawait,
where a thread can associate remote put/get operations with a work identifier,
then subsequently perform a \code{Wait} operation to wait for all put/get operations with
this identifier to finish executing. A similar functionality is
supported in \rdmatso via the \code{Poll} operation, which allows one
to check whether a remote operation towards a node has
completed. However, \code{Poll} (as provided by \rdmatso) is highly
brittle since it only waits for the {\em earliest} unpolled remote
operation in program order to complete. Thus, correct synchronisation
using \code{Poll} requires one to be certain of the number of remote
operations that have been called by the thread in question, so that
the correct number of $\mathtt{Poll}$s can be inserted whenever
synchronisation is required.
We discuss \rdmawait and its differences with \rdmatso in more detail in
\cref{sec:ov-rdmawait}.
Moreover, we prove the correctness of \rdmawait, as implemented by LOCO over the
existing \rdmatso model.

\paragraph{\textbf{MOWGLI}}

As a second contribution, we introduce a new compositional framework,
\framework
(\underline{MO}dular \underline{W}eak \underline{G}raph-based
\underline{LI}braries), for modelling and verifying weak libraries.
\framework is \emph{generic} in that it makes no assumptions about the
underlying memory model (\eg RDMA or TSO) in its core theory, but can
be instantiated to reason about \rdmawait programs and its abstractions.

Following LOCO's modular design, \framework supports modular proofs that
allow composition between client and library objects via so-called
``towers of abstraction''.  To this end, we build on a declarative
approach~\cite{DBLP:journals/pacmpl/RaadDRLV19,DBLP:conf/esop/StefanescoRV24},
where concurrent objects are specified using a
set of axioms (i.e., consistency predicates) over events. Each event
may represent a simple operation like a read or a write, or a more
complex operation such as a method call. However, as we shall see,
current approaches to declarative semantics are inadequate because
they are too coarse-grained to define RDMA methods, which may provide
intricate synchronisation guarantees.

In \framework, we introduce a novel notion of a \emph{subevent}, together
with axioms over subevents describing the allowable behaviours of each
program.  We distinguish between subevents using \emph{stamps}; each
event that is split into a subevent is paired with a stamp.
Stamps are meta-categories of behaviours, shared by all libraries, and are
independent from programs. Stamps are then used to define order between
(sub)events.
Within a thread they are used to define the
\emph{preserved program order}
(\ppo)~\cite{DBLP:journals/toplas/AlglaveMT14}, which relates
(sub)events executed by a thread that may not be reordered.
Across threads and nodes, stamps are used to define the
\emph{synchronisation order} (\so)~\cite{DBLP:conf/vmcai/DongolJRA18}
between methods calls of the same library.
Together \ppo and \so are used to define the happens-before relation.

\begin{wrapfigure}[9]{r}{0.28\textwidth}
  \centering
  \vspace{-22pt}
  \small
  \scalebox{0.95}{\begin{tikzpicture}[
    level 1/.style={sibling distance=9cm, level distance=0.7cm},
    level 2/.style={sibling distance=2cm, level distance=1cm},
    level 3/.style={sibling distance=1.4cm, level distance=1.2cm},grow'=up]
    \node (Root)
    {\rdmatso }
      child { node {\rdmawait (\cref{sec:rdmawait})}
        child { node {\intab{Shared \\ variables (\cref{sec:brlib})}}
          child { node {\intab{Barrier \\ (\cref{sec:balib})}} } %
          child { node {\intab{Ring buffer \\ (\cref{sec:rblib})}} }}
        child { node {\intab{Mixed-size \\writes (\cref{sec:msw})}}}
        }
    ;
  \end{tikzpicture}}
  \vspace{-20pt}
  \caption{Overview of proofs}
  \label{fig:dev}
\end{wrapfigure}
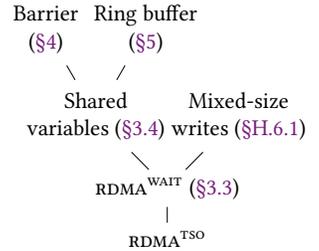
Our main result supporting modular proofs is a new locality result in
\framework for weak libraries, which enables one to decompose
soundness of a system into proofs about soundness of the individual
libraries that are used in the system. This is akin to the notion of
compositionality for
linearisability~\cite{DBLP:journals/toplas/HerlihyW90}, but
generalised to a partially ordered setting.
Note that \framework is more general than RDMA, \eg there is no notion of nodes,
thus it could be used to reason about other types of systems.
In our verification of LOCO, this allows us to verify a library, then use the
{\em specification} of the library in any program that uses the library.
Moreover, we show that our locality result supports both \emph{horizontal composition},
where a library is used within a client program, and \emph{vertical composition}, where
a library is developed from other libraries via a series of abstractions.

Within \framework, we verify key LOCO libraries and show that these
libraries can be used to also verify client programs. As a foundation
we specify the \rdmawait semantics, which contains operations for
local reads, writes, compare-and-swap, and memory fence (enabling TSO
programming), as well as remote put, get, wait and fence
primitives. These are then abstracted into a library that allows
mixed-size read/write operations and a library that supports broadcast
(which includes a global fence). The broadcast library is then used to
build both a barrier and a ring buffer, which we show can be used to
build client programs. An overview of the development is presented in
\cref{fig:dev}.

\paragraph{\textbf{Contributions}}

In summary, we make the following contributions:
\begin{itemize}[leftmargin=*]
\item We develop LOCO, a flexible, modular object library for RDMA,
  and demonstrate its compositionality by using simpler objects to
  build more advanced objects: \eg, a barrier, a ring buffer, a
  linearisable key-value store, a transactional locking scheme, and a
  distributed DC/DC converter. We will release the code under an
  open-source licence upon publication of this paper.

\item We define a new consistency model, \rdmawait, modelling LOCO's
  Wait operation, allowing the CPU to wait for the confirmation (by
  the NIC) for a \emph{specific} group of remote operations. We verify
  the correctness of the LOCO's Wait implementation w.r.t. the existing
  \rdmatso model.

\item We introduce a new modular formal framework, \framework,
  for specifying and verifying concurrent libraries over
  weakly consistent memory and distributed architectures.

\item We instantiate \framework to verify correctness of the aforementioned
  LOCO libraries.

\item We benchmark LOCO's barrier and ring buffer objects against
  the highly tuned OpenMPI implementations. LOCO's verified
  barrier is slower than the OpenMPI one, whereas LOCO's verified
  ring buffer consistently outperforms OpenMPI as the
  number of broadcasts increase.
\end{itemize}

%% file: overview.tex
\section{Overview of LOCO and MOWGLI}
\label{sec:ov}

In this section, we provide an informal, more detailed overview of LOCO and
\framework.
We present LOCO's base memory model, \rdmawait, in \cref{sec:ov-rdmawait},
then discuss the key libraries that we consider.
In \cref{sec:ov-framework}, we provide an overview of our \framework
verification framework.

\subsection{The \rdmawait Memory Model }
\label{sec:ov-rdmawait}

\input{poll_vs_wait}

We start by informally describing LOCO's base memory model, \rdmawait,
and contrast it to \rdmatso~\cite{OOPSLA-24}
via a set of simple examples.
The main difference between \rdmatso
and \rdmawait is the inter-node synchronisation technique, which is
achieved through polling in \rdmatso and via a wait command in
\rdmawait.

\begin{wrapfigure}[15]{r}{0.2\columnwidth}
  \vspace{-10pt}
  \centering
    \begin{minipage}[t]{.2\columnwidth}
      \small
        \centering
\begin{tabular}{|@{\hspace{3pt}}c @{\hspace{3pt}} || @{\hspace{3pt}}c@{\hspace{3pt}}|}
\hline
$y\!=\!0$
& $x\!=\!0$\\
\hline
$\inarr{
  \neqn{x} \assign^d 1 \\
  \rlwait{(d)} \\
  a \assign y
}$ &
$\inarr{
  \neqn{y} \assign^e 1 \\
  \rlwait{(e)} \\
  b \assign x
}$ \\
\hline
\end{tabular}
$$(a,b)=(0,0)\ \text{\checkyes}$$
\smallskip

\begin{tabular}{|@{\hspace{3pt}}c @{\hspace{3pt}} || @{\hspace{3pt}}c@{\hspace{3pt}}|}
\hline
$y,w\!=\!0$
& $x,z\!=\!0$\\
\hline
$\inarr{
  \neqn{x} \assign 1 \\
  c \assign^d \neqn{z} \\
  \rlwait{(d)} \\
  a \assign y
}$ &
$\inarr{
  \neqn{y} \assign 1 \\
  d \assign^e \neqn{w} \\
  \rlwait{(e)} \\
  b \assign x
}$ \\
\hline
\end{tabular}
$$(a,b)=(0,0)\ \text{\checkno}$$
\end{minipage}

\vspace{-5pt}

\caption{Preventing \\ RDMA store buffering}
\label{fig:wait-sb}
\end{wrapfigure}

To illustrate this difference, consider the \rdmatso programs in
\cref{fig:ex-rdmatso}. The program in \cref{subfig:rempollcpu}
comprises two nodes, with a variable $x$ in the left node (which we
call node $1$) and a variable $z$ in the right node (which we call
node $2$). Node 1 comprises a single thread that first puts the
value of $x$ to the remote location $z$ (located in node $2$).
After performing the put operation,
node $1$ performs a poll of node $2$, which effectively causes the
thread to wait until the put has been executed. Finally, it
updates $x$ to $1$. This means that the final value of $z$ is $0$, and
not $1$.  Note that in the absence of the \rtsopoll operation, the
final outcome $z = 1$ would be permitted since the instruction
$\neqn{z} \assign x$ could simply be offloaded to the NIC, followed by
the update of $x$ to $1$. When $\neqn{z} \assign x$ is later executed
by the NIC, it will load the value $1$ for $x$.

Synchronisation via \rtsopoll is however brittle, and sensitive to the
number of instructions occurring before the \rtsopoll. For example, as
shown in \cref{subfig:remrempollcpu} the final outcome $z = 1$ is once
again permitted because the \rtsopoll only waits for the earliest
remote operation that has not been polled to be executed in the remote
node. In particular, although \rtsopoll does wait for the first put
instruction, the second put may be offloaded to the NIC and the
local write $x \assign 1$ executed before the second put
($\neqn{z} \assign x$) is executed. To fix this, one requires a second
\rtsopoll operation as shown in \cref{subfig:remrempollpollcpu}.

Synchronisation via \rlwait proceeds as shown by the programs in
\cref{fig:ex-rdmawait}. In \rdmawait, puts are associated
with a work identifier, \eg $d$ in \cref{subfig:wait1}, which is
combined with a \rlwait{} operation to provide synchronisation. Thus,
unlike \rtsopoll, which waits for the first unpolled operation,
\rdmawait is able to wait for a specific put or get
operation.
Compared to \rdmatso, this improves robustness since the
\rlwait{} is independent of the number of instructions that have been
executed by each thread. For example, in \cref{subfig:wait2}, the
\rlwait{} can target the {\em second} put instruction using the work
identifier, which ensures that the unintended final outcome $z = 1$ is
not possible.

While \rlwait{} makes targeting a remote operation easier, it does not provide
more synchronisation guarantees than the \rtsopoll operation. In particular,
waiting for a put operation ($\neqn{z} \assign x$) does not guarantee that the
remote location $z$ has been modified, but only that the local value of $x$ has
been read. As shown at the top of \cref{fig:wait-sb}, it means the store
buffering behaviour across nodes is possible even if we wait for every remote
operation.

However, waiting for a get operation ($x \assign \neqn{z}$) does guarantee it
has fully completed, \ie that $z$ has been read and $x$ modified. This can be
exploited to prevent the store buffering behaviour. RDMA ordering rules ensure
that later gets execute after previous puts towards the same remote node. Thus,
waiting for a (seemingly unrelated) get operation can be used to ascertain the
completion of previous remote writes, as shown at the bottom of
\cref{fig:wait-sb}.

We present the formal definitions of \rdmawait in \cref{sec:rdmawait}
using a declarative style. Although, like \rdmatso, it is also
possible to derive an equivalent operational model, we elide these
details since the proof technique that we use (see
\cref{sec:ov-framework}) directly uses the declarative semantics.

\subsection{LOCO Libraries}
\label{sec:ov-rdma}

LOCO provides a set of commonly used distributed objects,
which we call \emph{channels}, built on top of \rdmawait.
Channels are \emph{named} and \emph{composable}.
To communicate over a channel, each participating node constructs a local
channel object, or \emph{channel endpoint}, with the same name. Each channel
endpoint allocates zero or more named local regions of network memory when
constructed, and delivers the metadata necessary to access these local memory
regions to the other endpoints during the setup process.
Moreover, each channel contains zero or more sub-channels, which are namespaced
under their parent. This feature is used to easily compose channel
functionality.

Channels make it significantly easier to develop RDMA applications and prove
their correctness, for minimal performance loss.
A LOCO application will usually consist of many channels (objects)
of many different channel types (classes).
In addition, each channel can itself instantiate member sub-channels.
For instance, a key-value store might include several mutexes as sub-channels
to synchronise access to its contents.

\begin{wrapfigure}[7]{r}{0.18\columnwidth}
  \vspace{-15pt}
  \begin{minipage}[t]{.18\columnwidth}
  \centering
  \small
  \begin{tabular}{|@{\hspace{2pt}}c @{\hspace{2pt}} || @{\hspace{2pt}}c @{\hspace{2pt}}|}
    \hline
    $y = 0$ & $x = 0$ \\
    \hline
    $\inarr{\phantom{a} \vspace{-6pt} \\
            \neqn{x} \assign 1 \\ \brlgf(2) \\ a \assign y \\
            \phantom{a} \vspace{-6pt}}$
    & $\inarr{\phantom{a} \vspace{-6pt} \\
            \neqn{y} \assign 1 \\ \brlgf(1) \\ b \assign x \\
            \phantom{a} \vspace{-6pt}}$
    \\
    \hline
  \end{tabular}

  $(a,b)=(0,0)\ \text{\checkno}$

  \vspace{-5pt}

  \caption{Using \brlgf}
  \label{fig:gf-example}
\end{minipage}
\end{wrapfigure}

\paragraph{\bfseries Shared Variable Library (\brl, \cref{sec:brlib})}

One of the most basic components of LOCO is the \emph{shared variable}
library. Each shared variable is replicated
across all (participating) nodes in the network and supports
$\brlwrite$ and $\brlread$ operations, which only access the
\emph{local} copy of the variable. Any updates to the variable may be
pushed to the other replicas by the modifying node via a $\brlbr$
operation.\footnote{It is also possible for replicas to pull the new
  value from a source node when a shared variable is modified, but we
  do not model this aspect because it is not used in the libraries we
  consider. Moreover, LOCO also defines a stronger form of a shared
  variable called an {\em owned variable}, which provides a mechanism
  for defining a variable's owner that provides a single authoritative
  version of the variable (describing its true value), defining a
  single-writer multi-reader register. } We provide
examples %
in \cref{sec:ov-framework}, \cref{fig:broadcast-sync}.

The shared variable library also provides a mechanism for synchronising
different nodes using a \emph{global fence} ($\brlgf$) operation. $\brlgf$ takes the
node(s) on which the fence should be performed as a parameter and causes the
executing thread to wait until all prior operations executed by the thread
towards the given nodes have fully completed. This is stronger than using the
\rlwait{} primitive, as the global fence also ensures the remote write parts
have completed. An example program using a $\brlgf$ is the store buffering
setting given in \cref{fig:gf-example}, which disallows the final outcome
$(a, b) = (0, 0)$, but allows all other combinations for $a$ and $b$ with values
from $\{0,1\}$. As can be guessed from the similarity with \cref{fig:wait-sb},
this global fence can be implemented by submitting get operations and waiting
for them.

\begin{wrapfigure}[7]{r}{0.22\columnwidth}
  \vspace{-20pt}
  \begin{minipage}[t]{.22\columnwidth}
  \centering
  \small
  \begin{tabular}{|@{\hspace{2pt}}c @{\hspace{2pt}} || @{\hspace{2pt}}c @{\hspace{2pt}}|}
    \hline
    $y = 0$ & $x = 0$ \\
    \hline
    $\inarr{\phantom{a} \vspace{-6pt} \\
            \neqn{x} \assign 1 \\ \balbar(z) \\ a \assign y \\
            \phantom{a} \vspace{-6pt}}$
    & $\inarr{\phantom{a} \vspace{-6pt} \\
            \neqn{y} \assign 1 \\ \balbar(z) \\ b \assign x \\
            \phantom{a} \vspace{-6pt}}$
    \\
    \hline
  \end{tabular}

  $(a,b)=(1,1)\ \text{\checkyes}$

  \vspace{-5pt}

  \caption{Using \balbar}
  \label{fig:bar-example}
\end{minipage}
\end{wrapfigure}
\paragraph{\bfseries Barrier Library (\bal, \cref{sec:balib})} A commonly
used object in distributed systems is a barrier, which provides a
stronger synchronisation guarantee than global fences. All threads
synchronising on a barrier must finish their operations before
execution continues. For example, consider the program in
\cref{fig:bar-example}, which only allows the final outcome
$(a, b) = (1, 1)$ and forbids all other outcomes. Here,
nodes~$1$ and $2$ synchronise on the barrier $z$, and hence nodes $1$
and $2$ both wait until both writes to $x$ and $y$ have completed.

\paragraph{\bfseries Ring Buffer Library (\rbl, \cref{sec:rblib})}
Similarly useful is a ring buffer, which allows one to develop
producer-consumer systems. LOCO's ring buffer supports a one-to-many
broadcast, and is the most sophisticated of the libraries that we
consider.

\paragraph{\bfseries Mixed-Size Writes (\msw, \cref{sec:msw})}
The final library we consider is the mixed-size write library,
which allows safe transmission of data spanning multiple words.
Here, due to the
asynchrony between the CPU and the NIC, it is possible for corrupted
data to be transmitted that does not correspond to any write performed
by the CPU. There are multiple solutions to this problem; we consider
a simple solution that transmits a hash alongside the data.

\paragraph{\bfseries LOCO API Example}
\label{ssec:chan-ov}

\begin{wrapfigure}[16]{r}{0.46\columnwidth}
 \vspace{-10pt}
  \begin{minipage}[t]{.46\columnwidth}
  \centering
\begin{lstlisting}[basicstyle={\ttfamily\scriptsize}]
class barrier : public loco::channel {
	unsigned count;
	loco::var_array<unsigned> arr;
	public:
	void waiting() {
		// complete outstanding RDMA ops
		loco::global_fence();
		count++; // increment our counter
		arr[loco::my_node()].store(count);
		arr[loco::my_node()].push_broadcast(); //and push
		bool waiting = true;
		while(waiting){  // wait for others
			waiting = false; // to match
			for (auto& i : arr) {
				if (i.load() < count){
					waiting = true;
					break;}
} } } };
\end{lstlisting}
\vspace{-10pt}
\caption{Complete C++ code for the LOCO barrier}
\label{lst:barrier-api}
\end{minipage}
\end{wrapfigure}

As an example of the LOCO C++ API,
Figure~\ref{lst:barrier-api} shows our implementation of a barrier object,
based on~\cite{gupta-clustr-2002}. The class uses an array (\code{arr}) of
shared variables as a sub-object~\cite{jha-socc-2017,jha-tocs-2019},
demonstrating composition. As with a traditional shared memory barrier, it is
used to synchronise all participants at a certain point in execution. For each
use of the barrier, participants increment their local \code{count} variable,
then broadcast the new value to others using their index in the array. They then
wait locally, leaving the barrier only when all participants have a count in the
array not less than their own. This code is a near-complete implementation of a
single-threaded barrier in LOCO, missing only a boilerplate constructor.

\paragraph{\bfseries LOCO-Based Applications} As mentioned earlier,
LOCO enables one to quickly build distributed applications. We
demonstrate this by using LOCO to construct a linearisable key-value store (\cref{sec:applications}), a
transactional locking scheme (\cref{sec:eval}), and a distributed DC/DC converter (\cref{app:powcon}).

\subsection{Towards a Modular Verification Framework for LOCO}
\label{sec:ov-framework}

To support libraries such as LOCO, we develop a modular
verification framework that supports \rdmawait programs.
Our point of departure is the Yacovet framework~\cite{DBLP:journals/pacmpl/RaadDRLV19,DBLP:conf/esop/StefanescoRV24}
that was used to reason about weak shared memory \emph{within} a single node.
Yacovet, however, is not expressive enough to model \rdmawait programs,
and so we need to develop a framework that can take into account 
both sources of weak consistency:
shared-memory concurrency (TSO) and distribution (RDMA).
This poses three main challenges.

\paragraph{\bfseries Lack of Causality}
\rdmawait assumes the TSO memory
model~\cite{DBLP:journals/toplas/AlglaveMT14,DBLP:conf/tphol/OwensSS09}
for each CPU within each node. This means that well-known effects such
as store buffering (see \cref{fig:SB}) are possible, where both reads
in the two threads read from the initial state.  Despite this
weakness, TSO guarantees causal consistency, \ie message passing (see
\cref{fig:MP}), where the right thread reading the new value $1$ for
$y$ guarantees that it also reads $1$ for $x$. Formally, this is due
to a relation known as \emph{preserved program order} (\ppo) between the read
of $w$, the write of this value to $x$, and the write to $y$.
However, under \rdmawait, when interacting with the NIC, causal
consistency is no longer guaranteed (see \cref{fig:RMP}). This leads
to our first modelling challenge: \rdmawait has a much weaker \ppo
relation than TSO~\cite{DBLP:journals/toplas/AlglaveMT14}.
Here, compositionality is critical to ensure proofs for scalability;
we offer this through our locality result (\cref{thm:locally-sound-impl}).

\begin{figure}[t]
  \centering
  \begin{subfigure}[b]{0.3\textwidth} \small
  \centering
  \vspace{0pt}
  \scalebox{1}{\begin{tabular}{|@{\hspace{2pt}}c @{\hspace{2pt}}|}
    \hline
    $x,y = 0$ \\
    \hline
    \phantom{a} \vspace{-6pt} \\
    $\inarrII{
      x \assign 1 \\ a \assign y
    }{
      y \assign 1 \\ b \assign x
    }$
    \vspace{-6pt} \\
    \phantom{a} \\
    \hline
  \end{tabular}}
  $$(a,b)=(0,0)\ \text{\checkyes}$$
  \vspace{-15pt}
  \caption{Store buffering}
  \label{fig:SB}
\end{subfigure}
\hfill
  \begin{subfigure}[b]{0.3\textwidth} \small
  \centering
  \vspace{0pt}
  \scalebox{1}{\begin{tabular}{|@{\hspace{2pt}}c @{\hspace{2pt}}|}
    \hline
    $x,y = 0, w = 1$ \\
    \hline
    \phantom{a} \vspace{-6pt} \\
    $\inarrII{
      x \assign w \\ y \assign 1
    }{
      a \assign y \\ b \assign x
    }$
    \vspace{-6pt} \\
    \phantom{a} \\
    \hline
  \end{tabular}}
  $$(a,b)=(1,0)\ \text{\checkno}$$
    \vspace{-15pt}
  \caption{Message passing}
  \label{fig:MP}
\end{subfigure}
\hfill
  \begin{subfigure}[b]{0.3\textwidth} \small
  \centering
  \vspace{0pt}
  \scalebox{1}{\begin{tabular}{|@{\hspace{2pt}}c @{\hspace{2pt}} || @{\hspace{2pt}}c@{\hspace{2pt}}|}
    \hline
    $x,y = 0$ & $w = 1$ \\
    \hline
    \phantom{a} & \phantom{a} \vspace{-6pt} \\
    $\inarrII{
    x \assign \neqn{w} \\ y \assign 1
    }{
      a \assign y \\ b \assign x
    }$
    &
    \vspace{-6pt} \\
    \phantom{a} & \phantom{a} \\
    \hline
  \end{tabular}}
  $$(a,b)=(1,0)\ \text{\checkyes}$$
    \vspace{-15pt}
  \caption{Remote message passing}
  \label{fig:RMP}
\end{subfigure}
\vspace{-5pt}
\caption{TSO effects of \rdmawait }
  \label{fig:TSO-effects}
\end{figure}

\begin{figure}[t]
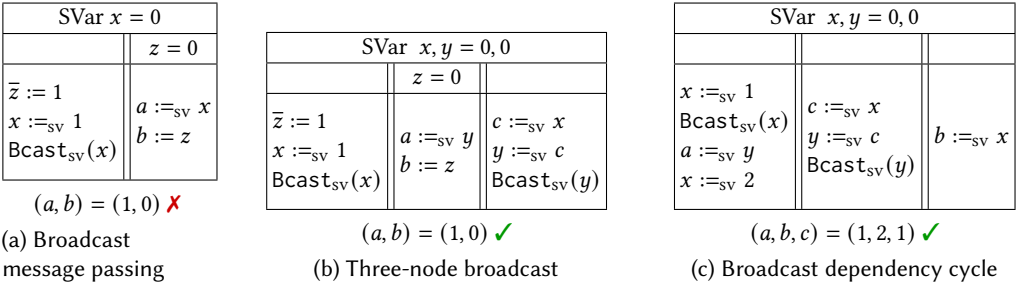


  \begin{subfigure}[b]{.2\textwidth}
  \centering
  \vspace{0pt} \small
  \scalebox{1}{\begin{tabular}{|@{\hspace{2pt}}c @{\hspace{2pt}} || @{\hspace{2pt}}c @{\hspace{2pt}}|}
    \hline
    \multicolumn{2}{|c|}{$\SVar$ $x = 0 $} \\
    \hline
    & $z = 0$ \\
    \hline
    $\inarr{\phantom{a} \vspace{-6pt} \\
            \neqn{z} \assign 1 \\ x \assignbrl 1 \\ \brlbr(x) \\
            \phantom{a} \vspace{-6pt}}$
    & $\inarr{a \assignbrl x \\ b \assign z}$
    \\
    \hline
  \end{tabular}}
  $$(a,b)=(1,0)\ \text{\checkno}$$
  \vspace{-15pt}
  \caption{Broadcast \\message passing}
   \label{subfig:bcast1}
\end{subfigure}
\qquad
\begin{subfigure}[b]{.33\textwidth}
  \centering
  \vspace{0pt} \small
  \scalebox{1}{\begin{tabular}{|@{\hspace{2pt}}c @{\hspace{2pt}} || @{\hspace{2pt}}c @{\hspace{2pt}} || @{\hspace{2pt}}c @{\hspace{2pt}}|}
    \hline
    \multicolumn{3}{|c|}{ \SVar $x, y = 0, 0$} \\
    \hline
    & $z = 0$ &   \\
    \hline
    $\inarr{\phantom{a} \vspace{-6pt} \\
            \neqn{z} \assign 1 \\ x \assignbrl 1 \\ \brlbr(x) \\
            \phantom{a} \vspace{-6pt}}$
    & $\inarr{a \assignbrl y \\ b \assign z}$
    & $\inarr{c \assignbrl x \\ y \assignbrl c \\ \brlbr(y)}$
    \\
    \hline
  \end{tabular}}
  $$(a,b)=(1,0)\ \text{\checkyes}$$
  \vspace{-15pt}
  \caption{Three-node broadcast}
  \label{subfig:bcast2}
\end{subfigure}
\qquad
\begin{subfigure}[b]{.35\textwidth}
  \centering
  \vspace{0pt} \small
  \scalebox{1}{\begin{tabular}{|@{\hspace{2pt}}c @{\hspace{2pt}} || @{\hspace{2pt}}c @{\hspace{2pt}} || @{\hspace{2pt}}c @{\hspace{2pt}}|}
    \hline
    \multicolumn{3}{|c|}{\SVar $x, y = 0, 0$} \\
    \hline
    & & \\
    \hline
    $\inarr{\phantom{a} \vspace{-6pt} \\
            x \assignbrl 1 \\ \brlbr(x) \\ a \assignbrl y \\ x \assignbrl 2 \\
            \phantom{a} \vspace{-6pt}}$
    & $\inarr{c \assignbrl x \\ y \assignbrl c \\ \brlbr(y)}$
    & $\inarr{b \assignbrl x}$
    \\
    \hline
  \end{tabular}}
  $$(a,b,c)=(1,2,1)\ \text{\checkyes}$$
  \vspace{-15pt}
  \caption{Broadcast dependency cycle}
  \label{subfig:bcast3}
\end{subfigure}
\hfill

\vspace{-5pt}
  \caption{Broadcast synchronisation}
  \label{fig:broadcast-sync}
  \vspace{-10pt}
\end{figure}

\paragraph{\bfseries Fine-Grained Synchronisation}

A second challenge in specifying RDMA libraries is that the \emph{same}
method call may interact with different library methods in different ways. To
make this problem concrete, consider a version of message passing in
\cref{subfig:bcast1}, where node $1$ updates the remote variable $z$ (located in
node $2$), and then broadcasts a new value of a shared variable $x$ to signal
that the remote value has changed. In \cref{subfig:bcast1}, when node $2$ sees
the new value of $x$, it means that the (earlier) write to $z$ must have also
taken effect. To represent this, we require that $\neqn{z} \assign 1$ happens
before $\brlbr(x)$ and that $\brlbr(x)$ happens before $a \assignbrl x$. These
orders {\em must} be part of the declarative semantics, in some shape or form,
to disallow the behaviour $(a,b)=(1,0)$.

However, naively specifying broadcast in this way is
problematic. Consider the example in \cref{subfig:bcast2}, where node
$1$ behaves as before, but the ``signal variable'' $x$ is picked up by
node $3$ and a new signal using $y$ is broadcast by node $3$. This
time, when node $2$ receives the signal on $y$ (\ie $a = 1$), there is
actually no guarantee that the write on $z$ has completed. The outcome
$(a,b)=(1,0)$ is allowed, as communication between each pair of nodes
is independent. Thus we {\em must not} have a happens-before
dependency between the write to $z$ (from node $1$) and the read on
$z$. %

For an even more precarious example, consider \Cref{subfig:bcast3},
which is a possible behaviour of LOCO's broadcast library. The final
outcome $(a,b,c)=(1,2,1)$ is only possible if node $1$
broadcasts $x = 1$ to node $2$, and $x = 2$ to node $3$ with a single
broadcast. %
The broadcast is allowed to pick up the later value $2$ since the CPU
might run the command $x \assignbrl 2$ before the NIC reads the value
of $x$.  %
As mentioned above, reading the result of a broadcast \emph{must} create %
happens-before order so that we can preclude behaviours %
like in \Cref{subfig:bcast1}. In this example, we thus need a sequence
of dependencies:
$x \assignbrl 1 \rightarrow \purple{\brlbr(x)} \rightarrow c \assignbrl x
\rightarrow y \assignbrl c \rightarrow \brlbr(y) \rightarrow a
\assignbrl~y \rightarrow x \assignbrl 2 \rightarrow \purple{\brlbr(x)}
\rightarrow b \assignbrl x$. %
This sequence seemingly contains a dependency cycle from $\purple{\brlbr(x)}$
to itself, and thus any reasonable system of dependencies on events
would not allow this valid behaviour.

We fix this apparent cycle by splitting the broadcast event into its four basic
components called subevents:
\begin{enumerate*}[label=\bfseries(\arabic*)]
\item reading $x$ to send to node $2$ (stamp \tagnlr[2]);
\item writing $x$ on node $2$ (stamp \tagnrw[2]);
\item reading $x$ to send to node $3$ (stamp \tagnlr[3]);
\item writing $x$ on node $3$ (stamp \tagnrw[3])
\end{enumerate*}.
With this we can create a more fine-grain sequence of dependencies:
$x \assignbrl 1 \rightarrow \purple{\tup{\brlbr(x), \tagnlr[2]}} \rightarrow
\purple{\tup{\brlbr(x), \tagnrw[2]}} \rightarrow c \assignbrl x \rightarrow \ldots
\rightarrow x \assignbrl 2 \rightarrow \purple{\tup{\brlbr(x), \tagnlr[3]}} \rightarrow
\purple{\tup{\brlbr(x), \tagnrw[3]}} \allowbreak \rightarrow b \assignbrl x$. For each remote node
the broadcast reads before writing, and we have a dependency between writing on
node $2$ and reading for node $3$, but this does not create a dependency cycle
at the level of the subevents and we can authorise the behaviour of
\cref{subfig:bcast3}.

Stamps are shared by all libraries and also allow us to precisely define \ppo,
\ie which pairs of effects are required to execute in order, even across
libraries. For instance in example \cref{subfig:bcast1} we have a dependency
$\tup{\neqn{z} \assign 1, \tagnrw[2]} \arr{\ppo} \tup{\brlbr(x), \tagnrw[2]}$
guaranteeing that the contents of $z$ and $x$ on node $2$ are modified in order.
However, note that this is more subtle than a dependency between events as the
location $x$ might still be read by the broadcast \emph{before} the content of $z$ is
modified,
\ie $\tup{\brlbr(x), \tagnlr[2]} \rightarrow \tup{\neqn{z} \assign 1, \tagnrw[2]} \rightarrow \tup{\brlbr(x), \tagnrw[2]}$,
as is allowed by the semantics of RDMA.

\paragraph{\bfseries Modularity}
A final challenge in developing \framework is to support modularity
through both horizontal composition
(the use of libraries in a client program) and vertical composition
(the development of libraries using other libraries as a
subcomponent). \framework presents a generic framework that is
independent of a memory model to support such proofs through a
locality theorem. It allows the simultaneous use of multiple libraries
within a single program, and defines a semantics when the
specification of a library is used in place of an
implementation. Finally, it provides local methods for proving that
a library implementation satisfies its specification.

%% file: poll_vs_wait.tex
\begin{figure}[t]
  \begin{minipage}[b]{0.63\textwidth}
  \begin{subfigure}[b]{0.27\textwidth}\small
  \vspace{0pt}
  \scalebox{1}{\begin{tabular}{|@{\hspace{3pt}}c @{\hspace{3pt}} || @{\hspace{2pt}}c@{\hspace{2pt}}|}
  \hline
  $x\!=\!0$
  & $z\!=\!0$\\
  \hline
  $\inarr{
    \neqn{z} \assign x \\ %

    \rtsopoll{(2)} \\
    x \assign 1 \\ %
    \phantom{a}\\
    \phantom{a}
  }$
  & \\
  \hline
  \end{tabular}}
  \caption{$z\!=\!0$\,\checkyes\; $z=1$\,\checkno}
  \label{subfig:rempollcpu}
  \end{subfigure}
  \quad
  \begin{subfigure}[b]{0.26\textwidth}\small
  \vspace{0pt}
  \scalebox{1}{\begin{tabular}{|@{\hspace{3pt}}c @{\hspace{3pt}} || @{\hspace{2pt}}c@{\hspace{2pt}}|}
  \hline
  $x\!=\!0$
  & $z\!=\!0$\\
  \hline
  $\inarr{
    \neqn{z} \assign x \\ %
    \neqn{z} \assign x \\ %
    \rtsopoll{(2)} \\
        x \assign 1\\
        \phantom{a}

  }$
  & \\
  \hline
  \end{tabular}}
  \caption{$z\!=\!0$\,\checkyes\;  $z\!=\!1$\,\checkyes}
  \label{subfig:remrempollcpu}
\end{subfigure} \quad
  \begin{subfigure}[b]{0.26\textwidth} \small
  \vspace{0pt}
  \scalebox{1}{\begin{tabular}{|@{\hspace{3pt}}c @{\hspace{3pt}} || @{\hspace{2pt}}c@{\hspace{2pt}}|}
  \hline
  $x\!=\!0$
  & $z\!=\!0$\\
  \hline
  $\inarr{
    \neqn{z} \assign x \\ %
    \neqn{z} \assign x \\ %
    \rtsopoll{(2)} \\
    \rtsopoll{(2)} \\
    x \assign 1
  }$
  & \\
  \hline
  \end{tabular}}
  \caption{$z\!=\!0$\,\checkyes\; $z\!=\!1$\,\checkno}
  \label{subfig:remrempollpollcpu}
\end{subfigure}
\vspace{-5pt}
\caption{Polling under \rdmatso}
\label{fig:ex-rdmatso}
\end{minipage}
\hfill
\begin{minipage}[b]{0.36\textwidth}
    \begin{subfigure}[b]{0.46\textwidth} \small
      \vspace{0pt}
      \scalebox{1}{
  \begin{tabular}{|@{\hspace{3pt}}c @{\hspace{3pt}} || @{\hspace{2pt}}c@{\hspace{2pt}}|}
  \hline
  $x\!=\!0$
  & $z\!=\!0$\\
  \hline
  $\inarr{
    \neqn{z} \assign^d x \\ %

    \rlwait{(d)} \\
    x \assign 1 \\ %
    \phantom{a}\\
    \phantom{a}
  }$
  & \\
  \hline
  \end{tabular}}
  \caption{$z\!=\!0$\,\checkyes\; $z=1$\,\checkno}
  \label{subfig:wait1}
  \end{subfigure}
  \quad
  \begin{subfigure}[b]{0.45\textwidth}\small
  \vspace{0pt}\scalebox{1}{
  \begin{tabular}{|@{\hspace{3pt}}c @{\hspace{3pt}} || @{\hspace{2pt}}c@{\hspace{2pt}}|}
  \hline
  $x\!=\!0$
  & $z\!=\!0$\\
  \hline
  $\inarr{
    \neqn{z} \assign^e x \\ %
    \neqn{z} \assign^d x \\ %
    \rlwait{(d)} \\
        x \assign 1\\
        \phantom{a}

  }$
  & \\
  \hline
  \end{tabular}}
  \caption{$z\!=\!0$\,\checkyes\; $z\!=\!1$\,\checkno}
  \label{subfig:wait2}
\end{subfigure} %

\vspace{-5pt}
\caption{Waiting under \rdmawait}
\label{fig:ex-rdmawait}
\end{minipage}

\end{figure}

%% file: metalanguage.tex


\section{The \framework Framework and the Shared Variable Library}
\label{sec:ml}

In this section we define \framework's meta-language and general
theory for modelling weak memory libraries, as well as its notion of
compositionality that enables modular proofs. We note that our
language and theory is generic and could be applied to other memory
models.  We present the syntax and semantics of \framework in
\cref{sec:syntax-semantics} and model for formalising libraries in
\cref{sec:libraries}. Throughout the section, we use the shared
variable library (\brl) as a running example and define its
consistency in \cref{sec:brlib}. Then we present library abstraction
in \cref{sec:abstraction} and our main locality result in
\cref{sec:locality}.



\subsection{Syntax and Semantics}
\label{sec:syntax-semantics}

In this section, we present the syntax and semantics of our basic
programming language. Our language is inspired by
Cminor~\cite{DBLP:conf/tphol/AppelB07} and
Yacovet~\cite{DBLP:conf/esop/StefanescoRV24}.

\paragraph{\bfseries Programs}

We assume a type $\Val$ of values, a type $\Loc \suq \Val$ of
locations\footnote{In \framework, every argument of a method call is a value.
  Thus identifiers $(x,y,\ldots)$ are called ``locations'' by the libraries but
  are seen as values by the meta-language.}, and a type $\Method$ of methods.
The syntax of sequential programs is given by the following grammar:
\begin{align*}
  v, v_i \in \Val \qquad \quad & m \in \Method  \qquad\quad \progf \in \Val \rightarrow \Progp \qquad\quad k \in \mathbb{N}^+
  \\
  \Progp \ni \progp & ::= v
  \mid m(v_1,\ldots,v_k)
  \mid \LetF{\progp}{\progf}
  \mid \code{loop} \ \progp
  \mid \code{break}_k \ v
\end{align*}
A method call is parameterised by a sequence of input values. In later
sections, we will instantiate $m$ to basic operations such as read and
write, as well as operations corresponding to method calls of a
high-level library.

For a function $\progf$ mapping values to sequential programs, the
syntax $\LetF{\progp}{\progf}$ denotes the execution of $\progp$ with
an output that is then used as an input for $\progf$.
This constructor is a generalisation of the more standard let-in syntax, and for
a program $\progp_2$ with a free meta-variable $x$ we can define
$\LetC{x}{\progp_1}{\progp_2}$ as
$\LetF{\progp_1}{(\lambda v. \progp_2[x := v])}$. We can also model sequential
composition, \ie $\progp_1 ; \progp_2$, as syntactic sugar for
$\LetF{\progp_1}{(\lambda \_.\ \progp_2)}$ using a constant function that
discards its input.
The syntax $\LetF{\progp}{\progf}$ also allows programs to perform branching and
pattern-matching, via a function mapping different kinds of values to different
continuations. In particular, $\ITE{v}{\progp_1}{\progp_2}$ can be taken as
syntactic sugar for
$\LetF{v}{ \set{\true \mapsto \progp_1 , \false \mapsto \progp_2 } }$.

Finally, our syntax includes $\code{loop} \ \progp$ that infinitely executes the
program $\progp$, as well as the $\code{break}_k \ v$ construct which exits $k$
levels of nested loops and returns $v$. While uncommon, these constructs can be
used to define usual \code{while} and \code{for} loops.

We assume top-level concurrency. We assume a fixed number $T$ of
threads and let $\Threads \defeq \{1, 2, \dots, T\}$ be the set of all
threads. A concurrent program is thus given by a tuple
$\progps = \tup{\progp_1,\ldots,\progp_T}$, where each thread $t$
corresponds to a program $\progp_t \in \Progp$. Note that we allow
libraries to discriminate threads, and so the position of a program in
$\progps$ matters, \eg the program $\tup{\progp_1,\ldots,\progp_T}$ is
\emph{not} equivalent to $\tup{\progp_T,\ldots,\progp_1}$. For
instance, a pair of RDMA threads have different interactions depending
on whether they run on the same node or not.

\begin{example}[Shared Variables]
  \label{ex:sv}
  For our RDMA libraries, we assume a set of nodes, $\Nodes$, of fixed
  size. Each thread $\threadt$ is associated to a node
  $\nodefun{\threadt}$. The \brl library uses the following methods:
  \begin{align*}
    m(\vect{v}) & ::=
                  \brlwrite(x,v)
                  \mid \brlread(x)
                  \mid \brlbr(x,d,\set{\node_1,\ldots,\node_k})
                  \mid \brlwait(d)
                  \mid \brlgf(\set{\node_1,\ldots,\node_k})
  \end{align*}

  $\brlwrite(x,v)$ writes a new value $v$ to the location $x$ of the current node.
  $\brlread(x)$ reads the location $x$ of the current node and returns its value.
  $\brlbr(x,d,\set{\node_1,\ldots,\node_k})$ broadcasts the local value of $x$ and
  overwrites the values of the copies of $x$ on the nodes $\set{\node_1,\ldots,\node_k}$, which might include the local node.
  $\brlwait(d)$ waits for previous broadcasts of the thread marked with the same
  work identifier $\wid \in \Wid$. As mentioned in the overview, this operation
  only guarantees that the local values of the broadcasts have been read, but not
  that remote copies have been modified. Finally, the global fence operation
  $\brlgf(\set{\node_1,\ldots,\node_k})$ ensures every previous operation of the thread
  towards one of the nodes in the argument is fully finished, including the writing
  part of broadcasts.
\end{example}
\paragraph{\bfseries Plain Executions}
The semantics of a program is given by an execution, which is a graph
over events. Each event has a label taken from the set
$\Act \defeq \Method\times \Val^* \times \Val$, \ie a triple
comprising the method, the input values, and the output value.
Labels are used to define events, which are elements of the set
$\Events \defeq \Threads \times \EId \times \Act$, where
$\EId \defeq \mathbb{N}$. For each event
$\tup{\threadt,\aident,\act} \in \Events$, we have that
$\threadt \in \Threads$ is the thread that executes the label
$\act \in \Act$, and $\aident$ is a unique identifier for the event.
For an event $\action = \tup{\threadt,\aident,\act}$, we note
$\threadfun{\action} \defeq \threadt$.

\begin{definition}
  We say that $\tup{E, \po}$ is a \emph{plain execution} iff $E \suq \Events$,
  $\po \subseteq E \times E$, and \linebreak
  $\po = \bigcup_{\threadt \in \Threads} \po\rst{t}$ where every $\po\rst{t}$ (\ie
  $\po$ restricted to the events of thread $\threadt$) is a total order.
\end{definition}
Here, $\po$ represents {\em program order} \ie
$\tup{\ev_1, \ev_2} \in \po$ iff $\ev_1$ is executed before $\ev_2$ by
the same thread.

We write $\emptyset_G \defeq \tup{\emptyset, \emptyset}$ for the empty
execution and $\set{\ev}_G \defeq \tup{\set{\ev}, \emptyset}$ for the
execution with a single event $\ev$.  Given two executions,
$G_1 = \tup{E_1, \po_1}$ and $G_2 = \tup{E_2 , \po_2}$, with disjoint
sets of events (\ie $E_1 \cap E_2 = \emptyset$), we define their
sequential composition, $G_1 ; G_2$ , by ordering all events of $G_1$
before those of $G_2$. Similarly, we define their parallel
composition, $G_1 {\parallel} G_2$, by taking the union of $G_1$ and
$G_2$. That is,
\begin{align*}
  G_1 ; G_2 & \defeq \tup{E_1 \cup E_2, \po_1 \cup \po_2 \cup (E_1 \times E_2 )} & G_1 {\parallel} G_2 & \defeq \tup{E_1 \cup E_2 , \po_1 \cup \po_2} 
\end{align*}
The plain semantics of a program $\progp$ executed by a thread $t$ is given by
$\interpt{\progp}$, which is a set of pairs of the form $\tup{r, G}$, where $r$
is the output and $G$ is a plain execution. This set represents all conceivable
unfoldings of the program into method calls, even those that will be rejected by
the semantics of the corresponding libraries. Each output is %
a pair
$\tup{v, k}$, where $v$ is a value and $k$ a break number, indicating the
program terminates by requesting to exit $k$ nested loops and returning the
value $v$.%
\begin{align*}
  \interpt{v} \defeq & \set{\tup{ \tup{v,0}, \emptyset_G }} \qquad \qquad \interpt{\texttt{break}_k \ v} \defeq \set{\tup{ \tup{v,k}, \emptyset_G }}
  \\
  \interpt{m(\vect{v})} \defeq & \setpred{\tup{ \tup{v',0}, \set{\tup{\threadt,\aident,\tup{m,\vect{v},v'}}}_G }}{v' \in \Val \ \land\  \aident \in \EId}
  \\
  \interpt{\LetF{\progp}{\progf}} \defeq & \inarrTTT{\setpred{\tup{r, G_1;G_2}}{\tup{\tup{v,0}, G_1} \in \interpt{\progp} \ \land \ \tup{r, G_2} \in \interpt{\progf \ v}} \\[2pt]
   \ \cup \setpred{\tup{\tup{v,k}, G_1}}{ \tup{\tup{v,k}, G_1} \in \interpt{\progp} \ \land \ k \neq 0}} \\
  \interpt{\texttt{loop} \ \progp} \defeq & \bigcup_{j \in \mathbb{N}} \setpred{\tup{\tup{v,k}, G_0;\ldots;G_j}}{\begin{matrix}
    (\forall 0 \leq i < j. \ \tup{\tup{\_,0}, G_i} \in \interpt{\progp})
\ \land \ \tup{\tup{v,k+1}, G_j} \in \interpt{\progp} 
  \end{matrix}}
\end{align*}

The execution of a value $v$ simply returns $\tup{v,0}$ with an empty graph.
Similarly, the execution of $\texttt{break}_k \ v$ returns $\tup{v,k}$ with a
non-zero break number and an empty graph.

The plain semantics of $\interpt{m(\vect{v})}$ considers every value $v'$ as a
possible output of the method call. For each, we can create a graph $G$ with a
single event $\tup{\threadt,\_,\tup{m,\vect{v},v'}}$, and the corresponding output
for the program is then $\tup{v',0}$ with a break number of $0$.

The execution of $\LetF{\progp}{\progf}$ has two kinds of plain semantics.
Either the execution of $\progp$ requests a break, \ie
$\tup{\tup{v,k}, G_1} \in \interpt{\progp}$ with $k \neq 0$, in which case
$\LetF{\progp}{\progf}$ breaks as well with the same output. Or $\progp$
terminates with a break number of zero, and the output value $v$ of $\progp$ is
given to $\progf$. In this second case, the plain execution of
$\LetF{\progp}{\progf}$ is the sequential composition of the plain executions
for $\progp$ and $(\progf\ v)$, and its output value is the one of
$(\progf\ v)$.

Finally, the execution of $\texttt{loop} \ \progp$ can be unfolded and
corresponds to the execution of $\progp$ any number $j+1$ of times. The first
$j$ times, $\progp$ returns without requesting a break and its output value is
ignored. The $(j+1)^\text{th}$ execution of $\progp$ returns a value $v$ and
break number $k+1$, and $\texttt{loop} \ \progp$ propagates $\tup{v,k}$ with a
decremented break number. The plain execution of the loop is then the sequential
composition of the plain executions of the $j+1$ iterations of $\progp$.

We lift the plain semantics to the level of concurrent programs and define
\[
  \interp{\progps} \defeq \setpred{\tup{\tup{v_1,\ldots,v_T},\parallel_{t \in \Threads} G_\threadt}}
{\begin{matrix}
  \forall t \in \Threads. \tup{\tup{v_\threadt,0},G_\threadt} \in
  \interpt{\progp_t}
\end{matrix}}
\]
Concurrent programs only properly terminate if each thread terminates with a
break number of $0$. In which case, the output of the concurrent program is the
parallel composition of the values and plain executions of the different
threads.

\paragraph{\bfseries Executions} We generate executions from plain executions by
\begin{enumerate*}
\item extending the model with subevents, then 
\item introducing additional relations describing synchronisation and
  happens-before order.
\end{enumerate*}
We will later define consistency conditions for executions in the
context of libraries.

We assume a fixed %
set of stamps, $\Tags = \set{\tagt_1, \ldots}$, and a relation
$\tagppo \suq \Tags \times \Tags$. We will use stamps to define
subevents and $\tagppo$ to define preserved program order over
subevents within an execution.

\begin{definition}
  \label[definition]{def:execution}
  We say that $\tup{E, \po, \gettags, \so, \hb}$ is an
  \emph{execution} iff each of the following holds:
  \begin{itemize}
    
  \item $\tup{E, \po}$ is a plain execution. 
  \item $\gettags : E \rightarrow \mathcal{P}(\Tags)$ is a function
    that associates each event with a non-empty set of stamps and
    induces a %
    set of \emph{subevents},
    $ \SEvents \defeq \setpred{\tup{\action,\tagt}}{\action \in E \land
      \tagt \in \gettags(\action)}$.

  \item $\so \suq \SEvents \times \SEvents$ and
    $\hb \suq \SEvents \times \SEvents$ are relations on $\SEvents$
    defining \emph{synchronisation order} and \emph{happens-before
      order}, respectively.
  \end{itemize}
\end{definition}

To define consistency, we must ultimately relate $\po$, $\so$, and $\hb$.
However, in many weak memory models such as RDMA, including all of \po into \hb
is too restrictive.
We therefore make use of a weaker relation called
\emph{preserved program order}, $\ppo \suq \SEvents \times \SEvents$, which we
derive from $\po$ and $\tagppo$ as follows:
$$\ppo \defeq
{\setpred{\tup{\tup{\action_1,\tagt_1},\tup{\action_2,\tagt_2}}}{
    \tup{\action_1,\action_2} \in \po \land \tagt_1 \in \gettags(\action_1)
    \land \tagt_2 \in \gettags(\action_2) \land \tup{\tagt_1,\tagt_2} \in
    \tagppo}}$$

\input{tags}

For our RDMA libraries, we define 11 kinds of stamps. We have \tagcread
representing a CPU read; \tagcwrite representing a CPU write; \tagcas for an
atomic read-modify-write operation; \tagmf for a TSO memory fence; \tagwait for
a \code{wait} operation; \tagnlr for a NIC local read; \tagnrw for a NIC remote
write; \tagnrr for a NIC remote read; \tagnlw for a NIC local write; \tagnf for
a NIC remote fence; and \taggf for a global fence operation. The last 6 are
families of stamps, as we create a different copy for each node
$\node \in \Nodes$.

The stamp order $\tagppo$ we use is defined in \cref{fig:to}. We note \cyes when
two stamps are ordered, \cno when they are not ordered, and \cqp when they are
ordered iff they have the same index node. For instance, the \cno in cell B1
indicates that when a CPU write is in program order before a CPU read, there is
no ordering guarantee between the two operations, as we assume the CPUs follow
the TSO memory model, and the read might execute first.

\begin{example}[\ppo for Shared Variables]
  For the \brl library, we use the stamping function $\gettagsbrl$:
  \begin{align*}
    \gettagsbrl(\tup{\_, \_, \tup{\brlwrite, \_, \_}}) & = \set{\tagcwrite}
    \\
    \gettagsbrl(\tup{\_, \_, \tup{\brlread, \_, \_}}) &= \set{\tagcread}
    \\
    \gettagsbrl(\tup{\_, \_, \tup{\brlwait, \_, \_}}) &= \set{\tagwait}
    \\
    \gettagsbrl(\tup{\_, \_, \tup{\brlgf, (\set{\node_1,\ldots,\node_k}), \_}}) &=
                                                                                  \set{\taggf[\node_1] ,\ldots ,\taggf[\node_k]}
    \\
    \gettagsbrl(\tup{\_, \_, \tup{\brlbr, (\_, \_, \set{\node_1,\ldots,\node_k}), \_}}) &=
                                                                                          \set{\tagnlr[\node_1] , \tagnrw[\node_1] , \ldots ,\tagnlr[\node_k] , \tagnrw[\node_k]}
  \end{align*}
  Broadcasts are associated with a NIC local read and NIC remote write stamp for
  each remote node they are broadcasting towards. Similarly, global fence
  operations are associated with a global fence stamp for each node.

  With this, the stamp order is enough to enforce the behaviour of the global
  fence. If we have a program
  $\brlbr(x,\wid,\set{\ldots,\node,\ldots}) ; \brlgf(\set{\ldots,\node,\ldots})$,
  the plain execution has two events $\action_{BR}$ and $\action_{GF}$, and the
  definitions of $\gettagsbrl$ and $\tagppo$ (cell G11 in \cref{fig:to}) imply
  $\tup{\action_{BR}, \tagnrw} \arr{\ppo} \tup{\action_{GF}, \taggf}$.
\end{example}


\subsection{Libraries}
\label{sec:libraries}

In this section, we describe how libraries and library consistency are
modelled in our framework.
\begin{definition}
  \label[definition]{def:library}
  We say that a triple $\tup{M,\loc,\cons}$ is a  \emph{library} iff each of the following holds. %
  \begin{enumerate}
  \item $M \subseteq \Method $ is a set of \emph{methods}. 
  \item $\loc : \Events\rst{M} \rightarrow \mathcal{P}(\Loc)$ is a function
    associating each method call to a set of locations accessed by the method
    call.
  \item $\cons$ is a {\em consistency predicate} over executions,
    respecting the following two properties.
    \begin{itemize}
    \item Monotonicity: If
      $\tup{E, \po, \gettags, \so, \hb} \in \cons$ (\ie is
      consistent), and $(\ppo \cup \so)^+ \suq \hb' \suq \hb$, then
      $\tup{E, \po, \gettags, \so, \hb'} \in \cons$.
    \item Decomposability: If
      $\tup{(E_1 \uplus E_2), \po, \gettags, \so, \hb} \in \cons$ and
      $\loc(E_1) \cap \loc(E_2) = \emptyset$, then
      $\tup{E_1, \po\rst{E_1}, \gettags\rst{E_1}, \so\rst{E_1},
        \hb\rst{E_1}} \in \cons$.
    \end{itemize}
  \end{enumerate}
\end{definition}

\gatxt{<updated>}

Usually, including for all of the examples considered in this paper, the
locations accessed by a method call are a subset of its arguments. \Eg, we say
that $\rlwrite(x,v)$ only accesses $x$. Monotonicity states that removing
constraints cannot disallow a behaviour; this is trivially respected by all
reasonable libraries. Decomposability states that method calls manipulating
different locations can be considered independently. Crucially, this means
combining independent programs \emph{cannot} create additional behaviours; a
prerequisite for modular verification. This holds for almost all libraries, and
usually only breaks when programs have access to meta-information (\eg the
number of instructions of the whole program).

However, decomposability does \emph{not} hold for \rdmatso. As show in
\cref{fig:ex-rdmatso}, the program
${\neqn{z} \assign x} ; \linebreak \rtsopoll{(2)}; {x \assign 1}$ does not allow
the outcome $z = 1$, while a combined program
$\progp ; \neqn{z} \assign x; \rtsopoll{(2)}; {x \assign 1}$ might, even when
$\progp$ seems independent (\ie does not use locations $z$ and $x$). This
composition problem fundamentally prevents modular verification of \rdmatso
programs. It is the reason we develop the alternative semantics of \rdmawait,
while ensuring the two semantics are as close as possible.

\gatxt{</updated>}

\paragraph{\bfseries Notation}
For a library $L$, we have
$\Events\rst{L.M} = \setpred{\tup{\_,\_,\tup{m,\_,\_}} \in \Events}{m \in
  L.M}$. We use $\Events\rst{L}$ to refer to
$\Events\rst{L.M}$. Moreover, $\loc(\action)$ is used to denote
$L.\loc(\action)$, where $L$ is the library containing $\action$ (\ie
$\action \in \Events\rst{L}$) and for $E \subseteq \Events$, we define
$\loc(E) \defeq \bigcup_{\action \in E} \loc(\action)$. From this, we
can also define the locations $\loc(\progps)$ of a program $\progps$
as
$\loc(\progps) \defeq \bigcup_{\tup{-,\tup{E,-}} \in \interp{\progps}}
\loc(E)$.

Given a relation $r$ and a set $A$, we write $r^+$ for the transitive closure of
$r$; $r^*$ for its reflexive transitive closure; $\inv r$ for the inverse of
$r$; $r\rst{A}$ for $r \cap (A \times A)$; and $[A]$ for the identity relation
on $A$, \ie $\set{\tup{a, a} \mid a \in A}$. We write $r_1; r_2$ for the
relational composition of $r_1$ and $r_2$:
$\{\tup{a, b} \mid \exists c.\, \tup{a, c} \in r_1 \land \tup{c, b} \in r_2\}$.

\paragraph{\bfseries Consistent Execution}

Two libraries are \emph{compatible} if their sets of methods are
disjoint. We use $\Lambda$ to denote a set of pairwise compatible
libraries.

\begin{definition}
  \label{def:lambda-cons}
  Let $\Lambda$ be a set of pairwise compatible libraries. An
  execution $\tup{E, \po, \gettags, \so, \hb}$ is {\em
    $\Lambda$-consistent} iff each of the following holds.
  \begin{itemize}
  \item $(\ppo \cup \so)^+ \suq \hb$ and $\hb$ is a strict partial order (\ie
    both irreflexive and transitive).
  \item $E = \bigcup_{L \in \Lambda} E\rst{L}$ and
    $\so = \bigcup_{L \in \Lambda} \so\rst{L}$.
  \item For all $L \in \Lambda$, we have
    $\tup{E\rst{L}, \po\rst{L}, \gettags\rst{L}, \so\rst{L}, \hb\rst{L}} \in
    L.\cons $. %
  \end{itemize}
\end{definition}

\noindent Although the definition of $\Lambda$-consistency allows
$\hb$ relations that are bigger than $(\ppo \cup \so)^+$, we usually
have $\hb = (\ppo \cup \so)^+$ for the program executions we are
interested in.

Given a concurrent program $\progps$ using libraries $\Lambda$, we note
$\outcome_\Lambda(\progps)$ the set of all output values of its
$\Lambda$-consistent executions.
$$\outcome_\Lambda(\progps) \defeq \setpred{\vect{v}}{ \exists \tup{E, \po,
    \gettags, \so, \hb} \ \Lambda\text{-consistent.\ }
  \tup{\vect{v},\tup{E,\po}} \in \interp{\progps} }$$

\subsection{The \rdmawait Library}
\label{sec:rdmawait}

\rdmawait is used as the lowest library of our tower of abstraction
(\cref{fig:dev}). As mentioned in \cref{sec:brlib}, it is the implementation
target for the shared variable library (\brl). It is an adaptation of \rdmatso
where the \code{poll} instruction is replaced by a more intuitive \code{wait}
operation.

The \rdmawait library uses the following 8 methods.
\begin{align*}
m(\vect{v}) & ::=
\rlwrite(x,v)
\mid \rlread(x)
\mid \rlcas(x,v_1,v_2)
\mid \rlmf() \\
& \quad
\mid \rlget(x,y,\wid)
\mid \rlput(x,y,\wid)
\mid \rlwait(\wid)
\mid \rlrf(\node)
\end{align*}

The first line covers usual TSO operations: $\rlwrite(x,v)$ is a CPU write;
$\rlread(x)$ is a CPU read; $\rlcas(x,v_1,v_2)$ is an atomic compare-and-swap
operation that overwrites $x$ to $v_2$ iff $x$ contained $v_1$, and returns the
old value of $x$; and $\rlmf()$ is a TSO memory fence flushing the store buffer.

The second line covers RDMA-specific operations: $\rlget(x,y,\wid)$ (noted
$x \assignwid \neqn{y}$ in our examples) is a get%
\footnote{In the RDMA specification, \rlget and \rlput are referred to as
  respectively ``RDMA Read'' and ``RDMA Write'' operations. We use the terms get
  and put to prevent confusion, as each of these perform both a read and a write
  subevents.}
operation with work identifier $\wid$ performing a NIC remote read on $y$ and
a NIC local write on $x$; similarly $\rlput(x,y,\wid)$ (noted
$\neqn{x} \assignwid y$) is a put operation with work identifier $\wid$
performing a NIC local read on $y$ and a NIC remote write on $x$; $\rlwait(d)$
waits for previous operations with work identifier $\wid$; and finally
$\rlrf(\node)$ is an RDMA remote fence for the communication channel towards
$\node$ that does not block the CPU.

We assume that each location $x$ is associated with a specific node
$\nodefun{x}$. From this, given $\tup{E, \po}$, there is a single valid stamping
function $\gettagsrl$. Notably we have
$\gettagsrl(\rlget(x,y,d)) = \set{\tagnrr[\nodefun{y}] ,
  \tagnlw[\nodefun{y}]}$ and
$\gettagsrl(\rlput(x,y,d)) = \set{\tagnlr[\nodefun{x}] ,
  \tagnrw[\nodefun{x}]}$. Put and get operations perform both a NIC read and a
NIC write, and as such are associated to two stamps, where the remote node can
be deduced from the location. Also, a succeeding \rlcas has a single stamp
\tagcas, while a failing \rlcas has stamps $\set{\tagmf , \tagcread}$, as it
behaves as both a memory fence (\tagmf) and a CPU read (\tagcread).

The formal semantics requires several functions and relations: $\vr$, $\vw$,
$\rf$, and $\mo$, with roles similar to the semantics of \brl (\cf
\cref{sec:brlib}), as well as the \emph{NIC-flush-order} relation \ro
representing the PCIe guarantees that NIC reads flush previous NIC writes. The
consistency predicate for \rdmawait is then stated from these relations and some
derived relations, similarly to \cref{sec:brlib}.

\subsection{Example: Consistency for Shared Variables}
\label{sec:brlib}

As mentioned in \cref{ex:sv}, \brl uses the methods
$M = \set{\brlwrite,\brlread,\brlbr,\brlwait,\brlgf}$. Since only the
method and arguments matter for the location function, we use
$\loc(m(\vect{v}))$ to denote
$\loc(\tup{\_, \_, \tup{m,\vect{v},\_}})$, where
$\loc(\brlwrite(x,\_)) = \loc(\brlread(x)) = \loc(\brlbr(x,\_,\_)) = \set{x}$
for events accessing a location $x$, and $\loc(\action) = \emptyset$ otherwise
for methods $\brlwait$ and $\brlgf$.

\paragraph{\bfseries Notation}

For a subevent $\saction$, we note $\saction.\action$ and $\saction.\tagt$ its
two components.
Given an execution $\exec = \tup{E, \po, \gettags, \so, \hb}$ and a stamp
$\tagt$, we write $\exec.\tagt$ for
$\setpred{\saction \in \exec.\SEvents}{\saction.\tagt = \tagt}$. For families,
by abuse of notation, we also write \eg $\exec.\tagnrr[]$ for
$\bigcup_{n \in \Nodes} \exec.\tagnrr$.
We extend the notation \loc to subevents by writing $\loc(\saction)$ for
$\loc(\saction.\action)$.
We define the set of \emph{reads} as
$\exec.\Read \defeq \exec.\tagcread \cup \exec.\tagcas \cup
\exec.\tagnlr[] \cup \exec.\tagnrr[]$ and \emph{writes} as
$\exec.\Write \defeq \exec.\tagcwrite \cup \exec.\tagcas \cup
\exec.\tagnlw[] \cup \exec.\tagnrw[]$.
We write
$\exec.\Write_x \defeq \setpred{\saction \in
  \exec.\Write}{\loc(\saction) = \set{x}}$ to constrain the set to
writes on a specific location $x$.
We also use $\rst{\threadt}$ to restrict a set or relation to a specific thread.
\Eg
$E\rst{\threadt} = \setpred{\action}{\action \in E \land \threadfun{\action} =
  \threadt}$ and
$\po\rst{\threadt} = [E\rst{\threadt}] ; \po ; [E\rst{\threadt}]$.

For the \brl library, we additionally define
$\exec.\Write^\node \defeq \setpred{\tup{\action,
  \tagcwrite}}{\nodefun{\threadfun{\action}} = \node} \cup \exec.\tagnrw$ as the set
of write subevents occurring on node $\node$. This includes CPU writes on the
node, as well as broadcast writes towards $\node$ from all threads. We also note
$\exec.\Write_x^\node \defeq \exec.\Write_x \cap \exec.\Write^\node$ as
expected. Similarly,
$\exec.\Read^\node \defeq \setpred{\saction}{\saction \in \exec.\Read \land
  \nodefun{\threadfun{\saction}} = \node}$ covers reads occurring on $\node$, either
by a CPU read or as part of a broadcast.

\paragraph{\bfseries Consistency} We now work towards a definition of
consistency for shared variables.
\begin{definition}  
  For an execution $\exec = \tup{E, \po, \gettagsbrl, \_, \_}$, we define
  the following:  %
\begin{itemize}
\item The \emph{value-read} function
  $\vr : \exec.\Read \rightarrow \Val$ that associates each read
  subevent with %
  the value returned, if available, \ie if
  $\action = \tup{\_, \_, \tup{\brlread, \_, v}}$, then $\vr(\action) = v$.
\item The \emph{value-written} function
  $\vw : \exec.\Write \rightarrow \Val$ that associates each write
  subevent with a value %
  $\exec$, \ie if $\action = \tup{\_, \_, \tup{\brlwrite, (\_, v), \_}}$, then
  $\vw(\action) = v$.
\item A \emph{reads-from} relation, $\rf \defeq \bigcup_{\node} \rf^{\node}$,
  where each $\rf^{\node} \subseteq \exec.\Write^\node \times \exec.\Read^\node$
  is a relation on subevents of the same location and node with matching values,
  \ie if $\tup{\saction_1, \saction_2} \in \rf^{\node}$ then
  $\loc(\saction_1) = \loc(\saction_2)$ and
  $\vw(\saction_1) = \vr(\saction_2)$.
\item A \emph{modification-order} relation
  $\mo \defeq \bigcup_{x,\node} \mo_x^{\node}$ describing the order in
  which writes on $x$ on node $\node$ reach memory.
\end{itemize}
\end{definition}

We define {\em well-formedness} for $\rf$ and $\mo$ as follows. For
each remote, a broadcast writes the corresponding read value: if
$\saction_1 = \tup{\action, \tagnlr} \in \exec.\SEvents$ and
$\saction_2 = \tup{\action, \tagnrw} \in \exec.\SEvents$, then
$\vr(\saction_1) = \vw(\saction_2)$. Each $\rf^{\node}$ is functional
on its range, \ie every read in $\exec.\Read^\node$ is related to at
most one write in $\exec.\Write^\node$. If a read is not related to a
write, it reads the initial value of zero, \ie if
$\saction_2 \in \exec.\Read^\node \land \tup{\_, \saction_2} \not\in
\rf^\node$ then $\vr(\saction_2) = 0$. Finally, each $\mo_x^{\node}$
is a strict total order on $\exec.\Write_x^{\node}$.

We further define the \emph{reads-from-internal} relation as
$\rfi \defeq [\tagcwrite] ; (\po \cap \rf) ; [\tagcread]$ (which
corresponds to CPU reads and writes using the same TSO store buffer),
and the \emph{reads-from-external} relation as
$\rfe \defeq \rf \setminus \rfi$.  As we shall see in
\cref{def:brl-consistency}, $\rfi$ does \emph{not} contribute to
synchronisation order, whereas $\rfe$ does. Moreover, given an
execution $\exec$ and well-formed %
\rf and \mo, we derive additional
relations.
\begin{align*}
  \pf & \defeq \setpred{\tup{\tup{\action_1, \tagnlr},\tup{\action_2, \tagwait}}}
{\tup{\action_1,\action_2} \in \po \wedge \left(\begin{matrix}
   \exists \wid.\ \action_1 = \tup{\_, \_, \tup{\brlbr, (\_, \_, \wid), \_}} \\
  \land \ \action_2 = \tup{\_, \_, \tup{\brlwait, (\wid), \_}}
\end{matrix}\right)} 
\\
\fr^\node & \defeq \setpred{\tup{r,w} \in \exec.\Read^\node \times \exec.\Write^\node} {\begin{matrix}
   \loc(r) = \loc(w) \\
  \land \ \left(\tup{r,w} \in (\inv{(\rf^\node)}; \mo^\node) \lor r \not\in
  \img{\rf^\node} \right)
\end{matrix}} \qquad \quad \fr \defeq \bigcup_{\node} \fr^{\node}
\\
\iso & \defeq \setpred{\tup{\tup{\action, \tagnlr},\tup{\action, \tagnrw}}} {\action = \tup{\_,
  \_, \tup{\brlbr, (\_, \_, \set{\ldots,\node,\ldots}), \_}} \in E}
\end{align*}

The \emph{polls-from} relation $\pf$ states that a $\brlwait$ operation
synchronises with the NIC local read subevents of previous broadcasts that use
the same work identifier. The \emph{reads-before} relation $\fr$ states that a read $r$ executes
before a specific write $w$ on the same node and location. This is either
because $r$ reads the initial value of~$0$, or because $r$ reads from a write
that is $\mo$-before $w$. Finally, the \emph{internal-synchronisation-order}
relation $\iso$ states that, within a broadcast, for each remote node the
reading part occurs before the writing part.

We can then define the consistency predicate $\brl.\cons$ as follows.
\begin{definition}[\brl-consistency]
  \label{def:brl-consistency}
  $\tup{E, \po, \gettags, \so, \hb}$ is \brl-consistent if:
  \begin{itemize}
  \item $\gettags = \gettagsbrl$ (defined in \cref{sec:syntax-semantics});
  \item there exists well-formed \vr, \vw, \rf, and \mo, such that
    $[\tagcread] ; (\inv{\po} \cap \fr) ; [\tagcwrite] = \emptyset$ and
    $\so = \iso \cup \rfe \cup \pf \cup \fr \cup \mo$.
  \end{itemize}
\end{definition}

It is straightforward to check that this consistency predicate
satisfies monotonicity and decomposability.  For CPU reads and writes,
we ask that \fr does not contradict the program order.  \Eg, a program
$\brlwrite(x,1) ; \brlread(x)$ must return $1$ and cannot return $0$,
even if the semantics of TSO allows for the read to finish before the
write.

There is no need to explicitly include conditions on \hb in the consistency of
the library, as the global consistency condition (\cf \cref{def:lambda-cons})
already enforces that $(\ppo \cup \so \cup \hb)^+$ is irreflexive.



\subsection{Library Implementations}
\label{sec:abstraction}

We now describe a mechanism for implementing the method calls of a
library by an implementation. Our ideas build on
Yacovet~\cite{DBLP:conf/esop/StefanescoRV24}, but have been adapted to
our setting, which comprises a much weaker happens-before relation
(based on \ppo instead of \po). In particular, \framework's notions of
implementation, soundness, and abstraction are similar to Yacovet (but
simpler), but the notion of ``local soundness'' is more complicated
due to the use of $\ppo$ and subevents.

An implementation for a library $L$ is a function
$I : (\Threads \times L.M \times \Val^*) \rightarrow \Progp$ associating every
method call of the library $L$ to a sequential program.
\begin{definition}
  We say that $I$ is \emph{well defined} for a library $L$ using $\Lambda$ iff
  for all $\threadt \in \Threads$, $m \in L.M$ and $\vect{v} \in \Val^*$, we
  have:
\begin{enumerate}
\item $L \not\in \Lambda$, and $I(\threadt,m,\vect{v})$ only calls methods of
  the libraries of $\Lambda$.
\item $\tup{\tup{-,k+1}, -} \not\in \interpt{I(\threadt,m,\vect{v})}$, \ie the
  implementation of a method call $m(\vect{v})$ cannot return with a non-zero
  break number, and thus cannot cause a loop containing a call to $m(\vect{v})$
  to break inappropriately.
\item if $\tup{\tup{v,0}, \tup{E, \po}} \in \interpt{I(\threadt,m,\vect{v})}$
  then $E \neq \emptyset$, \ie if an implementation successfully executes, it
  must contain at least one method call.
\end{enumerate}
\end{definition}

We note $\loc(I)$ the set of all locations that can be accessed by the
implementation of $I$:
$\loc(I) \defeq \bigcup_{\threadt,m,\vect{v}} \bigcup_{(-,\tup{E,-}) \in
  \interpt{I(\threadt,m,\vect{v})}} \loc(E)$.
We then define a function $\impli{\_}$ to map an implementation $I$ to a
concurrent program as follows.
\begin{align*}
  \implti{v} & \defeq v  & 
  \implti{m(v_1,\ldots,v_k)} &
    \defeq \begin{cases}
      I(\threadt,m, \tup{v_1,\ldots,v_k}) & \text{if } m \in L.M \\
      m(v_1,\ldots,v_k) & \text{otherwise} \\
    \end{cases} \\
  \implti{\texttt{loop} \ \progp} & \defeq
                                    \texttt{loop} \ \implti{\progp} & 
  \implti{\LetF{\progp}{\progf}} & \defeq
                                   \LetF{\implti{\progp}}{(\lambda v. \implti{\progf \ v})} \\
  \implti{\texttt{break}_k \ v} & \defeq
                                  \texttt{break}_k \ v & 
  \impli{\tup{\progp_1,\ldots,\progp_T}} & \defeq \tup{\impl{\progp_1}_{1,L},\ldots,\impl{\progp_T}_{T,L}} 
\end{align*}

As an example, we can define the implementation $\implbrl$ of the broadcast
library into \rdmawait. For each location $x$ of the broadcast library, we
create a location $x_\node$ for each node $\node \in \Nodes$. We also create a dummy
location per node, $\dumloc_\node$ for $\node \in \Nodes$, and we use an additional
dummy work identifier $\wid_0$.
\begin{align*}
  \implbrl(\threadt, \brlwrite, (x, v)) & \defeq \rlwrite(x_{\nodefun{\threadt}}, v) \\
  \implbrl(\threadt, \brlread, (x)) & \defeq \rlread(x_{\nodefun{\threadt}}) \\
  \implbrl(\threadt, \brlbr, (x, \wid, \set{\node_1, \ldots, \node_k})) & \defeq
  \rlput(x_{\node_1}, x_{\nodefun{\threadt}}, \wid) ; \ldots ;
  \rlput(x_{\node_k}, x_{\nodefun{\threadt}}, \wid) \\
  \implbrl(\threadt, \brlwait, (\wid)) & \defeq \rlwait(\wid) \\
  \implbrl(\threadt, \brlgf, (\set{\node_1, \ldots, \node_k})) & \defeq
  \rlget(\dumloc_{\nodefun{\threadt}}, \dumloc_{\node_1}, \wid_0) ; \ldots ;
  \rlget(\dumloc_{\nodefun{\threadt}}, \dumloc_{\node_k}, \wid_0) ; \rlwait(\wid_0)
\end{align*}
where $\set{\rlwrite,\rlread,\rlput,\rlget,\rlwait}$ are methods of the
\rdmawait library (see \cref{sec:rdmawait}).

A read/write on a thread $\threadt$ accesses the location of its node
$\nodefun{\threadt}$. A broadcast executes multiple \rlput operations. Each of them
reads the location of its node and overwrites the location of a designated node.
A wait operation works similarly to \rdmawait. Finally, a global fence executes
a \rlget operation towards each node requiring fencing, and waits for the
completion of all the \rlget operations. As mentioned in the overview, this
ensures that all previous NIC operations towards these nodes are completely
finished.

We can easily see that \implbrl is well defined, as it cannot return a break
number greater than zero, and every (succeeding) implementation generates at
least one event.

Using these definitions, we arrive at a notion of a sound
implementation, which holds whenever the implementation is a
refinement of the library specification.
\begin{definition}
  \label{def:sound}
  We say that $I$ is a \emph{sound implementation} of $L$ using
  $\Lambda$ if, for any program $\progps$ such that
  $\loc(I) \cap \loc(\progps) = \emptyset$, we have that
  $\outcome_\Lambda(\impli{\progps}) \suq
  \outcome_{\Lambda\uplus\set{L}}(\progps)$.
\end{definition}
For a concurrent program $\progps$ using methods of $(\Lambda\uplus\set{L})$,
$\impli{\progps}$ only uses methods of $\Lambda$. The implementation $I$ is
sound if the translation does not introduce any new outcomes. We can assume $I$
and $\progps$ use disjoint locations to avoid capture of location names.



\subsection{Abstractions and Locality}
\label{sec:locality}

We now work towards the modular proof technique for verifying
soundness of an implementation against a library in \framework. As is
common in proofs of refinement, we use an
{\em abstraction function}~\cite{DBLP:journals/tcs/AbadiL91} mapping
the concrete implementation to its abstract library specification.
For $f : A \rightarrow B$ and $r \suq A \times A$, we note
$f(r) \defeq \setpred{\tup{f(x),f(y)}}{\tup{x,y} \in r}$.

\begin{definition}
  \label{def:abstraction}
  Suppose $I$ is a well-defined implementation of a library $L$ using $\Lambda$,
  and that $G = \tup{E, \po}$ and $G'= \tup{E', \po'}$ are plain executions
  using methods of $\Lambda$ and $L$ respectively. We say that a surjective
  function $f : E \rightarrow E'$ abstracts $G$ to $G'$, denoted
  $\absf{L}{I}{G}{G'}$, iff
  \begin{itemize}
  \item $E\rst{L} = \emptyset$ (\ie $G$ contains no calls to the abstract
    library $L$) and $E'\rst{L} = E' $ (\ie $G'$ only contains calls to the
    abstract library $L$);
  \item $f(\po) \suq (\po')^*$ and
    $\forall \action_1,\action_2, \ \tup{f(\action_1),f(\action_2)} \in \po'
    \implies \tup{\action_1,\action_2} \in \po$; and
  \item if $\action' = \tup{\threadt,\aident,\tup{m,\vect{v},v'}} \in E'$ then
    $\tup{\tup{v',0}, G\rst{f^{-1}(\action')}} \in
    \interpt{I(\threadt,m,\vect{v})}$
  \end{itemize}
\end{definition}

Intuitively, $\absf{L}{I}{G}{G'}$ means there is some abstract concurrent
program $\progps$ on library $L$ such that $\tup{\_,G'} \in \interp{\progp}$ is
a plain execution of the abstract program,
$\tup{\_,G} \in \interp{\impli{\progp}}$ is a plain execution of its
implementation, and $G$ and $G'$ behave similarly. The abstraction function $f$
maps every event of the implementation to the abstract method call it was
created for. The second requirement states that the program order is preserved
in both directions.
The last requirement states that, for each abstract event $\action'$, its
implementation $G\rst{f^{-1}(\action')}$ behaves properly. We ask that this
subgraph be a valid plain execution of the implementation with the same output
value.

\begin{lemma}
  Given $\progps$ on library $L$ and a well-defined implementation $I$ of $L$,
  if $\tup{\vect{v}, G} \in \interp{\impli{\progps}}$ then there exists
  $\tup{\vect{v}, G'} \in \interp{\progps}$ and $f$ such that
  $\absf{L}{I}{G}{G'}$.
\end{lemma}

Finally, we can define a notion of local soundness for an implementation.
\begin{definition}
  \label[definition]{def:locally-sound}
  We say that a well defined implementation $I$ of a library $L$ is
  \emph{locally sound} iff, whenever we have a $\Lambda$-consistent execution
  $\exec = \tup{E, \po, \gettags, \so, \hb}$ and
  $\absf{L}{I}{\tup{E, \po}}{\tup{E', \po'}}$, then there exists $\gettags'$,
  $\so'$, and a concretisation function
  $g : \tup{E',\po',\gettags'}.\SEvents \rightarrow \exec.\SEvents$ such that:
  \begin{itemize}
  \item $g(\tup{\action', \tagt'}) = \tup{\action, \tagt}$ implies
    $f(\action) = \action'$ and
    \begin{itemize}
    \item For all $\tagt_0$ such that $\tup{\tagt_0, \tagt'} \in \tagppo$, there
      exists $\tup{\action_1, \tagt_1} \in \exec.\SEvents$ such that
      $f(\action_1) = \action'$, $\tup{\tagt_0, \tagt_1} \in \tagppo$, and
      $\tup{\tup{\action_1, \tagt_1}, \tup{\action, \tagt}} \in \hb^*$;
    \item For all $\tagt_0$ such that $\tup{\tagt', \tagt_0} \in \tagppo$, there
      exists $\tup{\action_2, \tagt_2} \in \exec.\SEvents$ such that
      $f(\action_2) = \action'$, $\tup{\tagt_2, \tagt_0} \in \tagppo$, and
      $\tup{\tup{\action, \tagt}, \tup{\action_2, \tagt_2}} \in \hb^*$.
    \end{itemize}
  \item $g(\so') \suq \hb$;
  \item For all $\hb'$ transitive such that $(\ppo' \cup \so')^+ \suq \hb'$ and
    $g(\hb') \suq \hb$, we have \\
    $\tup{E', \po', \gettags', \so', \hb'} \in L.\cons$, where
    $\ppo' \defeq \tup{E', \po', \gettags'}.\ppo$.
  \end{itemize}
\end{definition}

Unlike the notion of soundness (\cf \cref{def:sound}) expressed using an
arbitrary program, local soundness is expressed using an arbitrary abstraction.
It states that whenever we have an abstraction from $\tup{E, \po}$ to
$\tup{E', \po'}$ and we know the implementation $\tup{E, \po}$ has a
$\Lambda$-consistent execution $\exec$, then the abstract plain execution
$\tup{E', \po'}$ also has an $L$-consistent execution (third point) and the
implementation respects the synchronisation promises made by the abstract
library $L$ (first and second point).

To translate the synchronisation promises, we require a \emph{concretisation
  function} $g$ that maps every subevent of the abstraction to a subevent in
their implementation. The library $L$ makes two kinds of synchronisation
promises: $\tagppo$ (via stamps) and $\so'$. If we have
$\tup{\saction'_1, \saction'_2} \in \so'$ in the abstraction, then we require
that the concretisation of $\saction'_1$ synchronises with the concretisation of
$\saction'_2$, \ie we ask that $g(\so') \subseteq \hb$.

Whenever the abstraction contains a subevent of the form
$\tup{\action', \tagt'}$, the usage of the stamp $\tagt'$ carries an obligation.
The subevent promises to synchronise with \emph{any} earlier or later subevent,
not necessarily from library $L$, according to the \tagppo relation (\cf
\cref{fig:to} for RDMA). In most cases, the concretisation uses the same stamp,
\ie $g(\tup{\action', \tagt'}) = \tup{\action, \tagt}$ with $\tagt' = \tagt$,
and the property is trivially respected by the implementation with
$\tup{\action_1, \tagt_1} = \tup{\action_2, \tagt_2} = \tup{\action, \tagt}$.
Otherwise we have $\tagt' \neq \tagt$%
, and so for any earlier (\resp later) stamp $\tagt_0$ that
$\tagt'$ should synchronise with, we need to justify this synchronisation
happens in the implementation, \ie that we have
$\tup{\action_1, \tagt_1} \arr{\hb^*} \tup{\action, \tagt}$, where $\tagt_1$ can
perform the expected stamp synchronisation $\tup{\tagt_0, \tagt_1} \in \tagppo$.

An important point to note is that $\hb$ is potentially bigger than
$(\ppo \cup \so)^+$. In which case, we need to prove the result for any reasonable
$\hb'$ bigger than $(\ppo' \cup \so')^+$. Thus local soundness states that if the
implementation has a $\Lambda$-consistent execution \emph{with additional
  constraints}, then the abstraction similarly has an $L$-consistent execution
\emph{with these additional constraints}. This is required for the
implementation to work in any context, \ie for programs using $L$ in conjunction
to other libraries, as expressed by the following theorem.

\begin{theorem}
  \label{thm:locally-sound-impl}
  If a well-defined implementation is locally sound, then it is sound.
\end{theorem}
\begin{proof}
  See \Cref{thm:sound-sound}.
\end{proof}

In the case of the shared variable library, we can use this proof technique to
verify the implementation \implbrl.

\begin{theorem}
  \implbrl is locally sound, and hence \implbrl is sound.
\end{theorem}
\begin{proof}
  See \Cref{thm:brl-sound}.
\end{proof}

%% file: tags.tex
\begin{figure}

\setcounter{rowcounter}{1}
\setcounter{colcounter}{1}
\vspace{3pt}
\begin{center} \footnotesize
  \scalebox{0.95}{
\begin{tabular}{c|c|c|c|c|c|c|c|c?c|c|c|c|c|c|}
\multicolumn{3}{c}{} & \multicolumn{11}{c}{Second Stamp} \\
  \cline{2-15}
\multicolumn{1}{c|}{} & \multicolumn{3}{c|}{\multirow{3}{*}{\LARGE \tagppo}} & \multicolumn{5}{c?}{single} & \multicolumn{6}{c|}{families} \\
  \cline{5-15}
\multicolumn{1}{c|}{} & \multicolumn{3}{c|}{} & \colheader & \colheader & \colheader & \colheader
   & \colheader & \colheader & \colheader & \colheader & \colheader & \colheader & \colheader \\
\multicolumn{1}{c|}{} & \multicolumn{3}{c|}{}
                      & \tagcread & \tagcwrite & \tagcas & \tagmf & \tagwait & \tagnlr & \tagnrw & \tagnrr & \tagnlw & \tagnf & \taggf \\
  \cline{2-15}
\multirow{10}{*}{\rotatebox{90}{First Stamp}} & \multirow{5}{*}{\rotatebox{90}{single}}
   & \rowheader{\tagcread}& \cyes& \cyes & \cyes & \cyes & \cyes & \cyes & \cyes & \cyes & \cyes & \cyes & \cyes \\
  \cline{3-15}
&  & \rowheader{\tagcwrite}& \cno& \cyes & \cyes & \cyes & \cno  & \cyes & \cyes & \cyes & \cyes & \cyes & \cyes \\
  \cline{3-15}
&  & \rowheader{\tagcas} & \cyes & \cyes & \cyes & \cyes & \cyes & \cyes & \cyes & \cyes & \cyes & \cyes & \cyes \\
  \cline{3-15}
&  & \rowheader{\tagmf}  & \cyes & \cyes & \cyes & \cyes & \cyes & \cyes & \cyes & \cyes & \cyes & \cyes & \cyes \\
  \cline{3-15}
&  & \rowheader{\tagwait}& \cyes & \cyes & \cyes & \cyes & \cyes & \cyes & \cyes & \cyes & \cyes & \cyes & \cyes \\
  \Cline{1pt}{2-15}
& \multirow{6}{*}{\rotatebox{90}{families}}
   & \rowheader{\tagnlr} & \cno  & \cno  & \cno  & \cno  & \cno  & \cqp  & \cqp  & \cqp  & \cqp  & \cqp  & \cqp \\
  \cline{3-15}
&  & \rowheader{\tagnrw} & \cno  & \cno  & \cno  & \cno  & \cno  & \cno  & \cqp  & \cqp  & \cqp  & \cno  & \cqp \\
  \cline{3-15}
&  & \rowheader{\tagnrr} & \cno  & \cno  & \cno  & \cno  & \cno  & \cno  & \cno  & \cno  & \cqp  & \cqp  & \cqp \\
  \cline{3-15}
&  & \rowheader{\tagnlw} & \cno  & \cno  & \cno  & \cno  & \cno  & \cno  & \cno  & \cno  & \cqp  & \cno  & \cqp \\
  \cline{3-15}
&  & \rowheader{\tagnf}  & \cno  & \cno  & \cno  & \cno  & \cno  & \cqp  & \cqp  & \cqp  & \cqp  & \cqp  & \cqp \\
  \cline{3-15}
&  & \rowheader{\taggf}  & \cyes & \cyes & \cyes & \cyes & \cyes & \cyes & \cyes & \cyes & \cyes & \cyes & \cyes \\
  \cline{2-15}
\end{tabular}}
\end{center}
\vspace{-10pt}

\caption{Stamp order \tagppo for the RDMA libraries. Lines indicate the earlier
  stamp, columns the later. A cell marked \cyes indicates that the stamps are
  ordered, and that the \po ordering of subevents with these stamps is
  preserved. A cell marked \cno indicates that the stamps are not ordered, and
  that such subevents can execute out of order. Finally, \cqp indicates the
  stamps are ordered iff they have the same node index.}
\label{fig:to}
\end{figure}

%% file: barrier.tex
\section{Barrier Library}
\label{sec:balib}

As discussed informally in \cref{sec:ov-rdma}, LOCO implements a
barrier library (\bal), which supports synchronisation of threads
across multiple threads. Note that each barrier corresponds to a set
of threads, which we refer to as the ``participating threads'' of a
barrier. Each participating thread must wait for {\em all} operations
towards {\em all} participating threads (including its own) that are
\po-before each barrier to be completed. We first present a generic
specification for barriers with participating nodes in
\cref{sec:bal-specification}, and the LOCO barrier and its correctness
proof in \cref{sec:bal-implementation}. In
\cref{sec:strong-trans-semant} we discuss an issue with such a barrier
that only synchronises participating nodes and a possible fix.



\subsection{Generic Barrier Specification}
\label{sec:bal-specification}

The barrier library (\bal) only has the single method
$\balbar : \Loc \rightarrow ()$, taking a location as an input and producing no
output. Thus, we have $\loc(\balbar(x)) = \set{x}$.
The input location $x$ defines the set of threads that synchronise via
$\balbar(x)$. In our model, we assume a function
$\implbalb : \Loc \rightarrow \mathcal{P}(\Threads)$ associating each location
$x$ with a set of threads that perform a barrier synchronisation on $x$.

While the LOCO barrier implementation (see
\cref{sec:bal-implementation}) supports synchronisation across nodes
connected by RDMA, our specification is more general and abstracts
away the notion of nodes. Instead, our library defines synchronisation
between {\em threads}, providing freedom to implement different
synchronisation mechanisms depending on whether the threads are on the
same or on different nodes.

Since \framework allows libraries to be defined in isolation, we only
consider $E$ containing barrier calls. Let
$E_x \defeq \setpred{\action \in E}{\loc(\action) = \set{x}}$ denote
the set of barrier calls on the location
$x$.  %
\begin{definition}[\bal-consistency]
  We say that $\exec = \tup{E, \po, \gettags, \so, \hb}$ is
  \bal-consistent iff:
  \begin{itemize}
  \item $\gettags = \gettagsbal$, defined as
    $\gettagsbal(\tup{\_, \_, \tup{\balbar, (x), ()}}) =
    \bigcup_{\threadt \in \implbalb(x)} \set{\taggf[\nodefun{\threadt}]} \cup
    \set{\tagcread}$;
  \item for all $x$ and $\action \in E_x$, $\threadfun{\action} \in \implbalb(x)$;
    \ie non-participating threads do not participate;
  \item for all $x \in \Loc$, there is an integer $c_x$ such that for
    all thread $\threadt \in \implbalb(x)$ we have
    $\cardinal{E_x\rst{\threadt}} = c_x$; \ie each participating
    thread makes exactly $c_x$ calls to the barrier on
    $x$; %
  \item there is an ordering function $\ordero : E \rightarrow \mathbb{N}$ such
    that for all location $x$:
    \begin{itemize}
    \item if $\action \in E_x$ then $1 \leq \ordero(\action) \leq c_x$;
    \item if $\action_1, \action_2 \in E_x$ and
      $\tup{\action_1, \action_2} \in \po$ then
      $\ordero(\action_1) < \ordero(\action_2)$; and
    \end{itemize}
  \item
    $\so = \bigcup_{x \in \Loc} \bigcup_{1\leq i \leq c_x}
    \setpred{\tup{\tup{\action_1, \taggf}, \tup{\action_2,
          \tagcread}}}{\action_1,\action_2 \in (E_x \cap \inv{\ordero}(i))}$
  \end{itemize}
\end{definition}
\noindent 
This predicate clearly respects monotonicity (since \hb is
unrestricted) and decomposability (since each location is treated
independently).

The function $\ordero$ associates each barrier call to the number of times the
location has been used by this thread, in program order. We say that $\action_1$
and $\action_2$ \emph{synchronise together} iff
$\loc(\action_1) = \loc(\action_2)$ and
$\ordero(\action_1) = \ordero(\action_2)$. %
The stamps of the form $\taggf[]$ correspond to the \emph{entry
  points} of the barrier calls, waiting for previous operations to
finish before the synchronisation. The stamp $\tagcread$ represents
the \emph{exit point} of the barrier, after the synchronisation. The
synchronisation is then an \so ordering between $\taggf[]$ and
$\tagcread$ for barrier calls that synchronise together.



\subsection{LOCO Implementation} 
\label{sec:bal-implementation}

Given $\implbalb : \Loc \rightarrow \mathcal{P}(\Threads)$, for each location
$x$ with $\implbalb(x) = \set{\threadt_1,\ldots,\threadt_k}$ synchronising $k$
threads, we create a set of $k$ shared variables (\ie \brl locations)
$\set{x_{\threadt_1}, \ldots, x_{\threadt_k}}$. Each shared variable
$x_{\threadt}$ is used as a counter indicating how many times thread $\threadt$
has executed a barrier on $x$. The LOCO implementation decomposes the barrier
into three steps:
\begin{enumerate*}
\item wait for previous operations to finish;
\item increase your counter;
\item wait for the counters of other threads to increase.
\end{enumerate*}

We define the implementation \implbal in \cref{fig:loco-bar}. Clearly,
the implementation is well defined: it cannot return a break number
greater than zero, since all $\mathtt{break}$ commands have a break
number of~$1$ and are inside loops; and every succeeding
implementation generates at least one event.

\begin{wrapfigure}[16]{r}{0.45\linewidth}
  \vspace{-10pt}
  \scalebox{0.85}{
    \begin{minipage}{\linewidth}
      \addtolength{\jot}{-1mm}
\begin{align*} 
  \texttt{For } & \threadt \not\in \implbalb(x) \texttt{:} \ 
  \implbal(\threadt, \balbar, (x)) \defeq \texttt{loop} \ \{ () \} \\
  & \\
  \texttt{For } & \threadt \in \implbalb(x)=\set{\threadt_1,\ldots,\threadt_k} \texttt{:} \\
  \implbal(\threadt, &\ \balbar, (x)) \defeq \\
  & \LetC{s_n}{\setpred{\nodefun{\threadt_i}}{\threadt_i \in \implbalb(x)}}{} \\
  & \brlgf(s_n) ; \\
  & \LetC{v}{\brlread(x_{\threadt})}{} \\
  & \brlwrite(x_{\threadt}, v+1) ; \\
  & \brlbr(x_{\threadt}, \_, (s_n \setminus \set{\nodefun{\threadt}})) ; \\
  & \texttt{loop } \{ \\
  & \qquad \LetC{v'}{\brlread({x_{\threadt_1}})}{} \\
  & \qquad \ITE{v'>v}{\mathtt{break_1} ()}{()} \ \} ; \\
  & \ldots \\
  & \texttt{loop } \{ \\
  & \qquad \LetC{v'}{\brlread({x_{\threadt_k}})}{} \\
  & \qquad \ITE{v'>v}{\mathtt{break_1} ()}{()} \ \}
\end{align*}
\end{minipage}
} %
    \vspace{-10pt}
    \caption{\implbal implementation}
    \label{fig:loco-bar}
\end{wrapfigure}

If a method call is made by a non-participating thread, the call is
invalid and we implement it using a non-terminating loop. This is
necessary for soundness, as the outcomes of the implementation must be
valid, and in this situation the \bal specification does not allow any
valid outcomes.

If a method call is made by a participating thread $\threadt$, the
implementation starts with a global fence ensuring any previous operation
towards any relevant node is fully finished. Then, it increments its counter
$x_\threadt$ to indicate to other threads that the barrier has been reached and
executed. The value of $x_\threadt$ is immediately available to other threads on
the same node, and is made available to other participating nodes using a
broadcast. Note that the broadcast does not perform a loopback (\ie we exclude
$\nodefun{\threadt}$ from the targets), as asking the NIC to overwrite
$x_\threadt$ with itself might cause the new value of a later barrier call to be
reverted to the current value. Then, we repeatedly read the (local) values of
the other counters $x_{\threadt_i}$ and wait for each of them to indicate other
threads have reached their matching barrier call. Note that there is no reason
to wait for the broadcast to finish: the implementation on $\threadt$ might go
ahead before other threads are aware that $\threadt$ reached the barrier, but
that does not break the guarantees provided by the barrier.



\begin{theorem}
  The implementation \implbal is locally sound.
\end{theorem}
\begin{proof}
  See \Cref{thm:bal-sound}.
\end{proof}



\subsection{Supporting Transitivity}
\label{sec:strong-trans-semant}

\begin{wrapfigure}[5]{r}{0.38\columnwidth}
  \vspace{-20pt}
  \begin{minipage}[t]{.38\columnwidth}
    \small \centering
\begin{tabular}{|@{\hspace{3pt}}c @{\hspace{3pt}} || @{\hspace{3pt}}c@{\hspace{3pt}}|| @{\hspace{3pt}}c@{\hspace{3pt}}|}
\hline
& & $x = 0$ \\
\hline
$\inarr{
  \neqn{x} \assign 1 \\
  \balbar(b_1)
}$ &
$\inarr{
  \balbar(b_1) \\
  \balbar(b_2)
}$ &
$\inarr{
  \balbar(b_2) \\
  a \assign x \\
}$ \\
\hline
\end{tabular}
$$a=0\ \text{\checkyes}$$
\vspace{-20pt}

\caption{Allowed weak barrier behaviour}
\label{fig:weak-bar}
\end{minipage}
\end{wrapfigure}

The barrier semantics in \cref{sec:bal-specification} only performs a
global fence on nodes with participating threads. While this appears
intuitive and reduces assumptions about other nodes, barrier
synchronisation using such a library is {\em not} transitive. For
example, consider the program in \Cref{fig:weak-bar}. Since
$\neqn{x} \assign 1$ is an operation towards node $3$, the barrier
$\balbar(b_1)$ does not wait for it to finish, allowing $a=0$.

Such a transitive barrier can straightforwardly be obtained by
synchronising across \emph{all} nodes, instead of just
``participating'' threads. For the specification, we define
$\gettagsbal(\tup{\_, \_, \tup{\balbar, (x), ()}}) =
\bigcup_{\node \in \Nodes}\set{\taggf} \cup \set{\tagcread}$ and for the
implementation, we define
$\implbal(\threadt, \balbar, (x)) \defeq \LetC{s_n}{\Nodes}{\ldots}$.
This stronger version is the one implemented in LOCO (see
\cref{lst:barrier-api}).

%% file: ringbuffer.tex
\section{Ring Buffer Library}
\label{sec:rblib}

The ring buffer library (\rbl) provides methods for a
single-writer-multiple-reader FiFo queue for messages of any size,
where each message is duplicated as necessary and can be read once by
each reader.  Here, we present its specification
(\cref{sec:rbl-specification}), and an implementation and correctness
proof (\cref{sec:loco-implementation-1}). %



\subsection{Ring Buffer Specification}
\label{sec:rbl-specification}

The ring buffer library has two methods
$\rblsub : \Loc \times \Val^* \rightarrow \mathbb{B}$
and $\rblrec : \Loc \rightarrow \Val^* \uplus \set{\bot}$, with
$\loc(\rblsub(x,\_)) = \loc(\rblrec(x)) = \set{x}$.
$\rblsub(x,\vect{v})$ tries to add a new message $\vect{v}$ to the
ring buffer $x$. It can either fail if the ring buffer is full,
returning $\false$, or succeed returning $\true$.
$\rblrec(x)$ tries to read a message from the ring buffer $x$. It can either
succeed if there is at least one pending message, returning the next message, or
fail if there is no pending messages, returning $\bot$.

In our model, we assume two functions
$\rblthdw : \Loc \rightarrow \Threads$ and
$\rblthdr : \Loc \rightarrow \mathcal{P}(\Threads)$ associating each
location $x$ with a writing thread $\rblthdw(x)$ and a set of reader
threads $\rblthdr(x)$.  For subevents, we define the %
stamping function $\gettagsrbl$ as follows:
\begin{align*}
  \gettagsrbl(\tup{\threadt, \_, \tup{\rblsub, (x, \_), \true}}) & \defeq
  \setpred{\tagnrw[\nodefun{\threadt'}]}{\threadt' \in \rblthdr(x) \land
  \nodefun{\threadt'} \neq \nodefun{\threadt}} \cup \set{\tagcwrite}
  \\
  \gettagsrbl(\tup{\threadt, \_, \tup{\rblsub, (x, \_), \false}}) & \defeq
  \set{\tagwait} 
  \\
  \gettagsrbl(\tup{\_, \_, \tup{\rblrec, (x), \vect{v}}}) & \defeq \set{\tagcread}
  \\
  \gettagsrbl(\tup{\_, \_, \tup{\rblrec, (x), \bot}}) & \defeq \set{\tagwait}
\end{align*}
A successful call to $\rblsub$ (with return value $\true$) is denoted by a write
stamp for each relevant node: the stamp $\tagcwrite$ is used by the writer node,
and the stamps $\tagnrw[\nodefun{\threadt'}]$ are used by the corresponding
remote nodes. Failing calls (with return value $\false$ or $\bot$) are depicted by
the stamp $\tagwait$. Finally, a succeeding $\rblrec$ call uses the reading
stamp $\tagcread$.

We note different sets corresponding to calls to \rblsub succeeding
($\Write$) and calls to \rblrec failing ($\rblfailread$) or succeeding
($\Read$). Calls to \rblsub failing are ignored by the
specification.
\begin{align*}
  \Write^\node_x &\defeq \setpred{\tup{\action, \tagnrw}}{\action = \tup{\threadt, \_,
                   \tup{\rblsub,(x,\_), \true}} \in E \land \tagnrw \in \gettagsrbl(\action)}
  \\
                 & \quad \cup \setpred{\tup{\action, \tagcwrite}}{\action = \tup{\threadt, \_,
                   \tup{\rblsub,(x,\_), \true}}\in E \land \nodefun{\threadt} = \node}
  \\
  \rblfailread^\node_x &\defeq \setpred{\tup{\action, \tagwait}}{\action = \tup{\threadt,
                         \_, \tup{\rblrec,(x), \bot}} \in E \land \nodefun{\threadt} = \node}
  \\
  \Read^\node_x &\defeq \setpred{\tup{\action, \tagcread}}{\action = \tup{\threadt, \_,
                  \tup{\rblrec,(x), \vect{v}}} \in E \land \nodefun{\threadt} = \node}
\end{align*}

We then define the reads-from relation \rf matching successful \rblsub and
\rblrec events.

\begin{definition}
  Given $\exec = \tup{E, \po, \gettagsrbl, \_, \_}$, we say that \rf
  is \emph{well-formed} iff each of the following holds:
  \begin{enumerate}
  \item $\rf = \bigcup_{\node,x} \rf^\node_x$ with
    $\rf^\node_x \subseteq \Write^\node_x \times \Read^\node_x$
  \item $\rf^\node_x$ is total and functional on its range, \ie each read
    subevent in $\Read^\node_x$ is related to exactly one write subevent in
    $\Write^\node_x$.
  \item If
    $(\tup{\_, \_, \tup{\rblsub,(x,\vect{v}), \true}}, \tagt)
    \arr{\rf} (\tup{\_, \_, \tup{\rblrec,(x), \vect{v}'}}, \tagt')$
    then $\vect{v} = \vect{v}'$, \ie related events write and read the
    same tuple of values.
    
  \item If $\tup{\saction_1, \saction_2} \in \rf$,
    $\tup{\saction_1, \saction_3} \in \rf$, and $s_2 \neq s_3$, then
    $\threadfun{\saction_2} \neq \threadfun{\saction_3}$, \ie each thread can
    read each message at most once.
  \item If
    $\saction_1, \saction_2 \in \Write^\node_x$,
    $\tup{\saction_1, \saction_2} \in \po$, and $\tup{\saction_2, \saction_4} \in \rf$,
    then there is $\saction_3$ such that $\tup{\saction_1, \saction_3} \in \rf$, and
    $\tup{\saction_3, \saction_4} \in \po$, \ie threads cannot jump a message.
  \end{enumerate}
\end{definition}

We define the {\em fails-before} relation $\rblfr$ expressing that a
failing \rblrec occurs before a succeeding \rblsub as follows:
\begin{align*}
  \rblfr \defeq \bigcup_{\node,x} \left( \rblfailread^\node_x \times \Write^\node_x \setminus (\inv{\po} ; \inv{\rf}) \right)
\end{align*}
\noindent If $\saction_1 \in \Write^\node_x$ and $\saction_3 \in \rblfailread^\node_x$,
then the contents written by $\saction_1$ is not available when $\saction_3$ is
executed. Either there is $\saction_2$ such that
$\tup{\saction_1, \saction_2} \in \rf$ and $\tup{\saction_2, \saction_3} \in \po$, in
which case the message has been read; or there is no such $\saction_2$ and we have
$\tup{\saction_3, \saction_1} \in \rblfr$ to express that the message was not yet
written.

\begin{definition}[\rbl-consistency]
  \label{def:rbl-cons}
  We say that an execution
  $\exec = \tup{E, \po, \gettagsrbl, \so, \hb}$ is \bal-consistent
  iff:
  \begin{itemize}
  \item if
    $\tup{\threadt, \_, \tup{\rblsub, (x, \_), \_}} \in E$ then
    $\threadt = \rblthdw(x)$; and if $\tup{\threadt, \_, \tup{\rblrec, (x), \_}} \in E$
    then $\threadt \in \rblthdr(x)$; and 
  \item there exists a well-formed \rf such that $\so = \rf \cup \rblfr$.
  \end{itemize}
\end{definition}
Note that this definition allows the writer thread to also be a reader, and
nodes to have multiple reading threads. Moreover, the consistency predicate does
not tell us anything about failing writes; they may fail spuriously.

\paragraph{\bfseries Alternative weaker semantics}
\begin{wrapfigure}[5]{r}{0.43\columnwidth}
\vspace{-25pt}
  \centering
  \scalebox{0.9}{\begin{minipage}[t]{\linewidth}
    \begin{center}
      \begin{tabular}{|@{\hspace{2pt}}c @{\hspace{2pt}} || @{\hspace{2pt}}c @{\hspace{2pt}}|}
        \hline
        $\inarr{\phantom{a} \vspace{-10pt} \\
        a \assign \rblsub(x, 1) \\
        \brlgf(\set{\node_2}) \\
        c \assign \rblrec(y)\\
        \phantom{a} \vspace{-10pt}}$
        & $\inarr{
          b \assign \rblsub(y, 1) \\
        \brlgf(\set{\node_1})\\
        d \assign \rblrec(x)}$
        \\
        \hline
      \end{tabular}
    \end{center}
    $$(a,b,c,d)=(\true,\true,\bot,\bot)\ \text{\checkno}$$
  \end{minipage}}
  \vspace{-5pt}
\caption{Alternative ring buffer semantics}
\label{fig:alt-rb-sem}
\end{wrapfigure}
Instead of requiring $\so = \rf \cup \rblfr$, we could give an alternative %
specification with $\so = \rf$ and $\inv{\hb} \cap \rblfr = \emptyset$. The latter says
that you still cannot ignore (\rblfr) a write that you know (\hb) has finished;
but if you do ignore a write, you do not have to export the guarantee (\so) that
the write has not finished. For instance, take the litmus test in
\cref{fig:alt-rb-sem}. With the semantics in \cref{def:rbl-cons}, at least one
of the two $\rblrec$ has to succeed. With the weaker semantics, they are allowed
to both fail, even when both $\rblsub$ calls succeed.



\subsection{LOCO Implementation}
\label{sec:loco-implementation-1}

As before, we assume given the functions
$\rblthdw : \Loc \rightarrow \Threads$ and
$\rblthdr : \Loc \rightarrow \mathcal{P}(\Threads)$. We also assume an
integer $S$ representing the size of %
the ring buffer. %
We implement the ring buffer library (\rbl) using the shared variable
library (\brl). For each location $x$ with
$\rblthdr(x) = \set{\threadt_1,\ldots,\threadt_k}$ we create the
shared variable (\ie \brl locations) $x_0,\ldots,x_{S-1}$ for the
content of the buffer, as well as shared variables $h^x$ for the
writer and $h^x_{\threadt_1};\ldots;h^x_{\threadt_k}$ for the
readers. We also use a work identifier $\wid_x$.

Events that do not respect $\rblthdr$ or $\rblthdw$ are implemented using an
infinite loop (\ie $\texttt{loop } \{ () \}$), similarly to other implementations.
Otherwise, we use the implementation $\implrbl$ given in \cref{fig:implrbl},
where $\%$ represents the modulo operation.

\begin{figure}
  \scalebox{0.9}{
\begin{minipage}{.55\textwidth}\addtolength{\jot}{-0.5mm}
\begin{align*}
  & \texttt{For } \rblthdw(x) = \threadt \land \rblthdr(x) = \set{\threadt_1;\ldots;\threadt_k} \texttt{:} \\
  & \implrbl(\threadt, \rblsub, (x,\vect{v}=(v_1,\ldots,v_V)) \defeq \\
  & \LetC{s_n}{\setpred{\nodefun{\threadt_i}}{\threadt_i \in \rblthdr(x)} \setminus \set{\nodefun{\threadt}}}{} \\
  & \LetC{V}{\texttt{len}(\vect{v})}{} \\
  & \LetC{H}{\brlread(h^x)}{} \\
  & \LetC{H_1}{\brlread(h^x_{\threadt_1})}{} \\
  & \ldots \\
  & \LetC{H_k}{\brlread(h^x_{\threadt_k})}{} \\
  & \LetC{M}{\texttt{min}(\set{H_1, \ldots, H_k})}{} \\
  & \ITE{(H - M) + (V + 1) > S}{\false}{} \{ \\
  & \quad \brlwrite(x_{H \% S}, V) ; \ \brlbr(x_{H \% S}, \_, s_n) ; \\
  & \quad \brlwrite(x_{(H+1) \% S}, v_1) ; \ \brlbr(x_{(H+1) \% S}, \_, s_n) ; \\
  & \quad \ldots \\
  & \quad \brlwrite(x_{(H+V) \% S}, v_V) ; \ \brlbr(x_{(H+V) \% S}, \_, s_n) ; \\
  & \quad \brlwait(\wid_x) ; \\
  & \quad \brlwrite(h^x, H+V+1) ; \ \brlbr(h^x, \wid_x, s_n) ; \\
  & \quad \texttt{true} \ \};
\end{align*}
\end{minipage}}
\qquad
  \scalebox{0.9}{
\begin{minipage}{.43\textwidth}\addtolength{\jot}{-0.5mm}
\begin{align*}
  & \texttt{For } \threadt \not\in \rblthdr(x) \texttt{:} \\
  & \implrbl(\threadt, \rblrec, (x)) \defeq \texttt{loop} \  \{ () \} \\
  & \\
  & \texttt{For } \threadt \in \rblthdr(x) \texttt{:} \\
  & \implrbl(\threadt, \rblrec, (x)) \defeq \\
  & \LetC{H}{\brlread(h^x_\threadt)}{} \\
  & \LetC{H'}{\brlread(h^x)}{} \\
  & \ITE{H \geq H'}{\bot}{} \{ \\
  & \quad \LetC{V}{\brlread(x_{H \% S})}{} \\
  & \quad \LetC{v_1}{\brlread(x_{(H+1) \% S})}{} \\
  & \quad \ldots \\
  & \quad \LetC{v_V}{\brlread(x_{(H+V) \% S})}{} \\
  & \quad \brlwrite(h^x_\threadt, H+V+1) ; \\
  & \quad \ITE{\nodefun{\rblthdw(x)} = \nodefun{\threadt}}{()}{} \\
  & \quad \quad \{ \ \brlbr(h^x_\threadt, \_, \set{\nodefun{\rblthdw(x)}}) \ \} ; \\
  & \quad (v_1, \ldots, v_V) \ \};
\end{align*}
\end{minipage}}
\vspace{-10pt}
\caption{Implementation \implrbl of the ring buffer library into \brl}
\label{fig:implrbl}
\end{figure}

The value of $h^x$ represents the next place to write for the writing thread.
The value of $h^x_{\threadt_i}$ represents the next place thread $\threadt_i$
needs to read. If $h^x = h^x_{\threadt_i}$ then thread $\threadt_i$ is
up-to-date and needs to wait for the writer to send additional data. If the
difference between $h^x$ and $h^x_{\threadt_i}$ gets close to $S$, then the
buffer is full and the writer cannot send any more data.

In the implementation of $\rblsub$, the value $M$ represents the minimum of all
$h^x_{\threadt_i}$. As such, $(H - M)$ represents the amount of space currently
in use. Since $(V+1)$ represents the number of cells necessary to submit a new
message (the size $V$ itself is also submitted), we can proceed if
$H-M + V+1 \leq S$, \ie if there is enough free space.

Since, for a specific remote node, the broadcasts complete in order, when a reader
sees the new value of $h^x$ it means the written data is available. We need to
take care that the broadcast of $h^x$ must read from the write of the
\emph{same} function call, and not from the write of a later call to \rblsub.
Otherwise, the value of $h^x$ for the second submit might be available to
readers before the data of the second submit. For this, we simply need to wait
for the broadcast of previous function calls, using $\brlwait(\wid_x)$, before
modifying $h^x$.

When thread $\threadt_i$ wants to receive, it only proceeds if
$h^x > h^x_{\threadt_i}$, otherwise $\threadt_i$ is up-to-date and returns
$\bot$. After reading a message, the reader updates $h^x_{\threadt_i}$ to signal
to the writer the space of the message is no longer in use. If the reader is on
the same node as the writer, there is no need for a broadcast, otherwise the
reader broadcasts to the node of the writer.

With this implementation, each participating node possesses only one copy of the
data, and potentially multiple readers per node can read from the same memory
locations.



\begin{theorem}
  The implementation \implrbl is locally sound.
\end{theorem}
\begin{proof}
  See \Cref{thm:rbl-sound}.
\end{proof}



%% file: example-LOCO.tex
\section{Evaluation}

In this section, we explore the performance
of our LOCO primitives, then use them to build a
high performance key-value store.
Further applications %
can be found in \cref{sec:more applications}.

All results were collected using \code{c6525-25g} nodes on the Cloudlab
platform~\cite{cloudlab-machines}. These machines each have a 16-core AMD 7302P
CPU, running Ubuntu 22.04. Nodes communicate over a 25 Gbps Ethernet
fabric using Mellanox ConnectX-5 NICs. 

\subsection{LOCO Primitives}

\begin{figure*}
\centering
\begin{subfigure}{.45\textwidth}
\includegraphics[width=\textwidth]{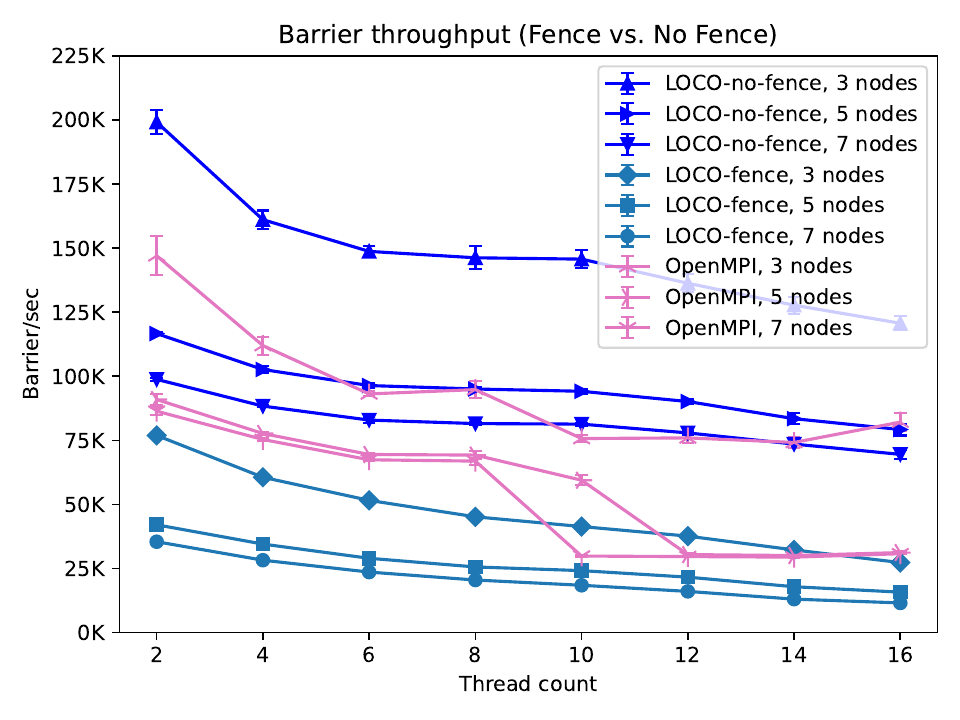}
\end{subfigure}%
\begin{subfigure}{.45\textwidth}
\includegraphics[width=\textwidth]{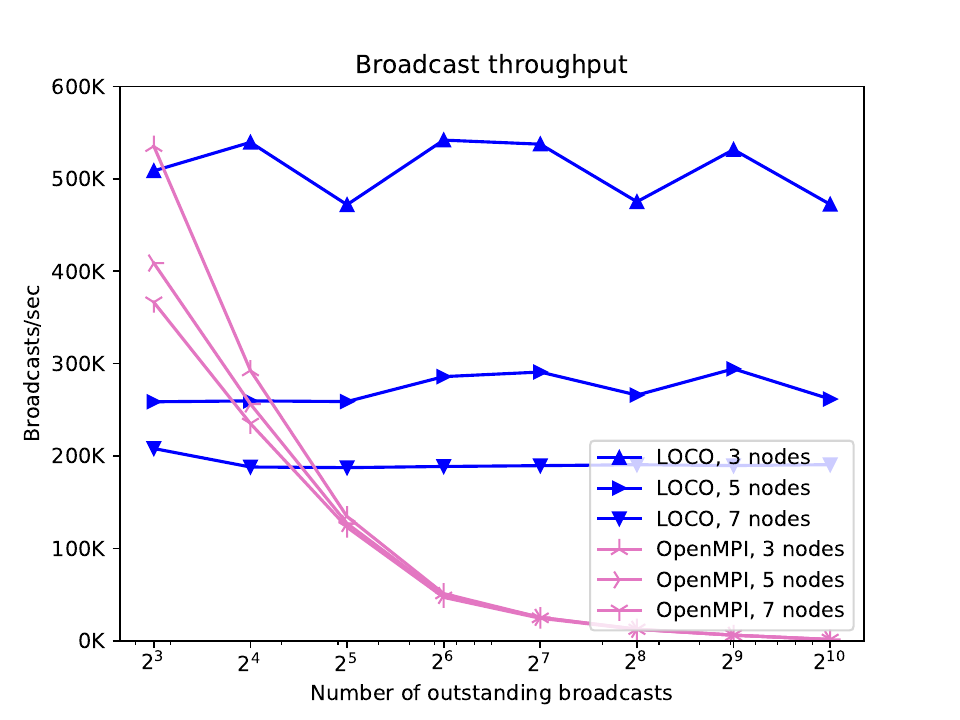}
\end{subfigure}
\caption{Comparison of barrier and broadcast operations for LOCO and OpenMPI.}
\label{fig:micro}
\end{figure*}

First, we compare the performance of the verified barrier (\bal{}) and
ring buffer (\rbl{}) primitives to equivalent operations in
OpenMPI~\cite{openmpi}, a message-passing library commonly used to build
distributed applications. We compare against OpenMPI 5.0.5, using the PML/UCX
backend for RoCE support. Results are shown in Figure~\ref{fig:micro}.

\begin{wrapfigure}[14]{r}{0.45\textwidth}
  \vspace{-20pt}
  \begin{center}
    \includegraphics[width=0.44\textwidth]{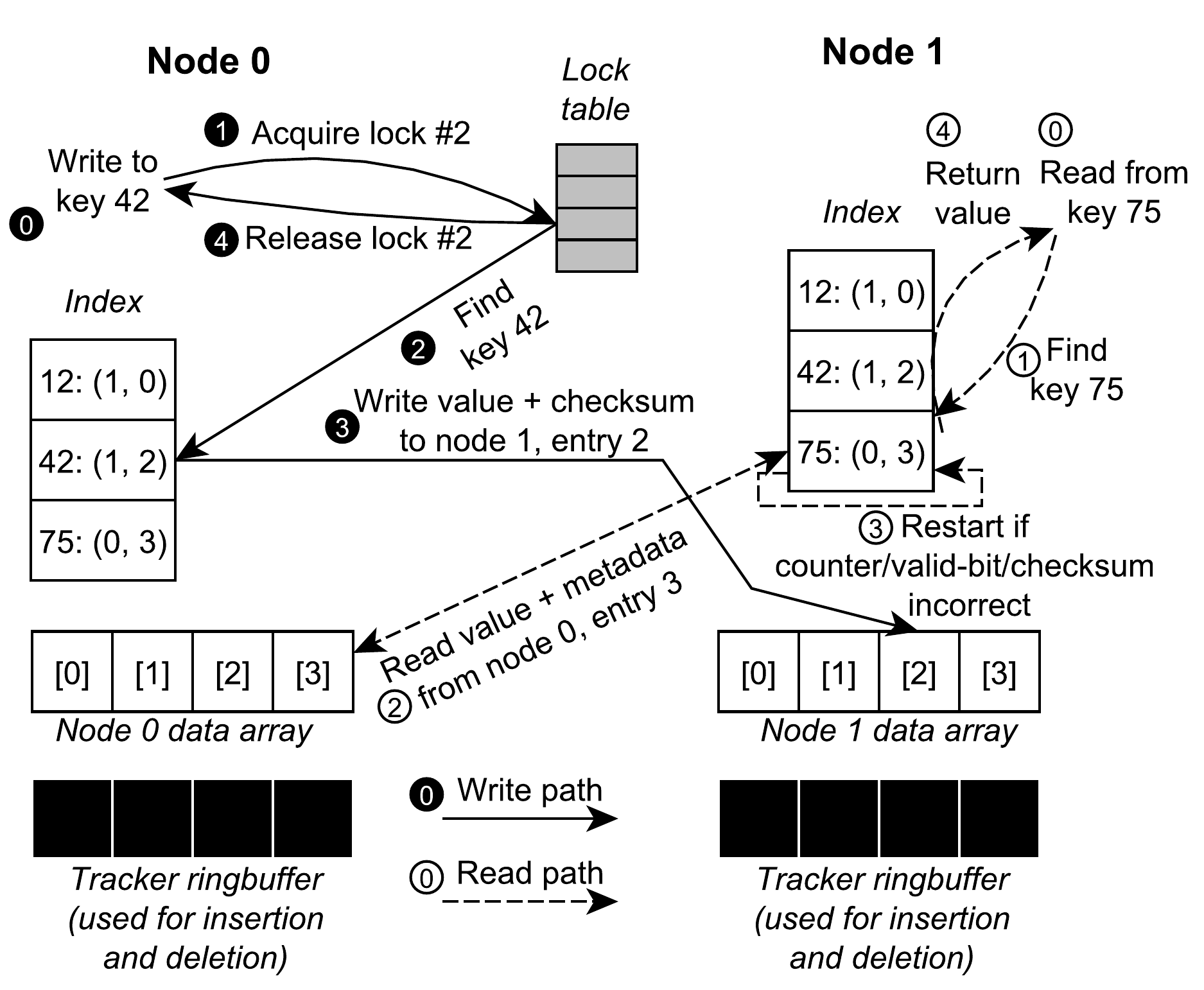}
  \end{center}
  \vspace{-10pt}
 \caption{\code{kvstore} read and write operations}
 \label{fig:kvstore-diag}
\end{wrapfigure}

For the barrier experiments, we compare to the \texttt{MPI\_Barrier} operation,
varying both thread count per node and node count. 
The MPI barrier does not actually provide synchronization, expecting
the user to instead appropriately track and fence operations
before using the primitive.
We compare the barrier to our LOCO barrier,
both with and without the synchronization fence,
and show that the LOCO barrier with equivalent semantics (no fence)
performs as well or better than the MPI barrier.
Note the MPI barrier dynamically switches between several internal algorithms adjusting to load
leading to non-smooth performance across the test domain.

For the ring buffer experiments, we compare a ring buffer broadcast to the
\texttt{MPI\_Ibcast} (non-blocking broadcast) operation. We measure across
different node counts and amounts of ``network load'', that is, the number $n$ of
outstanding broadcast operations in the network, along with total node count. A
single node acts as the sender: it starts by sending $n$ broadcasts, then sends
a new one every time a prior message completes. All other nodes wait to receive
and acknowledge messages.  Messages have a fixed size of 64 bytes. Here, we find
that the formally verified LOCO ring buffer provides better broadcast performance
than MPI in most configurations, with MPI performance falling drastically as
the number of outstanding messages rises.

\subsection{Example Application: A Key-Value Store}
\label{sec:applications}

Beyond our microbenchmarks,
we describe an example LOCO application: a key-value
store, built using composable LOCO primitives. 

Our \pcode{kvstore} object is a distributed key-value store with a
lookup operation that takes no locks, and insertion, deletion, and
update operations protected by locks. Lookup and update are depicted
in \cref{fig:kvstore-diag}. Each node allocates a
remotely-accessible memory region that is used to store
values and consistency metadata (a checksum for atomicity, a counter
for garbage collection, and a valid~bit).

Each node also maintains a local index (a C++ \code{unordered\_map}),
protected by a local reader-writer lock, which records the locations
of all keys in the \code{kvstore} as \code{(node\_id, array\_index)}
pairs, along with a counter matching the one stored with the data. The
\pcode{kvstore} is linearisable, with a proof given in
\cref{sec:proof} --- our proof is simplified by leveraging the
compositional properties of LOCO. Note that \rdmatso does not
have a semantics for locks or RDMA read-modify-write operations, which
means that this proof currently does not use \framework. We consider
an extension of \rdmatso with synchronisation operations (and hence a
full proof of \pcode{kvstore}) to be future work.  Almost all RDMA
maps~\cite{wang-sigmod-2022, li-fast-2023,
  kalia-sigcomm-2014,barthels-sigmod-2015,lu-taco-2024} lack any
formal safety specification (we are only aware of
two~\cite{dragojevic-nsdi-2014},\cite{alquraan-nsdi-2024}), likely due
to difficulties in encapsulation, which the LOCO philosophy solves.

\begin{figure*}
\centering
\begin{subfigure}{.5\textwidth}
\includegraphics[width=\textwidth]{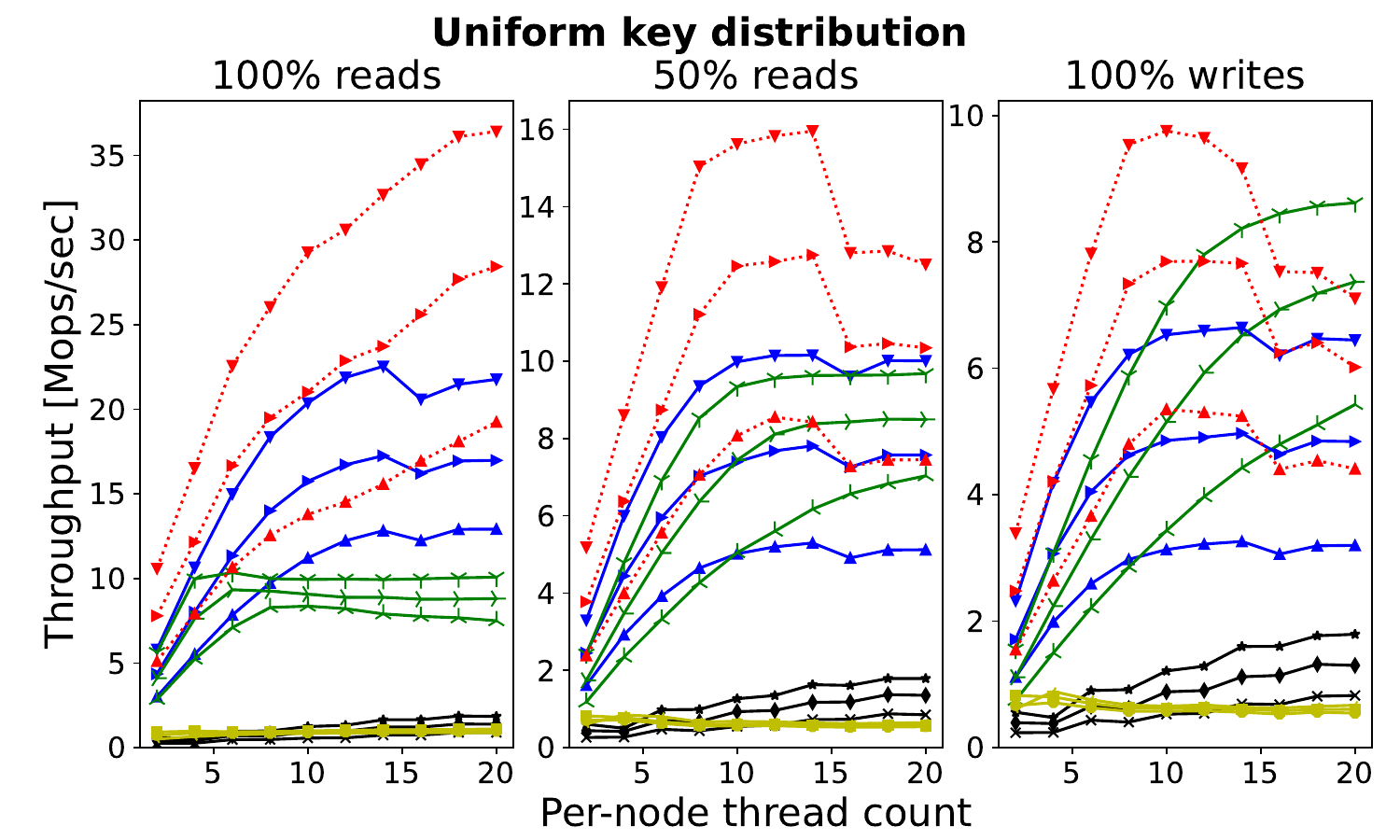}
\end{subfigure}%
\begin{subfigure}{.5\textwidth}
\includegraphics[width=\textwidth]{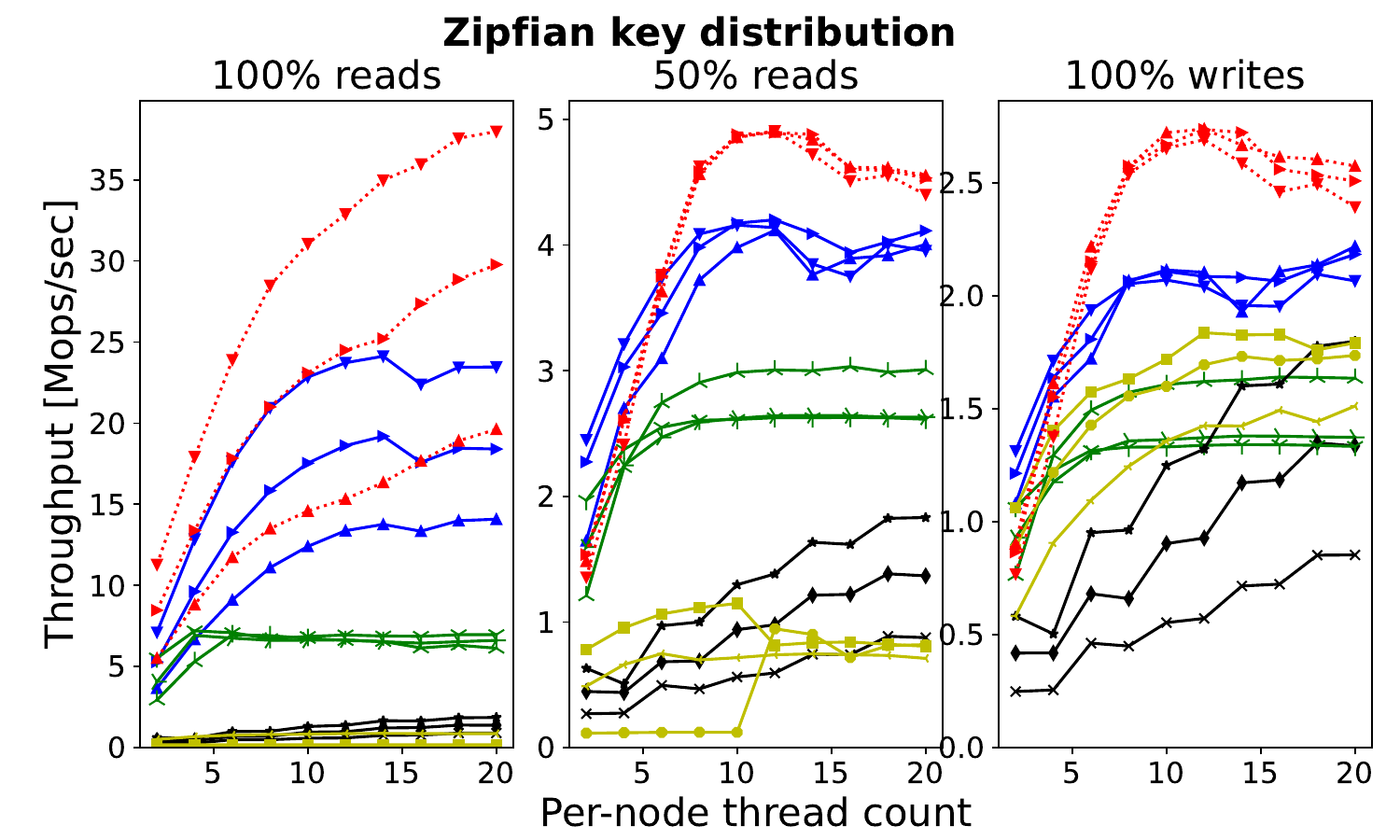}
\end{subfigure}
\includegraphics[width=.8\textwidth]{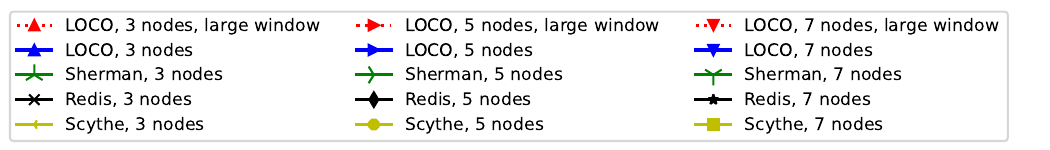}
\vspace{-5pt}
\caption{Throughput comparison of key-value stores.}
\label{fig:kvstore-perf}
\end{figure*}

We compared our key-value store design against Sherman~\cite{wang-sigmod-2022,
sherman-repo} and the MicroDB from Scythe~\cite{lu-taco-2024, scythe-repo}, two
state-of-the-art RDMA key-value stores. We also compare against
Redis-cluster~\cite{redis} as a non-RDMA baseline. Results are shown in Figure
\ref{fig:kvstore-perf}. We measured throughput on read-only, mixed read-write,
and write-only operation distributions, across both uniform and Zipfian
($\theta=0.99$) key distributions, and across different node counts and per-node
thread counts. Each data point is the geometric mean of 5 runs with a 20 second
duration, not including prefill.

All benchmarks use a 10MB keyspace, filled to 80\% capacity with 64-bit keys and
values. All benchmarks use the CityHash64 key hashing function~\cite{city-hash},
and the YCSB-C implementation of a Zipfian distribution~\cite{ycsb-c}.

We modified Sherman to issue a fence ($\brlgf$) between
lock-protected writes and lock releases to solve a bug related
to consistency issues.  %
Our
\code{kvstore} also issues a fence for the same reason. For both, 
this fence incurs a 15\% overhead.

For LOCO, Sherman, and Redis, write operations are updates. For Scythe, we found
that stressing update operations led to program
instability and very low throughput, so we use the performance of insertion
operations as an upper bound on write performance.
For Redis, we configure a cluster with no replication or persistence. Since each
Redis server instance uses 4 threads, we create \code{ceil(num\_threads/4)}
server instances for a given thread count. We use Memtier~\cite{memtier} as a
benchmark client. Each node runs
a single Memtier instance with threads equal to the thread count,
and 128 clients per thread (matching the LOCO large window size).

In addition, all systems expose a parameter we call the \emph{window
size}, which specifies the maximum number of outstanding operations per
application thread (note this is not a batch size -- each operation is
started and completed individually). 
Increasing LOCO's window size to 128 yielded
significant improvement (the ``large window''
series). However, increasing Sherman's and Scythe's window sizes appeared to
cause internal errors, so the main results for all systems except Redis (see
above) use a window size of 3 for accurate comparison.

LOCO outperforms Sherman on read-only configurations. We believe this is
because Sherman reads whole sections of the tree from remote memory, while the
LOCO design looks up the location locally and only remotely reads the value.
On the other hand, LOCO's advantage over Sherman for Zipfian writes likely
comes from the better performance under contention. %

Sherman outperforms LOCO (with a window size of 3) on mixed read-write and
write-only distributions on uniform keys, while the reverse is true for Zipfian
keys. Sherman's advantage here is likely due to the fact that, unlike LOCO,
Sherman colocates locks with data, allowing them to issue lock releases in a
batch with writes.

%% file: related.tex
\section{Related and Future Work}
\label{sec:related-work}

Although the formal semantics of RDMA has only recently been
established~\cite{OOPSLA-24}, our work is able to take advantage of
earlier results in weak memory hardware
\cite{DBLP:journals/toplas/AlglaveMT14,DBLP:conf/popl/FlurGPSSMDS16}
and programming languages
\cite{DBLP:conf/popl/BattyOSSW11,DBLP:conf/pldi/LahavVKHD17}. We do
not provide their details here since they are rather expansive.

\smallskip\noindent\emph{\textbf{RDMA Semantics.}}
Prior works on RDMA semantics include \corerma~\cite{rdma-sc} (which
formalises RDMA over the SC memory model) and
\rdmatso~\cite{OOPSLA-24}, a more realistic formal model that is very
close to the {Verbs} library~\cite{rdma-core}, describing the
behaviour of RDMA over TSO. These semantics are however low-level and
are difficult for programmers to use directly, as illustrated by
examples such as those in \cref{fig:ex-rdmatso}.

\smallskip\noindent\emph{\textbf{RDMA Libraries.}}
Much prior work in RDMA focuses on \emph{upper-level primitives}, e.g.\
consensus protocols~\cite{aguilera-osdi-2020,izraelevitz-icpp-2022,jha-tocs-2019,
poke-hpdc-2015, aguilera-podc-2019}, 
distributed maps or 
databases~\cite{dragojevic-sosp-2015,wang-sigmod-2022, dragojevic-nsdi-2014, li-fast-2023,
kalia-sigcomm-2014,barthels-sigmod-2015, alquraan-nsdi-2024, gavrielatos-ppopp-2020},
graph processing~\cite{wang-jpdc-2023}, distributed
learning~\cite{xue-eurosys-2019, ren-hpcc-2017}, 
stand-alone data structures~\cite{brock-icpp-2019,
devarajan-cluster-2020},
disaggregated scheduling~\cite{ruan-hotos-2023,ruan-nsdi-2023}
or 
file systems~\cite{yang-fast-2019,yang-nsdi-2020}.
These works focus on the final application,
rather than considering the
programming model as its own, partitionable problem. As a
result, the intermediate library between RDMA and the exported primitive is
usually ad-hoc and tightly coupled to the application, or effectively
non-existent. In general, these applied, specific, projects manage raw memory
explicitly statically allocated to particular nodes, use ad-hoc atomicity and
consistency mechanisms, and do not consider the possibility of primitive reuse.
This design is not a fundamentally flawed approach, but it does raise the
possibility of a better mechanism, which likely could underlie all the above solutions.

Some works have considered this intermediate layer explicitly,
however, the general approach for this
intermediate layer has been to encapsulate local and remote memory as 
\emph{distributed shared memory}, that is, a flat,
uniform, coherent, and consistent address space hiding 
the relaxed consistency and
non-uniform performance of the underlying RDMA network. 
These works generally focus on transparently
(or mostly-transparently~\cite{ruan-osdi-2020, zhang-sigmod-2022}) porting
existing shared memory applications.
We argue that this
technique, either with purely
software-based virtualisation~\cite{gouk-atc-2022, cai-vldb-2018,
wang-osdi-2020, zhang-sigmod-2022, ruan-osdi-2020}, or by extending
hardware~\cite{calciu-asplos-2021}, 
is unlikely to gain traction because the performance will always be
worse than an approach which takes into account the underlying memory network.

Other programming models have simply used RDMA
to implement existing distributed system abstractions.  For example,
both MPI~\cite{mpi} and NCCL~\cite{nccl} can use RDMA for inter-node communication.  
However, fundamentally, these are \emph{message passing programming models}
with explicit send and receive primitives.
While MPI does support some remote memory accesses,  
this support is best seen as a zero-copy send/receive
mechanism where synchronisation is either coarse-grained and inflexible,
or simply nonexistent.
While message-passing is well-suited for dataflow applications (e.g.\ machine
learning and signal processing) and highly parallel scale-out workloads
(e.g.\ physical simulation), it is less useful for workloads 
that exhibit data-dependent communication~\cite{liu-sigmodrec-2021}, such as transaction
processing or graph computations. In these applications, cross-node
synchronisation is unavoidable and unpredictable, so the ideal performance
strategy shifts from simply avoiding synchronisation to minimising contention,
accelerating synchronisation use, and reducing data movement.  

Compared to prior art, 
LOCO aims to build composable, reusable, and performant primitives for 
complicated memory networks, suitable for irregular workloads.
No such option currently exists in the literature.

\smallskip\noindent\emph{\textbf{Verification.}} Our proofs have followed the declarative
style~\cite{DBLP:journals/pacmpl/RaadDRLV19,DBLP:conf/esop/StefanescoRV24}
enabling modular verification. \rdmatso~\cite{OOPSLA-24} also includes
an operational model, which could form a basis for a program
logic~(e.g., \cite{pierogi,DBLP:conf/cav/LahavDW23}), ultimately
enabling operational abstractions and proofs of
refinement~\cite{DBLP:journals/pacmpl/DalvandiD22}. Other modular
approaches include modular proofs through separation
logics~\cite{DBLP:journals/jfp/JungKJBBD18}, but this additionally
requires a separation logic encoding of the \rdmawait memory model
(and an associated proof of soundness) before it can be applied to
verify libraries such as LOCO. We consider operational proofs and
those involving separation logic as a topic for future work.

\citet{DBLP:conf/nfm/NagasamudramBBMN24} have verified, in Rocq, key
properties of a coordination service known as
Derecho~\cite{DBLP:journals/tocs/JhaBGMSTRZB19}, which can be
configured to run over RDMA. However, their proofs start with a very
high-level model called a {\em shared-state table}, which is an array
of shared variables (cf \cref{lst:barrier-api}). Unlike our work,
these assumed shared state table semantics have not been connected to
any formal RDMA semantics. In future work, it would be interesting to
connect our work to middleware such as Derecho, ultimately leading to
a fully verified RDMA application stack. 

There is a rich literature of work around model checking under weak
and persistent memory
\cite{DBLP:conf/cav/Kokologiannakis21,DBLP:conf/tacas/AbdullaAKGT23}
including recent works that tackle refinement and linearisability
\cite{DBLP:conf/esop/RaadLWBD24,DBLP:journals/pacmpl/GolovinKV25}. It
would be interesting to know whether these techniques can be extended
to support \rdmatso (and by extension \rdmawait).

%% file: conc.tex
\section{Conclusion}
\label{sec:conc}

In this paper, we describe LOCO, a verified library for building composable and reusable
objects in network memory and its associated proof system \framework{}. 
Our results show that LOCO can expose the full
performance of underlying network memory to applications, while simultaneously
easing proof burden.

\begin{acks}
  This work was partially funded by industry partner Genuen, which
  provides hardware validation services using the harness described in
  Section~\ref{app:powcon}.  Izraelevitz also privately contracted
  with this company to assist with commercialization efforts of this
  harness.  Ambal is supported by the EPSRC grant EP/X037029/1 and
  Raad is supported by a UKRI fellowship MR/V024299/1, by the EPSRC
  grant EP/X037029/1, and by VeTSS.  Dongol and Chockler are supported
  by EPSRC grants EP/Y036425/1, EP/X037142/1 and EP/X015149/1 and
  Royal Society grant IES$\backslash$R1$\backslash$221226. Dongol is additionally supported
  by EPSRC grant EP/V038915/1 and VeTSS. Vafeiadis is supported by ERC
  Consolidator Grant for the project “PERSIST” (grant agreement
  No. 101003349).

\end{acks}

%% file: design.tex
\section{Further Details of LOCO's Design }
\label{sec:design}

Our Library of Composable Objects (LOCO) is functionally an extension
of the normal shared memory programming model, that is, an object-oriented
paradigm, onto the weak memory network of RDMA.
LOCO provides the ability to encapsulate
network memory access within special objects, which we call \emph{channels}. 
Channels are similar to traditional shared memory objects, in that
they export methods, control their own memory and members, and manage
their synchronisation. However, unlike traditional shared memory objects,
a single channel may use memory across multiple nodes, including
both network-accessible memory and private local memory. Examples of some
channel types (classes) in LOCO include cross-node mutexes, barriers, queues,
and maps. 

\subsection{Channel Overview}

A LOCO application will usually consist of many channels (objects) of many different
channel types (classes). In addition, each channel can itself instantiate member sub-channels
(for instance, a key-value store might include several mutexes as sub-channels
to synchronise access to its contents).  We argue that such a system of channels
makes it significantly easier to develop applications on network memory, without
sacrificing performance.

Figure~\ref{lst:barrier} shows our implementation of a barrier channel, based
on~\cite{gupta-clustr-2002}, using a SST sub-channel. As with a traditional
shared memory barrier, it is used synchronise all participants at a
certain point in execution. For each use of the barrier, participants increment
their local, private, count variable, then broadcast the new value to others
using their register in the SST. They then wait locally to leave the barrier until all
participants have a count in the SST not less than their own.

\begin{figure}[t]
\centering
\begin{subfigure}{0.45\textwidth}
\begin{lstlisting}[basicstyle={\ttfamily\scriptsize}]
class barrier : public loco::channel {
	unsigned count,num_nodes;
	loco::sst_var<unsigned> sst;
	public:
	void waiting() {
		// complete all outstanding RDMA operations 
		mgr()::fence(); 
		count++; // increment our counter
		sst.store_mine(count);
		sst.push_broadcast(); //and push
		bool waiting = true; 
		while(waiting){  // wait for others 
			waiting = false; // to match
			for (auto& row : sst) {
				if (row.load() < count){
					waiting = true; 
					break;} 
			}
		}
	}
	barrier(channel* parent, 
	 string name,manager& cm,int num):
	 channel(parent, name, cm, 
	 channel::expect_num(num-1)), *@\label{cd:bar-expect}@*
	 sst(this,"sst",cm)){ 
		count=0; num_nodes=num;
		channel::join(); *@\label{cd:chan-join}@*
	}
};
\end{lstlisting}
\caption{Complete C++ code for the network barrier, a simple channel object.}
\label{lst:barrier}
\end{subfigure}
\hfill
\begin{subfigure}{0.45\textwidth}
\centering
\begin{lstlisting}[basicstyle={\scriptsize\ttfamily}]
int main(int argc, char** argv) {
	map<uint32_t, string> hosts; 
	int node_id, num_nodes; *@\label{cd:host-map}@* 
	loco::parse_hosts(&hosts,
	 &node_id,&num_nodes,argv[1]);
	vector<timespec> lats;
	loco::manager cm(ip_addrs, node_id); *@\label{cd:mgr-init}@* 
	loco::barrier bar("bar", cm, num_nodes); *@\label{cd:bar-init}@*
	cm.wait_for_ready(); *@\label{cd:mgr-wait}@*
	for(int i=0; i<TEST_ITERS; ++i){
		timespec t0 = clock_now();
		bar.waiting(); *@\label{cd:bar-wait}@*
		timespec t1 = clock_now();
		lats.push_back(t1 - t0);
	}   
	cout<<"Avg latency:"<<
	 accumulate(lats.begin(),
	 lats.end(),0.0))/lats.size();
} *@\label{cd:end}@* 
\end{lstlisting}
\caption{A simple (complete) LOCO application measuring barrier latency.}
\label{lst:bar-lat}
\end{subfigure}
\caption{LOCO barrier code}
\end{figure}

\subsection{Channel Setup}
\label{ssec:chan-setup}

Figure~\ref{lst:bar-lat} shows a complete example LOCO application: a microbenchmark
which repeatedly waits on the barrier (Line~\ref{cd:bar-wait}) and measures its
latency. At line~\ref{cd:mgr-init}, we construct the \code{manager} object from
a set of (ID, hostname) pairs. The \code{manager} establishes connections with
peers and mediates access to per-node resources: peer connections, a shared
completion queue, and network-accessible memory.

The \code{manager} is then used to construct channel endpoints, in this case the
barrier and its sub-channels (Line~\ref{cd:bar-init}). Note that the barrier has
a name \code{"bar"}, which must match the name of the remote barrier endpoints
to complete the connection.  We use a `/' character to denote a sub-channel
relationship (e.g., the full name of the SST in the barrier object is
\code{"bar/sst"}, with component \code{owned\_var}s named \code{"bar/sst/ov0"} etc.), 
and a `.' character to denote a component memory region.

When a channel endpoint is constructed, it initialises its local state 
including subchannels, creates
local memory regions, and indicates by name what memory regions it expects other
participants to provide. %
Then, it sends a \emph{join} message
(Line~\ref{cd:chan-join}) to each peer with the channel name and the list of
memory regions it expects that peer to provide.

When a peer receives a join message, it first checks if a channel endpoint with
the same name exists locally, and ignores the message if not (in other words,
peers may not participate in all channels). If it finds a matching
endpoint, it verifies its allocated memory regions match those
requested, and returns
a \emph{connect} message containing metadata necessary to access the requested
regions. 
Channels can also register callbacks which run when join and/or
connect messages are received; these are used to create per-participant
sub-channels or memory regions. %

%% file: eval.tex
\section{Further LOCO-based Applications}
\label{sec:more applications}
\subsection{Transactional Locking}
\label{sec:eval}

\begin{figure}[t]
\begin{subfigure}{.43\textwidth}
\includegraphics[width=\textwidth]{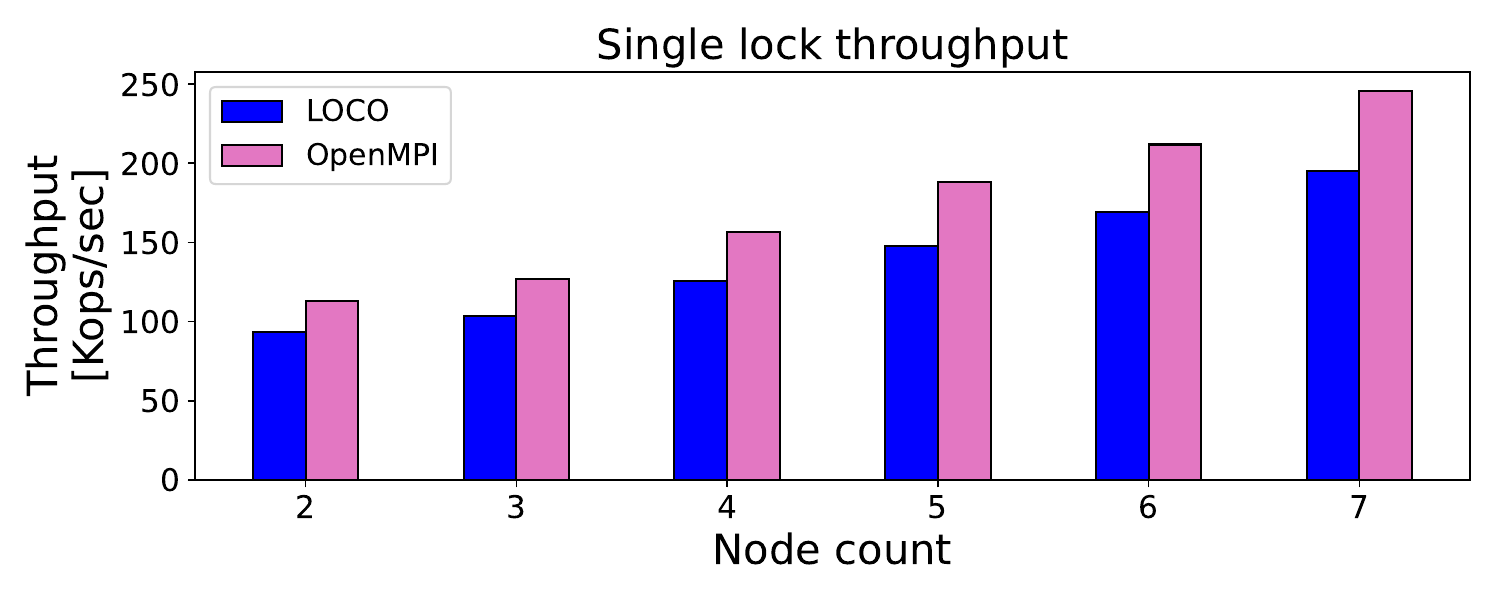}
\end{subfigure}
\begin{subfigure}{.43\textwidth}
\includegraphics[width=\textwidth]{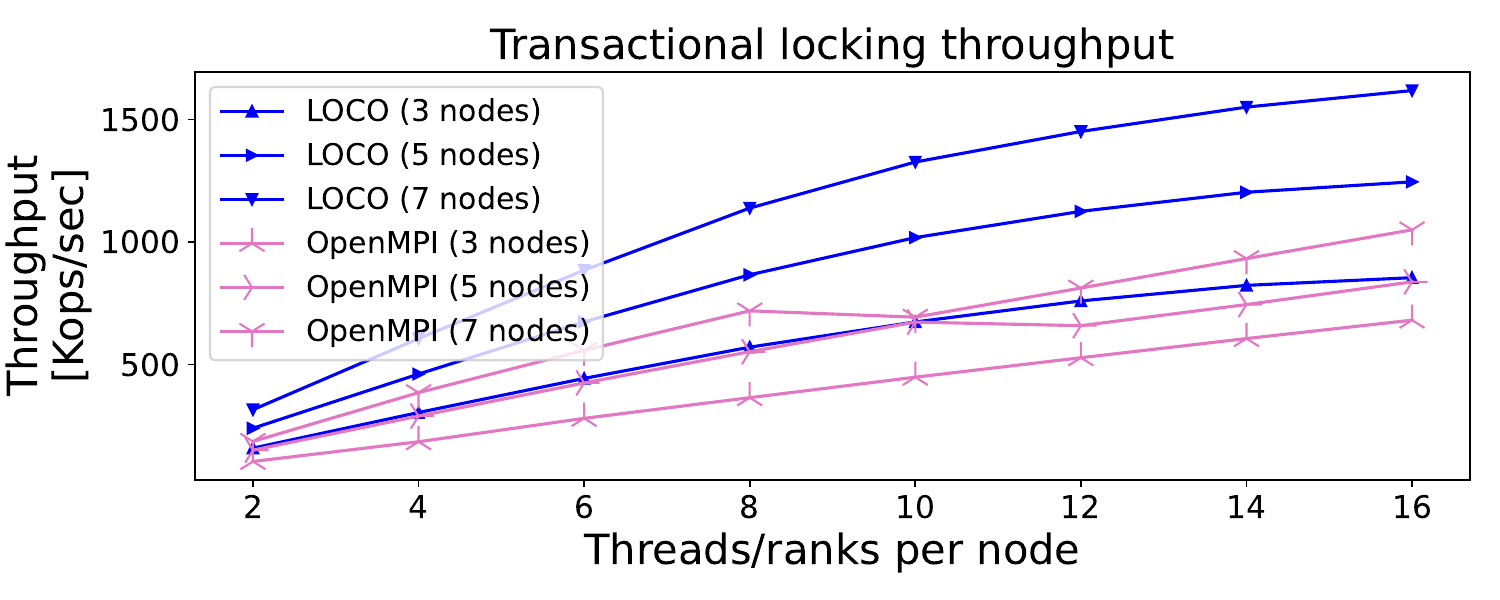}
\end{subfigure}
\caption{Throughput of single-lock and transactions
in OpenMPI and LOCO.}
\label{fig:lock-bench}
\end{figure}

In this section, we compare the performance of LOCO to the RDMA APIs provided by
OpenMPI~\cite{openmpi} on tasks involving contended synchronisation. We compare
against OpenMPI version 5.0.5, using RoCE support provided by the PML/UCX
backend. Results for both benchmarks are shown in Figure~\ref{fig:lock-bench}
(geomean of five 20-second runs). 

First, we measured the throughput of a contended single-lock critical section
(lock-protected read-modify-write) at different node counts, with one
rank/thread per node. Here, OpenMPI has a consistent advantage,
likely due to extensive optimisation and a more managed environment. 

Then, we measured the throughput of a transactional critical section, which
acquires the locks corresponding to two different accounts (array entries), and
transfers a randomly generated amount between them. We use 100 million accounts.
For intra-node scaling, LOCO creates multiple threads, while OpenMPI creates
separate ranks (MPI processes), due to MPI's limited support for multithreading
within a rank.

For LOCO, we create an array of \code{atomic\_vars} holding account values, striped
across participants. For OpenMPI, we distribute the accounts across 341 windows
(symmetrically allocated regions of remote memory, each associated with a
single lock per rank); 341 is the maximum supported.
To ensure a fair comparison, LOCO uses at most 341 locks per
thread. 

LOCO outperforms OpenMPI on transactional locking, despite the fact
that we use an equal number of locks and their lock performs better in
isolation. We believe this is due to the tight coupling between memory windows
and locks in MPI: 
windows likely have a one-to-one correspondence with RDMA
memory regions in the backend, and performing operations on many small memory
regions is slower than large ones due to NIC caching
structures~\cite{kong-nsdi-2023}.
LOCO avoids this penalty by disassociating regions and locks in its object system, while
also merging regions into 1 GB huge pages in the backend.

\input{dc}

%% file: dc.tex
\subsection{Distributed DC/DC Converter System}
\label{app:powcon}

As an additional application of LOCO, we implemented a
model of a hardware control loop which exploits its low latency.

\subsubsection{System Design}

\begin{figure}[t]
\includegraphics[width=0.45\textwidth]{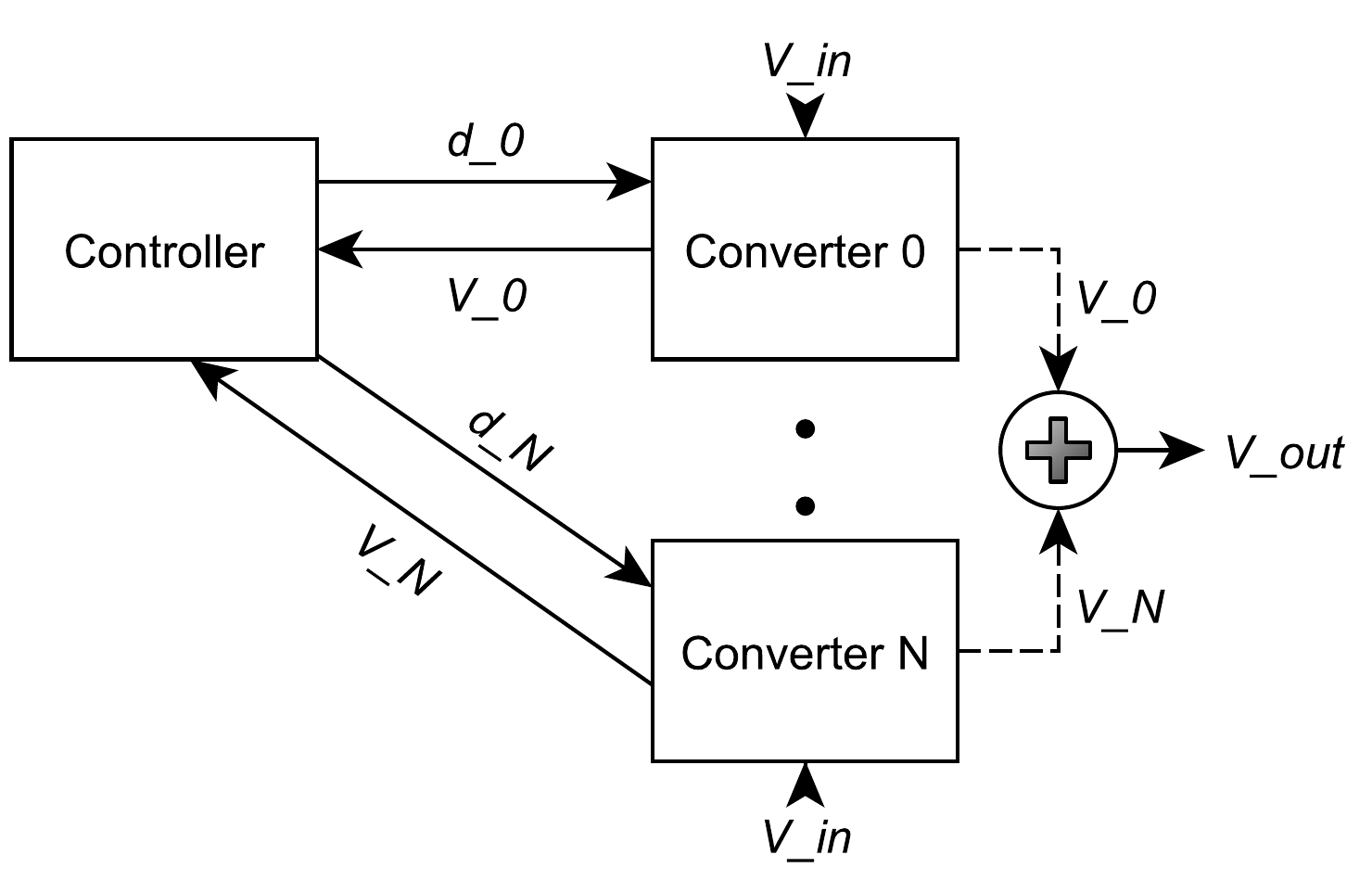}

\caption{Schematic of the system modeled by the \code{power\_controller} channel.
Solid arrows represent LOCO \code{owned\_vars}, and dashed arrows represent
electrical connections. \code{d_N} represents the duty cycle parameter used to
control converter \code{N}, and \code{V_N} represents the output voltage at
converter \code{N}.}

\label{fig:powcon}
\end{figure}

An additional application channel we have implemented is the
\code{power\_controller}, a real-time simulation of a distributed DC/DC converter
system controlled by a discrete-time control
loop~\cite{corradini-textbook-2015}. The simulation (Figure \ref{fig:powcon})
consists of a single machine which acts as a \emph{controller}, and an arbitrary
number of machines simulating the physical characteristics of a
\emph{converter}. The role of the controller is to regulate the duty cycles
($d$) of the converters, which are supplied with a steady input DC voltage, to
produce a target output voltage ($V_{ref}$). The converters return voltage
values ($V$) which are used to calculate the next setting of their duty cycles,
closing the control loop.

The \code{power\_controller} channel consists of two arrays of \code{owned\_vars}
representing the duty cycle (owned by the controller) and output voltage (owned
by the converter) for each converter. The participating machines run fixed-time
loops: each loop iteration at a converter calculates a new simulated $V$ and
pushes it to the controller, while each iteration at the controller calculates a
new $d$ for all controllers based on their most recent $d$ and $V$ values. The
overall output voltage of the system at each step (as seen by the controller) is
the sum of all converters' most recent output voltage.

Network memory is a good fit for this application because it is highly sensitive
to network latency; with the parameters we have chosen, the output will only
converge if latency of the control and feedback messages is consistently less
than 40 $\mu$s. This requirement would be difficult to meet with traditional
message-passing protocols: while a protocol such as UDP can easily achieve this
latency on an uncontended network, it would be difficult to manage the
scheduling jitter, copying, and cache contention in the software network
protocol stack. 

An extension of this control loop harness was developed in LOCO for validating
hardware components such as the power controller and converters within partially
simulated environments (hardware-in-loop testing). The system is currently in
beta testing for production use, with expected commercial release later this
year. 

\subsubsection{Evaluation}

\begin{figure}[t!]
\includegraphics[width=.45\textwidth]{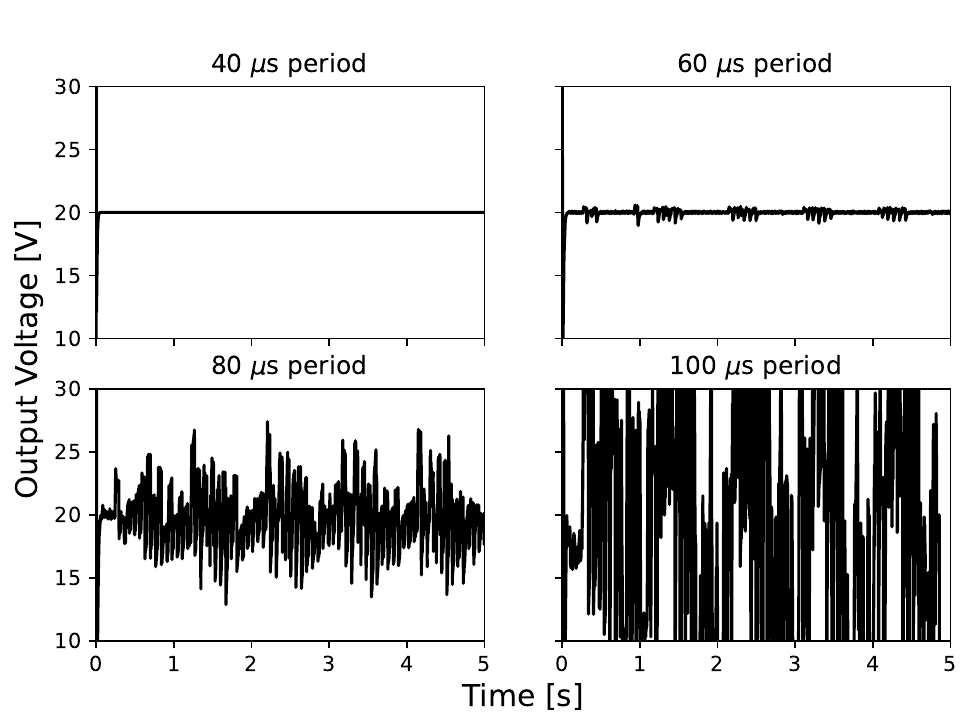}

\caption{Output voltage for the DC/DC converter simulation at various control
loop frequencies. 
}

\label{fig:powcon-perf}
\end{figure}

To evaluate whether LOCO meets the latency requirements of this system, we
instantiated a cluster with one controller and 20 converters and measured the
output voltage over time at various loop periods. The effect of changing the
loop period is to simulate higher link latency, since we cannot increase the
latency of the RDMA link. The loop period at the converters is fixed at 10
$\mu$s to approximate the continuous nature of their transfer function. We ran
each simulation for 5 seconds.

The system parameters are selected to maintain a stable output voltage with a
controller loop period of 40 $\mu$s or lower. The increasing instability in the
output resulting from increasing the loop period past this value is clearly
visible in Figure \ref{fig:powcon-perf}. The series with period greater than 40
$\mu$s also exhibit large transients at simulation start. These are
mostly invisible on the plots due to their brief duration, but would be
unacceptable in a real system.

%% file: backend.tex
\section{LOCO Backend}
\label{sec:backend}

In this section,
we briefly describe key features of the LOCO RDMA backend, which we have tuned
extensively to expose the full performance of RDMA to LOCO applications
(Section~\ref{sec:applications}).

The LOCO backend uses the \code{libibverbs} library for RDMA communication, and
the \code{librdmacm} library to manage RDMA connections. Both of these libraries
are components of the Linux \code{rdma-core} project~\cite{rdma-core2}. LOCO
currently supports only RoCE~\cite{rdma-rocev2-2014} as a link layer, although
the only element missing for InfiniBand support is an implementation of the
connection procedure. The current design assumes a reliable, static network of
IP-addressable peers specified at application startup (the \code{hostnames} map
declared at line \ref{cd:host-map} of Figure \ref{lst:bar-lat}).

\subsection{Local Scalability}

LOCO implements multiple features aimed at increasing the scalability of
performing RDMA operations across multiple local threads. First, each thread in
a LOCO application uses a private set of Queue Pairs (one RDMA communication channel per peer), to avoid
unnecessary synchronisation when multiple threads perform RDMA operations
simultaneously.  Second, all completions are delivered to a single completion
queue, which is monitored by a dedicated \textit{polling thread}, in order to
avoid contention on the completion queue.

Application code can monitor the progress of one or more operations by
registering an \code{ack\_key} object (modelled as work identifiers ``\wid'') with the polling thread, which provides
APIs for polling and waiting on completion of the operation. Internally, the
\code{ack\_key} is a lock-free bitset with bits mapped to in-progress
operations. As operations complete, the polling thread clears the corresponding
bits, so that checking for completion of an \code{ack\_key}'s registered
operations simply consists of testing whether the internal bitset is equal to
zero (i.e., empty). This approach avoids explicit synchronisation between the
polling thread and application threads waiting for operations to complete.

\subsection{Network Memory Management}

Another important service the backend provides is management of each node's
network memory. Memory must be registered with \code{libibverbs} before it can
be accessed remotely. Since registration of a memory region incurs
non-negligible latency, we aggregate all registered memory used by LOCO channels
into a series of 1GB huge pages, each of which corresponds to a single
\code{libibverbs} memory region. The named memory region objects constructed by
channels each correspond to a contiguous sub-range of one of these regions.
Using huge pages reduces TLB utilisation, which can have a significant
performance impact on multithreaded applications~\cite{chen-isca-1992}.

In addition to memory regions explicitly created by channels, we found it useful
to create a primitive for allocating temporary chunks of network memory used as
inputs and outputs of channel methods, which we call \code{mem\_refs}.  We
allocate of backing memory for these objects from a per-thread pool of
fixed-size block, which are in turn allocated from the larger pool of registered
memory described above.

Finally, LOCO also provides the capacity to allocate local memory regions backed
by \emph{device memory}, which resides on the network card.  RDMA accesses to
device memory are faster than those to system memory, since they are not
required to traverse the PCIe bus to main memory. However, since device memory
is not coherent with main memory, it is mainly useful for holding state
exclusively accessed through the network, such as mutex state.

%% file: bugs.tex
\section{LOCO Bugs Discovered (and Corrected)} 
\label{sec:loco-bugs-discovered}

The weak and asynchronous nature of the
  \rdmawait model means that developing correct RDMA programs is very
  difficult. During the course of our verification work, we discovered
  two critical bugs in LOCO, meaning that neither of the expected
  safety properties held.

  \begin{wrapfigure}[8]{r}{0.35\columnwidth}
    \vspace{-10pt}
    \begin{minipage}[t]{.35\columnwidth}
      \small \centering
      \begin{tabular}{|@{\hspace{3pt}}c @{\hspace{3pt}} || @{\hspace{3pt}}c@{\hspace{3pt}}|| @{\hspace{3pt}}c@{\hspace{3pt}}|}
        \hline
        & $x = 0$ & \\
        \hline
        $\inarr{
        \neqn{x} \assign 1 \\
        \balbar(z) \\
        \phantom{a}
        }$ &
             $\inarr{
             \phantom{a} \\
        \balbar(z) \\
        \phantom{a}
        }$ &
             $\inarr{
             \phantom{a} \\
        \balbar(z) \\
        a \assign \neqn{x}
        }$ \\
        \hline
      \end{tabular}
      $$a=0\ \text{\checkno} \quad \texttt{possible with bug}$$
      \vspace{-20pt}

      \caption{Possible incorrect behaviour with the buggy implementation}
      \label{fig:bug1}
    \end{minipage}
  \end{wrapfigure}
  \paragraph{The first bug:} the implementation of the barrier
  immediately notified every participating remote node that the barrier
  was reached. While some RDMA orderings made sure the notification
  would arrive after completion of previous RDMA operations towards the
  \emph{same} remote, it did not wait for operations towards
  \emph{other} nodes, even participating ones. So instead of doing a
  global synchronisation of a set of nodes, the previous buggy
  implementation is more akin to doing several pairwise synchronisation
  of nodes.  With this bug, examples such as \cref{fig:bar-example}
  behave correctly, but programs with three or more nodes could exhibit
  unwanted behaviours. For example, consider the program in
  \Cref{fig:bug1}. Node~1 modifies some location $x$ on node 2,
  synchronises with nodes 2 and 3, and then node 3 reads the location
  $x$. We would expect node 3 to necessarily see the new value of~$x$.
  However, with the buggy implementation, node 1 would immediately
  give the go-ahead to node 3 before the location $x$ is modified,
  allowing node 3 to read the outdated value. In \Cref{sec:balib}, we
  present and prove the corrected version of the barrier.

  \paragraph{The second bug:} LOCO has an implementation of mixed-size
  writes using the notion of a {\em guard} allowing transfer of data
  that cannot be read/written atomically by the CPU. In it, one byte
  (called a guard) is reserved on each side of the data and the writer
  proceeds by:
  \begin{enumerate*}
  \item updating the leading guard to a fresh value;
  \item writing the data;
  \item updating the trailing guard to the value of the leading guard
  \end{enumerate*}. The required invariant for the reader is that if the
  values of the leading and trailing guards match then the data is not
  corrupted. However, the buggy implementation of the mixed-size read
  operation was to
  \begin{enumerate*}
  \item read the leading guard;
  \item read the data;
  \item read the trailing guard; then
  \item return the data if the guards match
  \end{enumerate*}. This could lead to the following interleaving:
  $\neqn{g}_L \assign k \rightarrow \neqn{x}_1 \assign v_1 \rightarrow
  \teal{b_L \assign g_L} \rightarrow \teal{a_1 \assign x_1} \rightarrow
  \teal{a_2 \assign x_2} \rightarrow \neqn{x}_2 \assign v_2 \rightarrow
  \neqn{g}_T \assign k \rightarrow \teal{b_T \assign g_T}$, where $g_L$
  and $g_T$ are the leading and trailing guards and $x$ is split into
  two components $x_1$ and $x_2$. The \teal{teal} events are those of
  the reader, which reads the new value of $x_1$ and the old value of
  $x_2$ (leading to a corrupted value of $x$) yet accepts the write
  since $b_L = b_T = k$ at the end of computation.
  The correct implementation should read in
  the opposite direction:
  \begin{enumerate*}
  \item read the {\em trailing} guard;
  \item read the data;
  \item read the {\em leading} guard;
  \item accept the data if the guards match.
  \end{enumerate*} %

  In \Cref{sec:msw}, we model and prove correctness of a more general
  algorithm using hashes, which is also implemented in LOCO, that is
  valid for {\em any} size of data.

%% file: example.tex
\section{Clients using Multiple Libraries}
\label{sec:rbl-example}

\begin{wrapfigure}[8]{r}{0.42\columnwidth}
  \vspace{-10pt}
  \centering
  \scalebox{0.9}{\begin{minipage}[t]{\linewidth}
    \begin{center}
    \begin{tabular}{|@{\hspace{3pt}}c @{\hspace{3pt}} || @{\hspace{3pt}}c @{\hspace{3pt}}|}
      \hline
      \multicolumn{2}{|c|}{\text{Ring Buffer} $x$ \qquad \textrm{Barrier} $z$} \\
      \hline
      $\inarr{\phantom{a} \vspace{-6pt} \\
      a \assign \rblsub(x, 1) \\
      \balbar(z) \\
      \phantom{a} \\
      \phantom{a} \vspace{-6pt}}$
      & $\inarr{\phantom{a} \\
      \balbar(z) \\
      b := \rblrec(x)
      }$
      \\
      \hline
    \end{tabular}
    \end{center}
    $$(a,b)=(\true,\bot)\ \text{\checkno} \quad (a,b)=(\true,1)\ \text{\checkyes}$$
  \end{minipage}}
  \vspace{-5pt}
\caption{ring buffer + barrier example}
\label{fig:rbl-bal-example}
\end{wrapfigure}

Now that multiple libraries have been defined separately, we can write programs
combining methods from all of them. The synchronisation guarantees of each
library (\ie $\so$) are available to other libraries (via $\hb$) to restrict the
behaviours of the whole program. In the example of \cref{fig:rbl-bal-example}, a
first thread sends data to a second using a ring buffer, and the two threads
synchronise through a barrier, making sure the data is available. \Ie, if the
submit method succeeds, the receive method also has to succeed.

As an illustration, let us show we cannot have a $\set{\rbl,\bal}$-consistent
execution $\exec = \langle E, \po, \gettags, \linebreak \so, \hb \rangle$
corresponding to the disallowed behaviour $(a,b)=(\true,\bot)$. This result
corresponds to $E$ containing the four events
\begin{itemize}
\item $\action_S = \tup{\threadt_1, \_, \tup{\rblsub, (x,1), \true}}$,
\item $\action_{B1} = \tup{\threadt_1, \_, \tup{\balbar, (z), ()}}$,
\item $\action_{B2} = \tup{\threadt_2, \_, \tup{\balbar, (z), ()}}$, and
\item $\action_R = \tup{\threadt_2, \_, \tup{\rblrec, (x), \bot}}$
\end{itemize}
with
$\po = \set{\tup{\action_S, \action_{B1}} ; \tup{\action_{B2}, \action_R}}$.
$\set{\rbl,\bal}$-consistency (\cref{def:lambda-cons}) would imply:
\begin{enumerate}
\item $(\ppo \cup \so\rst{\rbl} \cup \so\rst{\bal})^+ \subseteq \hb$ is
  irreflexive;
\item
  $\tup{\set{\action_S ; \action_R}, \emptyset, \gettags\rst{\rbl},
    \so\rst{\rbl}, \_} \in \rbl.\cons$; and
\item
  $\tup{\set{\action_{B1} ; \action_{B2}}, \emptyset, \gettags\rst{\bal},
    \so\rst{\bal}, \_} \in \bal.\cons$.
\end{enumerate}

Assuming the consistency conditions of the two libraries, there is a single
possibility for $\gettags$:
\begin{itemize}
\item $\gettags(\action_S) = \set{\tagcwrite, \tagnrw[\node_2]}$;
\item $\gettags(\action_{B1}) = \gettags(\action_{B2}) = \set{\taggf[\node_1],
  \taggf[\node_2], \tagcread}$; and 
\item $\gettags(\action_R) = \set{\tagwait}$.
\end{itemize}
For the barrier library, we necessarily have $c_z =1$,
$\ordero(\action_{B1}) = \ordero(\action_{B2}) = 1$, and thus the two events
synchronise. Notably, we have
$\tup{\action_{B1}, \taggf[\node_2]} \arr{\so\rst{\bal}} \tup{\action_{B2},
  \tagcread}$.
For the ring buffer library, there is no succeeding receive and
$\rf = \emptyset$. We thus have
$\tup{\action_{R}, \tagwait} \arr{\rblfr} \tup{\action_{S}, \tagnrw[\node_2]}$,
with $\rblfr = \so\rst{\rbl}$.
From the definition of $\tagppo$ (\cref{fig:to}), we have
$\tup{\action_{S}, \tagnrw[\node_2]} \arr{\ppo} \tup{\action_{B1},
  \taggf[\node_2]}$ and
$\tup{\action_{B2}, \tagcread} \arr{\ppo} \tup{\action_{R}, \tagwait}$. This
implies an \hb cycle between the four subevents, and the execution cannot be
$\set{\rbl,\bal}$-consistent.

%% file: proofs.tex
\section{Correctness Proof of the \framework Framework}
\label{sec:proofs-framework}

As mentioned in the paper, for $f : A \rightarrow B$ and $r \suq A \times A$, we
note $f(r) \defeq \setpred{\tup{f(x),f(y)}}{\tup{x,y} \in r}$.
It is straightforward to show that $f(r_1 \cup r_2) = f(r_1) \cup f(r_2)$ and
$f(r\rst{A'}) \suq f(r)\rst{f(A')}$ for any subset $A' \subseteq A$.
When $r$ is a strict partial order, we write $\imm{r}$ for the \emph{immediate}
edges in $r$, \ie $r \setminus (r;r)$.

\subsection{Wide Abstraction}

First, we generalise the notion of abstractions (\cref{def:abstraction}) to
\emph{wide abstractions}.

\begin{definition}
  Suppose $I$ is a well-defined implementation of a library $L$ using $\Lambda$,
  and that $G = \tup{E, \po}$ and $G'= \tup{E', \po'}$ are plain executions
  using methods of $\Lambda$ and ($\Lambda \uplus \set{L}$) respectively. We say
  that a function $f : E \rightarrow E'$ is a wide abstraction of $G$ to $G'$,
  denoted $\wideabsf{L}{I}{G}{G'}$, iff
  \begin{itemize}
  \item $E' = f(E)$, \ie $f : E \rightarrow E'$ is surjective;
  \item $E\rst{L} = \emptyset$, \ie $G$ contains no calls to the abstract
    library $L$;
  \item $f(x) \not \in L \implies f(x) = x$, \ie events not part of an
    implementation of $L$ are kept unchanged;
  \item $f(\po) \suq (\po')^*$ and
    $\forall \action_1,\action_2, \ \tup{f(\action_1),f(\action_2)} \in \po'
    \implies \tup{\action_1,\action_2} \in \po$; and
  \item if $\action' = \tup{\threadt,\aident,\tup{m,\vect{v},v'}} \in E'$ then
    $\tup{\tup{v',0}, G\rst{f^{-1}(\action')}} \in
    \interpt{I(\threadt,m,\vect{v})}$
  \end{itemize}
\end{definition}
The difference with the normal abstraction is that $G'$ is not limited to
methods of $L$, but every method call not from $L$ is carried over to the
implementation $G$ (\ie in general $E \cap E' \neq \emptyset$) and the
abstraction function $f$ maps these events to themselves.



\subsection{Finding a Wide Abstraction}

\begin{lemma}
  \label{lem:find-abstraction}
  Given $\progps$ and an implementation $I$ of $L$ using $\Lambda$, if
  $\tup{\vect{v}, G} \in \interp{\impli{\progps}}$ then there is
  $\tup{\vect{v}, G'} \in \interp{\progps}$ and $f$ such that
  $\wideabsf{L}{I}{G}{G'}$.
\end{lemma}
\begin{proof}
  It is enough to show the following: for all $\threadt$ and $\progp$, if
  $\tup{\tup{v, k}, \tup{E, \po}} \in \interpt{\implti{\progp}}$ then there is
  $\tup{\tup{v, k}, \tup{E', \po'}} \in \interpt{\progp}$ and $f$ such that
  $\wideabsf{L}{I}{\tup{E, \po}}{\tup{E', \po'}}$. Indeed, if this holds, we
  can conclude by merging the results of each thread for the case $k=0$.

  For a given $\threadt$, we proceed by induction on $\progp$.

  \begin{itemize}

  \item If $\progp = v$ or $\progp = \texttt{break}_{k'} \ v$, then
    $\implti{\progp} = \progp$ and we have $\tup{E,\po} = \emptyset_G$. We
    simply take $\tup{E,\po} = \emptyset_G$ and we have
    $\wideabsf{L}{I}{\emptyset_G}{\emptyset_G}$ for the empty function $f$.

  \item If $\progp = m(\vect{v_0})$ with $m \not\in L.M$, then
    $\implti{\progp} = \progp$ and we have $\tup{E,\po} = \set{\action}_G$ for
    some event $\action$. We can choose $\tup{E',\po'} = \tup{E,\po}$. We have
    $\wideabsf[\Id]{L}{I}{\tup{E, \po}}{\tup{E', \po'}}$ for the identity
    function $\Id$ that maps $\action$ to itself:

    $\Id(\set{\action}) = \set{\action}$; $E\rst{L} = \emptyset$;
    $\Id(\action) = \action$; $\po' = \emptyset = \po$; and the last property
    holds since $E'\rst{L} = \emptyset$.

  \item If $\progp = m(\vect{v_0})$ with $m \in L.M$, then
    $\implti{\progp} = I(\threadt,m,\vect{v_0})$ and
    $\tup{\tup{v, k}, \tup{E, \po}} \in \interpt{I(\threadt,m,\vect{v_0})}$. By
    definition of the implementation, we have $k=0$ and $E \neq \emptyset$. Let
    $\action' = (\threadt,\aident,\tup{m,\vect{v_0},v})$ for some ident
    $\aident$, we take $\tup{E',\po'} = \tup{\set{\action'},\emptyset}$ and we
    indeed have $\tup{\tup{v, 0}, \tup{E', \po'}} \in \interpt{m(\vect{v_0})}$.
    We choose the function $f$ that maps every element of $E$ to $\action'$, and
    we need to check $\wideabsf{L}{I}{\tup{E, \po}}{\tup{E', \po'}}$.

    We do have $\set{\action'} = f(E)$ since $E \neq \emptyset$. We have
    $E\rst{L} = \emptyset$ since $I$ does not use $L$. Forall $x \in E$,
    $f(x) = \action' \in L$. If $(\action_1,\action_2) \in \po$ then
    $f(\action_1)=\action'=f(\action_2)$. $\po' = \emptyset$. Finally, for
    $\action' \in \set{\action'}$, we have $f^{-1}(\action') = E$ and
    $\tup{\tup{v, 0}, \tup{E, \po}} \in \interpt{I(\threadt,m,\vect{v_0})}$
    holds.

  \item If $\progp = \texttt{let} \ \progp_1 \ \progp_2$, then
    $\implti{\texttt{let} \ \progp_1 \ \progp_2} \defeq \texttt{let} \
    \implti{\progp_1} \ (\lambda v. \implti{\progp_2 \ v})$, and
    $\tup{\tup{v, k}, \tup{E, \po}} \in \interpt{\texttt{let} \
      \implti{\progp_1} \ (\lambda v. \implti{\progp_2 \ v})}$ has two possible
    sources.
    \begin{itemize}
    \item If $\tup{\tup{v, k}, \tup{E, \po}} \in \interpt{\implti{\progp_1}}$
      (and $k \neq 0$), then by induction hypothesis we have $E'$, $\po'$, and
      $f$ such that $\tup{\tup{v, k}, \tup{E', \po'}} \in \interpt{\progp_1}$
      and $\wideabsf{L}{I}{\tup{E, \po}}{\tup{E', \po'}}$. Then
      $\tup{\tup{v, k}, \tup{E', \po'}} \in \interpt{\progp}$ also holds and we
      are done.
    \item Else there is $E_1,E_2,\po_1,\po_2,v'$ such that $E = E_1 \cup E_2$,
      $\po = \po_1 \cup \po_2 \cup (E_1 \times E_2)$,
      $\tup{\tup{v',0}, \tup{E_1,\po_1}} \in \interpt{\implti{\progp_1}}$, and
      $\tup{\tup{v,k}, \tup{E_2,\po_2}} \in \interpt{\implti{\progp_2 \ v}}$. By
      induction hypothesis, there is $E'_1,E'_2,\po'_1,\po'_2,f_1,f_2$ such that
      $\tup{\tup{v',0}, \tup{E'_1,\po'_1}} \in \interpt{\progp_1}$,
      $\tup{\tup{v,k}, \tup{E'_2,\po'_2}} \in \interpt{\progp_2 \ v}$,
      $\wideabsf[f_1]{L}{I}{\tup{E_1, \po_1}}{\tup{E'_1, \po'_1}}$, and
      $\wideabsf[f_2]{L}{I}{\tup{E_2, \po_2}}{\tup{E'_2, \po'_2}}$. We choose
      $E' = E'_1 \cup E'_2$, $\po' = \po'_1 \cup \po'_2 \cup (E'_1 \times E'_2)$
      and we have
      $\tup{\tup{v, k}, \tup{E', \po'}} \in \interpt{\texttt{let} \ \progp_1 \
        (\lambda v. \progp_2 \ v)}$ by definition. We define
      $f : (E_1 \cup E_2) \rightarrow (E'_1 \cup E'_2)$ as the sum of $f_1$ (on
      $E_1$) and $f_2$ (on $E_2$). We are left to show
      $\wideabsf{L}{I}{\tup{E, \po}}{\tup{E', \po'}}$.

      \begin{itemize}
      \item
        $E' = E'_1 \cup E'_2 = f_1(E_1) \cup f_2(E_2) = f(E_1 \cup E_2) = f(E)$
      \item $(E_1 \cup E_2)\rst{L} = E_1\rst{L} \cup E_2\rst{L} = \emptyset$
      \item If $x \in E_i$ and $f(x) = f_i(x) \not \in L$, then
        $f(x) = x$ since the property holds for $f_i$
      \item
        $f(\po) = f(\po_1 \cup \po_2 \cup (E_1 \times E_2)) = f(\po_1) \cup
        f(\po_2) \cup (f(E_1) \times f(E_2)) \suq \po'_1 \cup \Id\rst{E'_1} \cup
        \po'_2\rst{E'_2} \cup (E'_1 \times E'_2) = \po' \cup \Id$.
      \item If
        $(f(\action_1),f(\action_2)) \in \po' = \po'_1 \cup \po'_2 \cup (E'_1
        \times E'_2)$, then we have three cases. If
        $(f(\action_1),f(\action_2)) \in \po'_1$, then
        $(\action_1,\action_2) \in \po_1 \suq \po$. If
        $(f(\action_1),f(\action_2)) \in \po'_2$, then
        $(\action_1,\action_2) \in \po_2 \suq \po$. Finally, if
        $f(\action_1) \in E'_1$ and $f(\action_2) \in E'_2$, then
        $\action_1 \in E_1$ and $\action_2 \in E_2$, and so
        $(\action_1,\action_2) \in (E_1 \times E_2) \suq \po$.
      \item If
        $\action' = (\threadt,\aident,\tup{m,\vect{v},v'}) \in E'_i\rst{L}$,
        from our hypothesis we know
        $\tup{\tup{v',0}, \tup{E_i, \po_i}\rst{f_i^{-1}(\action')}} \in
        \interpt{I(\threadt,m,\vect{v})}$. We simply have
        $\tup{E_i, \po_i}\rst{f_i^{-1}(\action')} = \tup{E,
          \po}\rst{E_i}\rst{f_i^{-1}(\action')} = \tup{E,
          \po}\rst{f^{-1}(\action')}$, so
        $\tup{\tup{v',0}, \tup{E, \po}\rst{f^{-1}(\action')}} \in
        \interpt{I(\threadt,m,\vect{v})}$ holds
      \end{itemize}
    \end{itemize}

  \item Similarly for $\progp$ of the shape $\texttt{loop} \ \progp'$.
  \end{itemize}

  From this, we can conclude the lemma. Given $\progps$ and an implementation
  $I$ of $L$ using $\Lambda$, if
  $\tup{(v_1,\ldots,v_T), G} \in \interp{\impli{\progps}}$ then by definition
  $G$ is of the form $G = \parallel_{1 \leq t \leq T} G_t$ and
  $\forall 1 \leq t \leq T. \tup{\tup{v_t,0},G_t} \in
  \interpt{\implti{\progps(t)}}$. Using the result above, we have
  $G'_1,\ldots,G'_T$ and $f_1,\ldots,f_T$ such that
  $\tup{\tup{v_t,0},G'_t} \in \interpt{\progps(t)}$ and
  $\wideabsf[f_t]{L}{I}{G_t}{G'_t}$. We define
  $G' = \parallel_{1\leq t \leq T} G'_t$ and $f : G'.E \rightarrow G.E$ the sum
  of $f_1,\ldots,f_T$. We have $\tup{\vect{v}, G'} \in \interp{\progps}$ by
  definition, and we can easily show $\wideabsf{L}{I}{G}{G'}$ similarly to the
  proof above.
\end{proof}



\subsection{Locally Sound Implies Sound}

\begin{theorem}
  \label{thm:sound-sound}
  If a well-defined implementation is locally sound, then it is sound.
\end{theorem}
\begin{proof}
  Let $I$ be a locally sound implementation of $L$ using $\Lambda$. Let
  $\progps$ such that $\loc(I) \cap \loc(\progps) = \emptyset$. We need to show
  that
  $\outcome_\Lambda(\impli{\progps}) \suq
  \outcome_{\Lambda\uplus\set{L}}(\progps)$.

  Let $\tup{E, \po, \gettags, \so, \hb}$ $\Lambda$-consistent such that
  $\tup{\vect{v}, \tup{E, \po}} \in \interp{\impli{\progps}}$. From
  \Cref{lem:find-abstraction}, there is $E',\po',f$ such that
  $\tup{\vect{v}, \tup{E', \po'}} \in \interp{\progps}$ and
  $\wideabsf{L}{I}{\tup{E, \po}}{\tup{E', \po'}}$.

  Let $E_L \defeq E'\rst{L}$ and $E_p \defeq E' \setminus E_L$. We also note
  $\po_L \defeq \po'\rst{E_L}$ and $\po_p \defeq \po'\rst{E'_p}$. By definition
  of $\wideabsf{L}{I}{\tup{E, \po}}{\tup{E', \po'}}$, we have $E_p \suq E$ and
  $f\rst{E_p} = \Id\rst{E_p}$: by surjectivity if $\action \in E_p$ then there
  is $\action_0 \in E$ such that $f(\action_o) = \action$, but since
  $\action \not\in L$ we have $f(\action_o) = \action_o = \action$ and thus
  $\action \in E$.

  We note $E_i = E \setminus E_p$, and create notations such that
  $\tup{E_i, \po_i, \gettags_i, \so_i, \hb_i} = \tup{E, \po, \gettags, \so,
    \hb}\rst{E_i}$ and
  $\tup{E_p, \po_p, \gettags_p, \so_p, \hb_p} = \tup{E, \po, \gettags, \so,
    \hb}\rst{E_p}$. Thus $E' = E_L \cup E_p$ and $E = E_i \cup E_p$.
  Intuitively, $E_i$ is the implementation of $E_L$ while the common part $E_p$
  is not modified.

  We note $f_i = f\rst{E_i}$. We can easily check that
  $\absf[f_i]{L}{I}{\tup{E_i, \po_i}}{\tup{E_L, \po_L}}$ holds:
  \begin{itemize}
  \item $E_L = f_i(E_i)$: Let $\action \in E_L$, since $f$ is surjective there
    is $\action' \in E$ such that $f(\action') = \action$. Since
    $f\rst{E_p} = \Id\rst{E_p}$, for $\action_0 \in E_p$ we have
    $f(\action_0) = \action_0 \not\in E_L$, so $\action' \in E_i$.
  \item $E_i\rst{L} = \emptyset$ since $E_i \suq E$ and $E\rst{L} = \emptyset$.
  \item $E_L = E_L\rst{L}$ by definition.
  \item Let $\action_1,\action_2 \in E_i$, if
    $f_i(\action_1) \neq f_i(\action_2)$ then by
    $\wideabsf{L}{I}{\tup{E, \po}}{\tup{E', \po'}}$ we have
    $(f_i(\action_1), f_i(\action_2)) \in \po'\rst{E_L} = \po_L$.
  \item Let $\action_1,\action_2 \in E_i$ such that
    $(f_i(\action_1),f_i(\action_2)) \in \po_L \suq \po$. By
    $\wideabsf{L}{I}{\tup{E, \po}}{\tup{E', \po'}}$ we have
    $(\action_1,\action_2) \in \po\rst{E_i} = \po_i$.
  \item Let $\action' = (\threadt,\aident,\tup{m,\vect{v},v'}) \in E_L$. From
    $\wideabsf{L}{I}{\tup{E, \po}}{\tup{E', \po'}}$ we have
    $\tup{\tup{v',0}, \tup{E, \po}\rst{f^{-1}(\action')}} \in
    \interpt{I(\threadt,m,\vect{v})}$. Since $\action' \in E_L$ and
    $f\rst{E_p} = \Id\rst{E_p}$, we have
    $f^{-1}(\action') = f_i^{-1}(\action') \suq E_i$ and thus
    $\tup{E, \po}\rst{f^{-1}(\action')} = \tup{E_i,
      \po_i}\rst{f_i^{-1}(\action')}$. So
    $\tup{\tup{v',0}, \tup{E_i, \po_i}\rst{f_i^{-1}(\action')}} \in
    \interpt{I(\threadt,m,\vect{v})}$.
  \end{itemize}

  Next, let us show that $\tup{E_i, \po_i, \gettags_i, \so_i, \hb_i}$ is
  $\Lambda$-consistent. The first two points are trivial, and we need to show
  that for any library $L' \in \Lambda$ we have
  $\tup{E_i, \po_i, \gettags_i, \so_i, \hb_i}\rst{L'}$ $L'$-consistent. By
  hypothesis, we already know that
  $\tup{(E_i \cup E_p), \po, \gettags, \so, \hb}\rst{L'}$ is $L'$-consistent.
  Thus, by the decomposability property, it would be enough to show that
  $\loc(E_i) \cap \loc(E_p) = \emptyset$. Since $E_p \suq E'$ and
  $\tup{\vect{v}, \tup{E', \po'}} \in \interp{\progps}$, we know that
  $\loc(E_p) \suq \loc(E') \suq \loc(\progps)$. Since
  $\loc(I) \cap \loc(\progps) = \emptyset$, it would be enough to show
  $\loc(E_i) \suq \loc(I)$. Let $\action \in E_i$, we have
  $f_i(\action) \in E_L$ of the form $(\threadt,\aident,\tup{m,\vect{v_0},v'})$.
  By definition of local abstraction,
  $\tup{\tup{v',0}, \tup{E_L, \po_L}\rst{f_i^{-1}(f_i(\action))}} \in
  \interpt{I(\threadt,m,\vect{v_0})}$ and
  $\action \in E_L\rst{f_i^{-1}(f_i(\action))}$. By definition of implementation,
  we have
  $\loc(\action) \suq \loc(E_L\rst{f_i^{-1}(f_i(\action))}) \suq \loc(I)$.

  Since $I$ is locally sound, we can use
  $\tup{E_i, \po_i, \gettags_i, \so_i, \hb_i}$ $\Lambda$-consistent and
  $\absf[f_i]{L}{I}{\tup{E_i, \po_i}}{\tup{E_L, \po_L}}$ to produce
  $\gettags_L$, $g_i$, and $\so_L$ such that:
  \begin{itemize}
  \item $g_i(\action', \tagt') = (\action, \tagt)$ implies
    $f_i(\action) = \action'$ and
    \begin{itemize}
    \item Forall $\tagt_0$ such that $(\tagt_0, \tagt') \in \tagppo$, there exists
      $(\action_1, \tagt_1) \in \SEvents_i$ such that
      $f_i(\action_1) = \action'$, $(\tagt_0, \tagt_1) \in \tagppo$, and
      ($(\action_1, \tagt_1), (\action, \tagt)) \in (\hb_i \cup \Id)$;
    \item Forall $\tagt_0$ such that $(\tagt', \tagt_0) \in \tagppo$, there
      exists $(\action_2, \tagt_2) \in \SEvents_i$ such that
      $f_i(\action_2) = \action'$, $(\tagt_2, \tagt_0) \in \tagppo$, and
      ($(\action, \tagt), (\action_2, \tagt_2)) \in (\hb_i \cup \Id)$.
    \end{itemize}
  \item $g_i(\so_L) \suq \hb_i$;
  \item Forall $\hb_L$ transitive such that $(\ppo_L \cup \so_L)^+ \suq \hb_L$
    and $g_i(\hb_L) \suq \hb_i$, we have \\
    $\tup{E_L, \po_L, \gettags_L, \so_L, \hb_L} \in L.\cons$, where
    $\ppo_L \defeq \tup{E_L, \po_L, \gettags_L}.\ppo$.
  \end{itemize}

  We define $\gettags'$ on $E'$ by the sum of $\gettags_L$ and $\gettags_p$. We
  define $\so' \defeq \so_L \cup \so_p$, as well as
  $\ppo' \defeq \tup{E', \po', \gettags'}.\ppo$, and
  $\hb' \defeq (\ppo' \cup \so')^+$. We extend
  $g_i : \tup{E_L, \gettags_L}.\SEvents \rightarrow \tup{E_i,
    \gettags_i}.\SEvents$ into a function
  $g : \tup{E', \po', \gettags'}.\SEvents \rightarrow \tup{E, \po,
    \gettags}.\SEvents$ using the identity function. I.e., for
  $(\action', \tagt') \in \tup{E_p, \gettags_p}.\SEvents$ we have
  $g(\action', \tagt') = (\action', \tagt')$. The first property above on $g_i$
  and $f_i$ carries over to $g$ and $f$, by taking, for any stamp, the
  intermediary subevents to be the output of $g$ itself.

  As an important intermediary result, let us show $g(\hb') \suq \hb$. Since
  $\hb' = (\ppo' \cup \so')^+$, we need to show the inclusion for each
  component.
  $g(\so') = g_i(\so_L) \cup \Id(\so_p) \suq \hb_i \cup \so_p \suq \hb$ is
  immediate, and we are left with proving $g(\ppo') \suq \hb$. Let
  $(\action'_1, \tagt'_1), (\action'_2, \tagt'_2) \in \SEvents'$ such that
  $(\action'_1, \action'_2) \in \po'$ and $(\tagt'_1, \tagt'_2) \in \tagppo$.
  Let $(\action_i, \tagt_i) \defeq g(\action'_i, \tagt'_i)$ ($i \in \set{1,2}$),
  we need to show that $((\action_1, \tagt_1), (\action_2, \tagt_2)) \in \hb$.
  From the properties of $g$, using $(\action_1, \tagt_1)$ and the stamp
  $\tagt_2$, there is $(\action^g_1, \tagt^g_1)$ such that
  $f(\action^g_1) = \action'_1$, $(\tagt^g_1, \tagt_2) \in \tagppo$, and
  $((\action_1, \tagt_1), (\action^g_1, \tagt^g_1)) \in \hb$. From the
  properties on $g$, using $(\action_2, \tagt_2)$ and the stamp $\tagt^g_1$, there
  is $(\action^g_2, \tagt^g_2)$ such that $f(\action^g_2) = \action'_2$,
  $(\tagt^g_1, \tagt^g_2) \in \tagppo$, and
  $((\action^g_2, \tagt^g_2), (\action_2, \tagt_2)) \in \hb$.
  We know that $\action'_1 = f(\action^g_1)$ and $\action'_2 = f(\action^g_2)$,
  so from $\wideabsf{L}{I}{\tup{E, \po}}{\tup{E', \po'}}$ and
  $(\action'_1, \action'_2) \in \po'$ we have
  $(\action^g_1, \action^g_2) \in \po$. Thus
  $((\action^g_1, \tagt^g_1), (\action^g_2, \tagt^g_2)) \in \ppo$, and by
  transitivity we have $((\action_1, \tagt_1), (\action_2, \tagt_2)) \in \hb$.
  This finishes the intermediary result $g(\hb') \suq \hb$.

  To conclude the theorem, we want to show that
  $\tup{E', \po', \gettags', \so', \hb'}$ is $(\Lambda\cup\set{L})$-consistent.
  \begin{itemize}
  \item The first few points hold because $\hb'$ is irreflexive, since
    $g(\hb') \suq \hb$ and $\hb$ is irreflexive.

  \item For $L' \in \Lambda$, we need to show that
    $\tup{E_p, \po_p, \gettags_p, \so_p, \hb'}\rst{L'}$ is $L'$-consistent. We
    know that $\tup{E_p, \po_p, \gettags_p, \so_p, \hb_p}\rst{L'}$ is
    $L'$-consistent, so by monotonicity it is enough to show
    $\hb'\rst{E_p} \suq \hb\rst{E_p}$, which holds because
    $\hb'\rst{E_p} = g(\hb')\rst{E_p} \suq \hb\rst{E_p}$.

  \item For $L$, we need to show that
    $\tup{E_L, \po_L, \gettags_L, \so_L, \hb'\rst{L}}$ is $L$-consistent. Using
    our hypothesis, it is enough to show that $\hb_L \defeq \hb'\rst{L}$, which
    is transitive and includes $(\ppo_L \cup \so_L)^+$, satisfies
    $g_i(\hb_L) \suq \hb_i$. Once again, this holds because
    $g_i(\hb_L) = g(\hb'\rst{E_L}) \suq g(\hb')\rst{E_i} \suq \hb\rst{E_i} =
    \hb_i$.
  \end{itemize}
\end{proof}

\input{waittotso}

\section{Correctness Proofs of the Core LOCO Libraries}
\label{sec:proofs-rdma}

\subsection{\brl Library}

\begin{theorem}
  \label{thm:brl-sound}
  The implementation \implbrl of the \brl library into \rdmawait given in the
  paper is locally sound.
\end{theorem}

\begin{proof}
  We assume an $\set{\rdmawait}$-consistent execution
  $\exec = \tup{E, \po, \gettags, \so, \hb}$ which is abstracted via $f$ to
  $\tup{E',\po'}$ that uses the \brl library, \ie
  $\absf{\brl}{\implbrl}{\tup{E, \po}}{\tup{E', \po'}}$ holds. We need to
  provide $\gettags'$, $\so'$, and
  $g : \tup{E',\po',\gettags'}.\SEvents \rightarrow \exec.\SEvents$ respecting some
  conditions. From $\tup{E',\po'}$, we simply take $\gettags' = \gettagsbrl$. We
  note $\SEvents'$ for $\tup{E',\po',\gettags'}.\SEvents$.

  Since $\exec$ is $\set{\rdmawait}$-consistent, it means
  $(\ppo \cup \so)^+ \suq \hb$, $\hb$ is transitive and irreflexive, and $\exec$ is
  \rdmawait-consistent. Firstly, it means that for all thread $\threadt$ we have
  $\po\rst{\threadt}$ is a strict total order. From the properties of
  $\absf{\brl}{\implbrl}{\tup{E, \po}}{\tup{E', \po'}}$, we can easily see
  that it implies $\po'\rst{\threadt}$ is also a strict total order. Secondly,
  there exists well-formed \vr, \vw, \rf, \mo, and \ro such that $\ib$ is
  irreflexive, $\gettags = \gettagsrl$, and
  $\so = \iso \cup \rfe \cup \pfget \cup \ro \cup \fr \cup \mo \cup
  ([\Inst];\ib)$.

  We define $g$ as follows.
  \begin{itemize}
  \item For an event $\action' = (\threadt, \_, (\brlread, (x), v))$, the only
    subevent is $(\action', \tagcread) \in \SEvents'$. By definition of the
    abstraction $f$, the set
    $\interpt{\implbrl(\threadt, \brlread, (x))} =
    \interpt{\rlread(x_{\nodefun{\threadt}})} = \linebreak \setpred{\tup{
        \tup{v',0},
        \set{(\threadt,\aident,\tup{\rlread,x_{\nodefun{\threadt}},v'})}_G }}{v'
      \in \Val \land \aident \in \AId}$ contains
    $\tup{\tup{v,0}, \tup{E, \po}\rst{f^{-1}(\action')}}$, so there is an event
    $\action = (\threadt, \_, (\rlread, (x_{\nodefun{\threadt}}), v)) \in E$ with
    $f(\action) = \action'$. From the definition of $\gettagsrl$, it is
    associated to a single subevent $(\action, \tagcread) \in \exec.\SEvents$, and
    we define $g(\action', \tagcread) = (\action, \tagcread)$. The first
    condition of $g$ trivially holds for this input since the output uses the
    same stamp: for any stamp $\tagt_0$ we can choose
    $(\action_1,\tagt_1) = (\action_2,\tagt_2) = (\action, \tagcread)$ using the
    $\Id$ function, and the \tagppo order is preserved for any previous or later
    stamp.

  \item For an event $\action' = (\threadt, \_, (\brlwrite, (x,v), ()))$, a
    similar reasoning allows us to choose
    $g(\action', \tagcwrite) = (\action, \tagcwrite)$ with
    $\action = (\threadt, \_, (\rlwrite, (x_{\nodefun{\threadt}},v), ())) \in
    f^{-1}(\action')$.

  \item For an event $\action' = (\threadt, \_, (\brlwait, (\wid), ()))$, a
    similar reasoning allows us to choose
    $g(\action', \tagwait) = (\action, \tagwait)$ with
    $\action = (\threadt, \_, (\rlwait, (\wid), ())) \in f^{-1}(\action')$.

  \item For an event
    $\action' = (\threadt, \_, (\brlbr, (x, \wid, \set{n_1;\ldots;n_k}), ()))$
    and a subevent $(\action', \tagnlr)$, since the implementation of $\action'$
    contains $\rlput(x_{\node}, x_{\nodefun{\threadt}}, \wid)$, the abstraction $f$
    similarly implies an event
    $\action = (\threadt, \_, (\rlput, (x_{\node}, x_{\nodefun{\threadt}}, \wid),
    ())) \in f^{-1}(\action')$. As before, given \gettagsrl, we can choose
    $g(\action', \tagnlr) = (\action, \tagnlr)$ and the first condition on $g$
    holds using the identity function.

  \item Similarly for an event
    $\action' = (\threadt, \_, (\brlbr, (x, \wid, \set{n_1;\ldots;n_k}), ()))$
    and a subevent $(\action', \tagnrw)$, we can choose
    $g(\action', \tagnrw) = (\action, \tagnrw)$ with
    $\action = (\threadt, \_, (\rlput, (x_{\node}, x_{\nodefun{\threadt}}, \wid),
    ())) \in f^{-1}(\action')$.

  \item For an event
    $\action' = (\threadt, \_, (\brlgf, (\set{n_1;\ldots;n_k}), ()))$ and a
    subevent $(\action', \taggf)$, the relevant part of the implementation
    $f^{-1}(\action')$ of $\action'$ contains an event of label
    $\rlget(\dumloc_{\nodefun{\threadt}}, \dumloc_{n}, \wid_0)$ (with stamps
    $\tagnrr$ and $\tagnlw$) followed by one of label $\rlwait(\wid_0)$ (with
    stamp $\tagwait$). Since the restrictive stamp $\taggf$ needs to be implemented
    using weaker stamps, the choice of $g$ is more delicate. We choose for $g$ to
    map to the last subevent, \ie $g(\action', \taggf) = (\action, \tagwait)$
    with
    $\action = (\threadt, \_, (\rlwait, (\wid_0), ())) \in f^{-1}(\action')$,
    and we need to check the stamp ordering is preserved. For a later stamp
    $\tagt_0$ such that $(\taggf, \tagt_0) \in \tagppo$, we can simply use
    $(\action_2, \tagt_2) = (\action, \tagwait)$ using the $\Id$ function. We
    have $(\tagwait, \tagt_0) \in \tagppo$ by definition (in~\Cref{fig:to},
    lines E and K are identical). For an earlier stamp $\tagt_0$ such that
    $(\tagt_0, \taggf) \in \tagppo$, we use the entry point
    $(\action_1, \tagt_1) = ((\threadt, \_, (\rlget,(\dumloc_{\nodefun{\threadt}},
    \dumloc_{n}, \wid_0),())), \tagnlw)$. As previously, we clearly have
    $\action_1 \in f^{-1}(\action')$. We have $(\tagt_0, \tagnlw) \in \tagppo$
    by definition (in~\Cref{fig:to}, columns 9 and 11 are identical). We also
    need to check that $((\action_1, \tagnlw), (\action, \tagwait)) \in \hb$.
    Since $\exec$ is $\set{\rdmawait}$-consistent, this is simply because
    $((\action_1, \tagnlw), (\action, \tagwait)) \in \pfget \subseteq \so
    \subseteq \hb$ as the two events are in \po and use the same identifier
    $\wid_0$.
  \end{itemize}

  Now we need to find $\so'$ such that $g(\so') \suq \hb$ and such that
  $\exec' = \tup{E', \po', \gettags', \so', \hb'}$ is \brl-consistent for any
  reasonable $\hb'$. Actually, since $\hb'$ does not appear in the consistency
  predicate, we can ignore the properties of $\hb'$ and we need to check that
  $\tup{E', \po', \gettags', \so', \_}$ is \brl-consistent. For this, we need to
  choose well-formed $\vr'$, $\vw'$, $\rf'$, and $\mo'$.

  For $\vr'$ and $\vw'$, we simply take
  $\vr'(\saction') \defeq \vr(g(\saction'))$ and
  $\vw'(\saction') \defeq \vw(g(\saction'))$. For the methods $\brlwrite$ and
  $\brlread$, these new functions $\vr'$ and $\vw'$ respect the value
  read/written, since $\vr$ and $\vw$ do so in $\exec.\SEvents$. Similarly, if
  $\saction'_1 = (\action', \tagnlr) \in \SEvents'$ and
  $\saction'_2 = (\action', \tagnrw) \in \SEvents'$ (so $\action'$ calls the
  \brlbr method), then by definition of $g$ they are mapped to
  $\saction_1 = (\action, \tagnlr)$ and $\saction_2 = (\action, \tagnrw)$ using
  the same \rlput event and so
  $\vr'(\saction'_1) = \vr(\saction_1) = \vw(\saction_2) = \vw'(\saction'_2)$
  since $\vr$ and $\vw$ are well-formed.

  We define $\rf' \defeq \bigcup_{\node} \rf^{\node}$ and
  $\mo' \defeq \bigcup_{x,\node} \mo_x^{\node}$ from $\rf$ and $\mo$ as follows:
  \begin{align*}
    \rf^n & \defeq \setpred{(w,r)}{r \in \exec'.\Read^n \land (g(w),g(r)) \in \rf}\\
    \mo_x^n & \defeq \setpred{(w_1,w_2)}{(g(w_1),g(w_2)) \in \mo_{n(x)}}
  \end{align*}

  It is straightforward to check that
  $\rf^{\node} \subseteq \exec'.\Write^\node \times \exec'.\Read^\node$. If
  $(\saction'_1, \saction'_2) \in \rf^{\node}$, then
  $(g(\saction'_1), g(\saction'_2)) \in \rf$ and
  $\vw'(\saction'_1) = \vw(g(\saction'_1)) = \vr(g(\saction'_2)) =
  \vr'(\saction'_2)$.

  We argue that if $\saction'_2 \not\in \img{\rf^\node}$ then
  $g(\saction'_2) \not\in \img{\rf}$. This might not be obvious since \rf is
  bigger, as it has for instance statements about the $\dumloc_{n}$ locations.
  The reason is that, for each node $\node$ and \brl location $x$, the relation
  $\inv{g}$ is total and functional on $\exec.\Write_x^\node$, \ie every write
  subevent in the implementation (outside those on the dummy locations
  $\dumloc_{n}$) is associated with a write subevent of the \brl library. This
  can be checked by considering $\implbrl$ and the different cases of the
  definition of $g$. Thus if $(\saction_1, g(\saction'_2)) \in \rf$ there is
  $\saction'_1$ such that $g(\saction'_1) = \saction_1$ and
  $\saction'_2 \in \img{\rf^{\node}}$. So for a subevent
  $\saction'_2 \not\in \img{\rf^\node}$ we have
  $g(\saction'_2) \not\in \img{\rf}$ and
  $\vr'(\saction'_2) = \vr(g(\saction'_2)) = 0$.

  We also need to check that each $\mo_x^{\node}$ is a strict total order on
  $\exec'.\Write_x^{\node}$. This is simply because for all
  $\saction' \in \exec'.\Write_x^{\node}$ we have
  $g(\saction') \in \exec.\Write_{\nodefun{x}}$, and we know $\mo$ is a strict total
  order on $\exec.\Write_{\nodefun{x}}$.

  We now prove that $g(\so') \subseteq \hb$, which can be checked component by
  component.
  \begin{itemize}
  \item If $(\saction'_1, \saction'_2) \in \iso'$, then there is $\node$ and
    $\action' = (\threadt, \_, (\brlbr, (x, \_, \set{\ldots;\node;\ldots}), \_))
    \in E'$ such that $\saction'_1 = (\action', \tagnlr)$ and
    $\saction'_2 = (\action', \tagnrw)$. By definition of $g$, there is
    $\action = (\threadt, \_, (\rlput, (x_{\nodefun{\threadt}}, \_, \_), \_)) \in
    f^{-1}(\action')$ such that $g(\saction'_1) = (\action, \tagnlr)$ and
    $g(\saction'_2) = (\action, \tagnrw)$. And by definition of $\exec.\iso$, we
    have
    $(g(\saction'_1), g(\saction'_2)) \in \exec.\iso \subseteq \so \subseteq
    \hb$.
  \item By definition $g(\rf') \subseteq \rf$. We want to show
    $g(\rfe') \subseteq \rfe \subseteq \so \subseteq \hb$. Note that for all
    node $\node$ and subevent
    $\saction' \in \exec'.\Read^\node \cup \exec'.\Write^\node$, the function $g$ maps
    to a subevent using the same stamp: $\saction'.\tagt = g(\saction').\tagt$.
    Also, from the abstraction $f$, we know that $g$ preserves the program
    order: if $(\saction'_1, \saction'_2) \in \po'$, then
    $(g(\saction'_1), g(\saction'_2)) \in \po$. Thus $g$ preserves the
    internal/external distinction: $g(\rfi') \subseteq \rfi$ and
    $g(\rfe') \subseteq \rfe$, which implies $g(\rfe') \subseteq \hb$.
  \item If $(\saction'_1, \saction'_2) \in \pf'$, then by definition there is
    $\wid$, $\node$,
    $\action'_1 = (\_, \_, (\brlbr, (\_, \wid, \set{\ldots;\node;\ldots}),
    \_))$, and $\action'_2 = (\_, \_, (\brlwait, (\wid), \_))$ such that
    $(\action'_1, \action'_2) \in \po'$, $\saction'_1 = (\action'_1, \tagnlr)$,
    and $\saction'_2 = (\action'_2, \tagwait)$.

    From the abstraction $f$ and the definition of $g$, there is
    $\action_1 = (\_, \_, (\rlput, (\_, \_, \wid), \_))$ and
    $\action_2 = (\_, \_, (\rlwait, (\wid), \_))$ such that
    $(\action_1, \action_2) \in \po$, $g(\saction'_1) = (\action_1, \tagnlr)$,
    and $g(\saction'_2) = (\action_2, \tagwait)$. We have
    $(\action_1, \tagnlr) \arr{\exec.\iso} (\action_1, \tagnrw) \arr{\exec.\pfput}
    (\action_2, \tagwait)$, and so
    $((\action_1, \tagnlr), (\action_1, \tagnlr)) \in \exec.\ib$. Since
    $(\action_1, \tagnlr) \in \exec.\tagnlr \subseteq \exec.\Inst$, we have
    $(g(\saction'_1), g(\saction'_2)) \in ([\exec.\Inst];\exec.\ib) \subseteq \so
    \subseteq \hb$.

  \item We can check that $g(\fr') \subseteq \fr$. If
    $(\saction'_1, \saction'_2) \in \fr^\node$, then by definition there is $x$
    such that $\saction'_1 \in \exec'.\Read^\node$,
    $\saction'_2 \in \exec'.\Write_x^\node$,
    $\loc(\saction'_1) = x = \loc(\saction'_2)$, and either
    $(\saction'_1,\saction'_2) \in (\inv{(\rf^\node)}; \mo_x^\node)$ or
    $\saction'_2 \not\in \img{\rf^\node}$. Since
    $\exec'.\Read^\node \cap \exec'.\Write^\node = \emptyset$, as the library does not
    have any read-modify-write method, we also know
    $\saction'_1 \neq \saction'_2$ and by definition of $g$ that
    $g(\saction'_1) \neq g(\saction'_2)$.
    \begin{itemize}
    \item If there is $\saction'_3$ such that
      $(\saction'_3,\saction'_1) \in \rf^\node$ and
      $(\saction'_3,\saction'_2) \in \mo_x^\node$, then by definition
      $(g(\saction'_3),g(\saction'_1)) \in \rf$ and
      $(g(\saction'_3),g(\saction'_2)) \in \mo_{\nodefun{x}}$, so
      $(g(\saction'_1), g(\saction'_2)) \in \fr$.
    \item If $\saction'_2 \not\in \img{\rf^\node}$ then
      $g(\saction'_2) \not\in \img{\rf}$ (proved earlier) and
      $(g(\saction'_1), g(\saction'_2)) \in \fr$.
    \end{itemize}
    And so $g(\fr') \subseteq \fr \subseteq \so \subseteq \hb$.
  \item Finally we have $g(\mo') \subseteq \mo \subseteq \so \subseteq \hb$
    by definition.
  \end{itemize}
  Thus $g(\so') \subseteq \hb$.

  Lastly, we are left to prove that
  $[\tagcread] ; (\inv{\po'} \cap \fr') ; [\tagcwrite] = \emptyset$. This comes
  from the fact that
  $[\tagcread] ; (\inv{\po} \cap \fr) ; [\tagcwrite] \subseteq \fri \subseteq
  \ib$, $[\tagcwrite] ; \po ; [\tagcread] \subseteq \ippo \subseteq \ib$, and
  $g(\fr') \subseteq \fr$ (proved earlier).
  So if
  $(\saction'_1, \saction'_2) \in [\tagcread] ; (\inv{\po'} \cap \fr') ;
  [\tagcwrite]$, we have
  $(g(\saction'_1), g(\saction'_2)) \in [\tagcread] ; (\inv{\po} \cap \fr) ;
  [\tagcwrite] \subseteq \ib \cap \inv{\ib} = \emptyset$ which is not possible,
  since we know $\ib$ is transitive and irreflexive.

  Thus $\exec'$ is \brl-consistent and the implementation \implbrl is locally sound.
\end{proof}

\begin{corollary}
  The implementation \implbrl is sound.
\end{corollary}

\subsection{\msw Library}

\begin{theorem}
  \label{thm:msw-sound}
  Given a function $\mswsize$, the implementation \implmsw of the \msw library
  into \rdmawait given in the paper is locally sound.
\end{theorem}
\begin{proof}
  We assume an $\set{\rdmawait}$-consistent execution
  $\exec = \tup{E, \po, \gettags, \so, \hb}$ which is abstracted via $f$ to
  $\tup{E',\po'}$ that uses the \msw library, \ie
  $\absf{\msw}{\implmsw}{\tup{E, \po}}{\tup{E', \po'}}$ holds. We need to
  provide $\gettags'$, $\so'$, and
  $g : \tup{E',\po',\gettags'}.\SEvents \rightarrow \exec.\SEvents$ respecting some
  conditions. From $\tup{E',\po'}$, we simply take $\gettags' = \gettagsmsw$.

  Since the implementation \implmsw maps events that do not respect the \mswsize
  function to non-terminating loops, the abstraction $f$ tells us that every
  event in $E'$ does respect the \mswsize.

  Since $\exec$ is $\set{\rdmawait}$-consistent, it means
  $(\ppo \cup \so)^+ \suq \hb$, $\hb$ is transitive and irreflexive, and $\exec$ is
  \rdmawait-consistent. Firstly, it means that $\tup{E, \po}$ respects nodes.
  From the properties of $\absf{\msw}{\implmsw}{\tup{E, \po}}{\tup{E', \po'}}$,
  we can easily see that it implies $\tup{E, \po}$ respects nodes, as the
  implementation locations are mapped to the same nodes:
  $\nodefun{x_1} = \ldots = \nodefun{x_{\mswsize(x)}} = \nodefun{x}$. Secondly, there
  exists well-formed \vr, \vw, \rf, \mo, and \ro such that $\ib$ is irreflexive,
  $\gettags = \gettagsrl$, and
  $\so = \iso \cup \rfe \cup \pfget \cup \ro \cup \fr \cup \mo \cup ([\Inst];\ib)$.

  We define $g$ as follows.
  \begin{itemize}
  \item For an event $\action' = (\threadt, \_, (\mswwrite, (x,\vect{v}), ()))$,
    we choose $g(\action', \tagcwrite) = (\action, \tagcwrite)$ with
    $\action = (\threadt, \_, (\rlwrite, (x_0,\mswhash(\vect{v})), ())) \in
    \inv{f}(\action')$.
  \item For an event $\action' = (\threadt, \_, (\mswread, (x), \vect{v}))$, we
    choose $g(\action', \tagcread) = (\action, \tagcread)$ with
    $\action = (\threadt, \_, (\rlread, (x_0), v_0))$ and
    $v_0 = \mswhash(\vect{v})$.
  \item For an event $\action' = (\threadt, \_, (\mswread, (x), \bot))$,
    we choose $g(\action', \tagwait) = (\action, \tagcread)$ with
    $\action = (\threadt, \_, (\rlread, (x_0), v_0)))$.
  \item For an event $\action' = (\threadt, \_, (\mswput, (x, y, \wid), ()))$,
    we choose $g(\action', \tagnlr[\nodefun{x}]) = (\action, \tagnlr[\nodefun{x_0}])$
    and $g(\action', \tagnrw[\nodefun{x}]) = (\action, \tagnrw[\nodefun{x_0}])$ with
    $\action = (\threadt, \_, (\rlput, (x_0, y_0, \wid), ())))$.
  \item For an event $\action' = (\threadt, \_, (\mswget, (x, y, \wid), ()))$,
    we choose $g(\action', \tagnrr[\nodefun{y}]) = (\action, \tagnrr[\nodefun{y_0}])$
    and $g(\action', \tagnlw[\nodefun{y}]) = (\action, \tagnlw[\nodefun{y_0}])$ with
    $\action = (\threadt, \_, (\rlget, (x_0, y_0, \wid), ())))$.
  \item For an event $\action' = (\threadt, \_, (\mswwait, (\wid), ()))$, we
    choose $g(\action', \tagwait) = (\action, \tagwait)$ with
    $\action = (\threadt, \_, (\rlwait, (\wid), ())))$.
  \end{itemize}
  This definition of $g$ clearly preserves \tagppo (first property to check)
  using the identity function, since $\tagcread$ and $\tagwait$ have the same
  relation to other stamps.

  Note that for every location $x$, every write subevent on $x_0$ in the
  implementation is in the image of $g$.

  Now we need to find $\so'$ such that $g(\so') \suq \hb$ and such that
  $\exec' = \tup{E', \po', \gettags', \so', \_}$ is \msw-consistent. For this, we
  need to choose well-formed $\vr'$, $\vw'$, $\rf'$, $\mo'$, and $\ro'$. We
  define $\vr'(\saction') = \inv{\mswhash}(\vr(g(\saction')))$, and similarly
  for $\vw'$. \Eg, when the implementation of $\mswput(x, y, \wid)$ reads (and
  writes) the values
  $\mswhash((v_1, \ldots, v_{\mswsize(x)})), v'_1, \ldots, v'_{\mswsize(x)}$, we
  pretend $\mswput(x, y, \wid)$ actually reads $(v_1, \ldots, v_{\mswsize(x)})$,
  even if the following data is corrupted and does not correspond to the hash.
  For the sake of simplicity, we assume that $\mswhash(\vect{0}) = 0$, or
  equivalently that the hash locations can be initialised to
  $\mswhash(\vect{0})$. For \mswwrite events, the $\vw'$ function matches the
  values written, as required. For a succeeding \mswread event, the if-then-else
  construct ensures that the value returned is the inverse of the hash, matching
  the $\vr'$ function as required.

  We then define
  $\rf' = \setpred{(\saction'_1, \saction'_2)}{(g(\saction'_1), g(\saction'_2))
    \in \rf}$,
  $\mo' = \setpred{(\saction'_1, \saction'_2)}{(g(\saction'_1), g(\saction'_2))
    \in \mo}$, and
  $\ro' = \setpred{(\saction'_1, \saction'_2)}{(g(\saction'_1), g(\saction'_2))
    \in \ro}$, and they are well-formed:

  \begin{itemize}
  \item If $(\saction'_1, \saction'_2) \in \rf'$, then we have
    $\vw'(\saction'_1) = \inv{\mswhash}(\vw(g(\saction'_1))) =
    \inv{\mswhash}(\vr(g(\saction'_2))) = \vr'(\saction'_2)$. If
    $\saction'_2 \not\in \img{\rf'}$ on location $x$, then since every write
    subevent on $x_0$ in the implementation is in the image of $g$ we have
    $g(\saction'_2) \not\in \img{\rf}$ and
    $\vr'(\saction'_2) = \inv{\mswhash}(\vr(g(\saction'_2))) = \inv{\mswhash}(0)
    = \vect{0}$.
  \item $\mo'_x$ is total on $\exec'.\Write_x$ since $\mo_{x_0}$ is total on
    $\exec.\Write_{x_0}$ and every write on $x_0$ is in the image of $g$.
  \item If $\threadfun{\saction'_1} = \threadfun{\saction'_2}$ and
    $(\saction'_1, \saction'_2) \in \exec'.\tagnlr \times \exec'.\tagnlw$ (\resp
    $\exec'.\tagnrr \times \exec'.\tagnrw$) then
    $\threadfun{g(\saction'_1)} = \threadfun{\saction'_1} = \threadfun{\saction'_2} =
    \threadfun{g(\saction'_2)}$ and
    $(g(\saction'_1), g(\saction'_2)) \in \exec.\tagnlr \times \exec.\tagnlw$. So
    $(g(\saction'_1), g(\saction'_2)) \in \ro \cup \inv\ro$ and we also have
    $(\saction'_1, \saction'_2) \in \ro' \cup \inv{\ro'}$.
  \end{itemize}

  It is straightforward to see that $g(\rf') \subseteq \rf$,
  $g(\mo') \subseteq \mo$, $g(\ro') \subseteq \ro$,
  $g(\pfget') \subseteq \pfget$, $g(\pfput') \subseteq \pfput$,
  $g(\ippo') \subseteq \ippo$, $g(\ppo') \subseteq \ppo$,
  $g(\rfe') \subseteq \rfe$, and $g(\iso') \subseteq \iso$. The only non-obvious
  relation might be $g(\fr') \subseteq \fr$. Let
  $(\saction'_1, \saction'_2) \in \fr'$:
  \begin{itemize}
  \item If $\saction'_2 \not\in \img{\rf'}$, as mentioned earlier we have
    $g(\saction'_2) \not\in \img{\rf}$ and thus
    $(g(\saction'_1), g(\saction'_2)) \in \fr$.
  \item If there is $\saction'_3$ such that
    $(\saction'_3, \saction'_1) \in \rf'$ and
    $(\saction'_3, \saction'_2) \in \mo'$, then we have
    $(g(\saction'_3), g(\saction'_1)) \in \rf$ and
    $(g(\saction'_3), g(\saction'_2)) \in \mo$, and so
    $(g(\saction'_1), g(\saction'_2)) \in \fr$.
  \end{itemize}
  And of course $g(\fri') \subseteq \fri$ also holds since the stamps are
  preserved.

  Thus we have $g(\ib') \subseteq \ib$, implying $\ib'$ is irreflexive,
  $g(\so') \subseteq \so \subseteq \hb$, and we have
  $\exec' = \tup{E', \po', \gettags', \so', \_}$ is \msw-consistent.
\end{proof}

\begin{corollary}
  The implementation \implmsw is sound.
\end{corollary}

\subsection{\bal Library}

\begin{theorem}
  \label{thm:bal-sound}
  Given a function $\implbalb$, the implementation \implbal of the \bal library
  into \brl given in the paper is locally sound.
\end{theorem}
\begin{proof}
  We assume an $\set{\brl}$-consistent execution
  $\exec = \tup{E, \po, \gettags, \so, \hb}$ which is abstracted via $f$ to
  $\tup{E',\po'}$ that uses the \bal library, \ie
  $\absf{\bal}{\implbal}{\tup{E, \po}}{\tup{E', \po'}}$ holds. We need to
  provide $\gettags'$, $\so'$, and
  $g : \tup{E',\po',\gettags'}.\SEvents \rightarrow \exec.\SEvents$ respecting some
  conditions. From $\tup{E',\po'}$, we simply take $\gettags' = \gettagsbal$.

  Since $\exec$ is $\set{\brl}$-consistent, it means $(\ppo \cup \so)^+ \suq \hb$,
  $\hb$ is transitive and irreflexive, and $\exec$ is \brl-consistent. Firstly, it
  means that for all thread $\threadt$ we have $\po\rst{\threadt}$ is a strict
  total order. From the properties of
  $\absf{\bal}{\implbal}{\tup{E, \po}}{\tup{E', \po'}}$, we can easily see
  that it implies $\po'\rst{\threadt}$ is also a strict total order. Secondly,
  $\gettags = \gettagsbrl$ and there exists well-formed \vr, \vw, \rf, and \mo
  such that $[\tagcread] ; (\inv{\po} \cap \fr) ; [\tagcwrite] = \emptyset$ and
  $\so = \iso \cup \rfe \cup \pf \cup \fr \cup \mo$.

  By definition of the abstraction, for each event
  $\action' = (\threadt, \_, (\balbar, (x), ())) \in E'$ we have
  $\tup{\tup{(),0}, \tup{E, \po}\rst{f^{-1}(\action')}} \in
  \interpt{\implbal(\threadt, \balbar, (x))}$. Since by definition
  $\interpt{\texttt{loop} \{ () \}} = \emptyset$, we necessarily have
  $\threadt \in \implbalb(x)$.
  We note $s_n = \setpred{\nodefun{\threadt_i}}{\threadt_i \in \implbalb(x)}$ the
  nodes involved in the barrier. The size of $E\rst{f^{-1}(\action')}$ depends
  on how many times the loops read the locations of other threads, but this
  subgraph contains at least the global fence
  $\action_{GF} = (\threadt, \_, (\brlgf, (s_n), ()))$, the first read
  $\action_{FR} = (\threadt, \_, (\brlread, (x_\threadt), (v)))$, the write
  $\action_{W} = (\threadt, \_, (\brlwrite, (x_\threadt, {v+1}), ()))$, and the
  last read $\action_{LR} = (\threadt, \_, (\brlread, (x_{\threadt_k}), (v')))$
  with $v' > v$, such that for any other event
  $\action_0 \in E\rst{f^{-1}(\action')}$ besides these four, we have
  $\action_{GF} \arr{\po} \action_{FR} \arr{\po} \action_{W} \arr{\po}
  \action_{0} \arr{\po} \action_{LR}$. If all threads are not on the same nodes,
  we also have a broadcast event
  $\action_{BR} = (\threadt, \_, (\brlbr, (x_{\threadt}, \_, (s_n \setminus
  \set{\nodefun{\threadt}})), ()))$ with $\action_{W} \arr{\po} \action_{BR}$.

  We define $g$ as expected: $g(\action', \taggf) \defeq (\action_{GF}, \taggf)$
  for $\node \in s_n$, and
  $g(\action', \tagcread) \defeq (\action_{LR}, \tagcread)$. This clearly
  preserves \tagppo (first property of $g$) using the identity function.

  We also define $\ordero(\action') \defeq {v+1}$, \ie the value written by
  $\action_{W}$. For a location $x$, we note
  $c_x \defeq \texttt{max}_{(\action' \in E'_x)} \ordero(\action')$ the maximum
  value attributed to a barrier call on $x$. We are forced to take the only
  valid synchronisation order
  $\so' = \bigcup_{x \in \Loc} \bigcup_{1\leq i \leq c_x} \setpred{((\action'_1,
    \taggf),(\action'_2, \tagcread))}{\action'_1,\action'_2 \in (E'_x \cap
    \inv{\ordero}(i))}$ and we need to show that $g(\so') \subseteq \hb$ and
  that $\exec' = \tup{E', \po', \gettags', \so', \_}$ is \bal-consistent.

  Let us start with the conditions on $\exec'$, where we need to check that $c_x$
  and $\ordero$ respect some properties. By definition, for $\action' \in E'_x$
  we have $1 \leq \ordero(\action) \leq c_x$. For a thread
  $\threadt \not\in \implbalb(x)$, we have seen that the implementation prevents
  any event on $x$. For a thread $\threadt \in \implbalb(x)$, we will show
  $\cardinal{E'_x\rst{\threadt}} = c_x$ by checking that every number from $1$
  to $c_x$ is attributed once.

  Note that, for a given thread $\threadt$ and location $x$, since $E'$ only
  contains barrier calls, only the events $\action_{W}$ of the form
  $(\threadt, \_, (\brlwrite, (x_\threadt, {v+1}), ()))$ are able to modify the
  value of $x_\threadt$ on node $\nodefun{\threadt}$\footnote{This is why the
    broadcast event must \emph{not} overwrite $x_\threadt$ with itself on node
    $\nodefun{\threadt}$.}. Similarly, the value of $x_\threadt$ on another node
  can only be modified by a broadcast event from the thread $\threadt$, thus
  copying the value written by an $\action_{W}$ event.

  Firstly, let us show $c_x$ is attributed on every participating thread
  $\threadt$. By definition of $c_x$ there is $\action'_0$ on thread
  $\threadt_0$ writing $c_x$. From the definition of the implementation of
  $\action'_0$, there is a loop that only finishes when reading $x_\threadt$
  with value $v' \geq c_x$. This value $v'$ can only be is created by an event
  $\action' \in E'_x\rst{\threadt}$, and we have
  $\ordero(\action') = v' \geq x_k$. Since $c_x$ is defined as the maximum of
  such values, we have $v' = c_x$ and $c_x$ is attributed on $\threadt$.

  Secondly, let us show that if $v+2$ is attributed, then $v+1$ is attributed.
  This is simply because if $\ordero(\action'_2) = v+2$, \ie the implementation
  of $\action'_2$ writes $v+2$, then the initial read events
  $\action_{FR} = (\threadt, \_, (\brlread, (x_\threadt), (v+1))) \in
  \inv{f}(\action'_2)$ reads the value $v+1$. As before, this value can only be
  is created by an event $\action'_1 \in E'_x\rst{\threadt}$, and we have
  $\ordero(\action'_1) = v+1$.

  Thirdly, let us show that if $\action'_1, \action'_2 \in E'_x$ and
  $(\action'_1, \action'_2) \in \po'$ then
  $\ordero(\action'_1) < \ordero(\action'_2)$. By contradiction, let us assume
  $(\action'_1, \action'_2) \in \imm{\po'}$ the first pair (in $\po'\rst{E'_x}$
  order) such that $\ordero(\action'_1) = i+1 \geq j+1 = \ordero(\action'_2)$.
  As previously, their implementations have events
  $\action^1_{FR} = (\threadt, \_, (\brlread, (x_\threadt), (i))) \in
  \inv{f}(\action'_1)$,
  $\action^1_{W} = (\threadt, \_, (\brlwrite, (x_\threadt, i+1), ())) \in
  \inv{f}(\action'_1)$, and
  $\action^2_{FR} = (\threadt, \_, (\brlread, (x_\threadt), (j))) \in
  \inv{f}(\action'_2)$, with
  $\action^1_{FR} \arr{\po} \action^1_{W} \arr{\po} \action^2_{FR}$. Let
  $s_2 = (\action^2_{FR}, \tagcread)$ and $s_1 = (\action^1_{W}, \tagcwrite)$.
  Since we know $[\tagcread] ; (\inv{\po} \cap \fr) ; [\tagcwrite] = \emptyset$,
  showing $(s_2,s_1) \in \fr$ would be a contradiction.
  \begin{itemize}
  \item If $(\_,s_2) \not\in \rf$ (\ie $j = 0$), then $(s_2,s_1) \in \fr$ is a
    contradiction.
  \item If $(s_3,s_2) \in \rf$ with $(s_2,s_3) \in \po$, then since $s_3$ uses
    the stamp $\tagcwrite$ we have $(s_2,s_3) \in \ppo \subseteq \hb$ and
    $(s_3,s_2) \in \rfe \subseteq \so \subseteq \hb$. Thus we have an \hb cycle,
    which is a contradiction.
  \item If $(s_3,s_2) \in \rf$ with $(s_3,s_2) \in \po$, since
    $(\action'_1, \action'_2) \in \imm{\po'}$ there is no write in-between $s_1$
    and $s_2$ and thus $(s_3,s_1) \in \po$. Since
    $\mo^{\nodefun{\threadt}}_{x_\threadt}$ is included in $\hb$ and only uses the
    stamp $\tagcwrite$, it coincides with \po and so $(s_3,s_1) \in \mo$ and
    $(s_2,s_1) \in \fr$ is a contradiction.
  \end{itemize}
  Thus $(\action'_1, \action'_2) \in \po'$ implies
  $\ordero(\action'_1) < \ordero(\action'_2)$.

  By combining the pieces above, every number from $1$ to $c_x$ is attributed
  exactly once and $\cardinal{E_x\rst{\threadt}} = c_x$. This concludes the
  properties on $\exec'$ and we have that $\exec' = \tup{E', \po', \gettags', \so', \_}$
  is \bal-consistent.

  Finally, the last part of the proof is to check that $g(\so') \subseteq \hb$.
  Let us assume $s'_1 = (\action'_1, \taggf)$, $s'_2 = (\action'_2, \tagcread)$,
  and $(s'_1,s'_2) \in \so'$ for some $x$ and $i$ on threads $\threadt_1$ and
  $\threadt_2$. So $\action'_1,\action'_2 \in E'_x$ and
  $\ordero(\action'_1) = \ordero(\action'_2) = i$. By definition of the
  implementation and $g$ we have $\action^1_{GF} \arr{\po} \action^1_{W}$ in
  $\inv{f}(\action'_1)$ with $g(s'_1) = (\action^1_{GF}, \taggf)$, as well as
  $\action^2 \arr{\po} \action^2_{LR}$ in $\inv{f}(\action'_2)$ with
  $g(s'_2) = (\action_{LR}, \tagcread)$ and
  $\action^2 = (\threadt_2, \_, (\brlread, (x_{\threadt_1}), (v')))$ is the last
  read of the loop for thread $\threadt_1$ reading a value $v' \geq i$.
  In the very specific case where $\threadt_1 = \threadt_2$, we have
  $(g(s'_1),g(s'_2)) \in \ppo \subseteq \hb$. Otherwise, let
  $s_1 = (\action^1_{W}, \tagcwrite)$ and $s_2 = (\action^2, \tagcread)$. By
  definition of $\tagppo$ we have $(g(s'_1), s_1) \in \ppo \subseteq \hb$ and
  $(s_2, g(s'_2)) \in \ppo \subseteq \hb$, so we are left to prove
  $(s_1, s_2) \in \hb$.

  If $\threadt_1$ and $\threadt_2$ are on the same node, there is no broadcast
  involved. If $s_2$ reads from $s_1$ (\ie $v' = i$), then we immediately have
  $(s_1, s_2) \in \rfe \subseteq \so \subseteq \hb$. Else (\ie $v' > i$) $s_2$
  reads from some subevent $(\action^3, \tagcwrite)$ on thread $\threadt_1$ in
  the implementation of a later barrier, with $\ordero(f(\action^3)) = v'$. From
  the properties of $\ordero$ and the abstraction we have
  $(\action^1_{W}, \action^3) \in \po$. So
  $(s_1, s_2) \in (\ppo ; \rfe) \subseteq \hb$.

  If $\threadt_1$ and $\threadt_2$ are on different nodes $\node_1$ and
  $\node_2$, the reasoning is similar except a broadcast from $\threadt_1$
  bridges the gap. There is
  $\action_{B} = (\threadt_1, \_, (\brlbr, (x_{\threadt_1}, \_,
  \set{\ldots;\node_2;\ldots}), ()))$ such that \linebreak
  $((\action_B, \tagnlr[\node_2]), (\action_B, \tagnrw[\node_2])) \in \iso
  \subseteq \hb$, $((\action_B, \tagnrw[\node_2]), s_2) \in \rfe \subseteq \hb$,
  and $(\action_B, \tagnrw[\node_2])$ reads the same value $v'$. We fall back to
  the previous case: if $(\action_B, \tagnrw[\node_2])$ reads from $s_1$ we have
  $(s_1, (\action_B, \tagnrw[\node_2])) \in \rfe \subseteq \hb$; if it reads
  from a later write we have
  $(s_1, (\action_B, \tagnrw[\node_2])) \in (\ppo ; \rfe) \subseteq \hb$. In all
  cases, we have $(s_1, s_2) \in \hb$.
\end{proof}

\begin{corollary}
  The implementation \implbal is sound.
\end{corollary}



\subsection{\rbl Library}

\begin{theorem}
  \label{thm:rbl-sound}
  Given the functions $\rblthdw$ and $\rblthdr$ and a size $S$, the
  implementation \implrbl of the \rbl library into \brl given in the paper is
  locally sound.
\end{theorem}
\begin{proof}
  We assume an $\set{\brl}$-consistent execution
  $\exec = \tup{E, \po, \gettags, \so, \hb}$ which is abstracted via $f$ to
  $\tup{E',\po'}$ that uses the \rbl library, \ie
  $\absf{\rbl}{\implrbl}{\tup{E, \po}}{\tup{E', \po'}}$ holds. We need to
  provide $\gettags'$, $\so'$, and
  $g : \tup{E',\po',\gettags'}.\SEvents \rightarrow \exec.\SEvents$ respecting some
  conditions. From $\tup{E',\po'}$, we simply take $\gettags' = \gettagsrbl$.

  Since the implementation \implrbl maps events that do not respect $\rblthdr$
  or $\rblthdw$ to non-terminating loops, the abstraction $f$ tells us that
  every event in $E'$ does respect these functions.

  Since $\exec$ is $\set{\brl}$-consistent, it means $(\ppo \cup \so)^+ \suq \hb$,
  $\hb$ is transitive and irreflexive, and $\exec$ is \brl-consistent. Firstly, it
  means that for all thread $\threadt$ we have $\po\rst{\threadt}$ is a strict
  total order. From the properties of
  $\absf{\rbl}{\implrbl}{\tup{E, \po}}{\tup{E', \po'}}$, we can easily see
  that it implies $\po'\rst{\threadt}$ is also a strict total order. Secondly,
  $\gettags = \gettagsbrl$ and there exists well-formed \vr, \vw, \rf, and \mo
  such that $[\tagcread] ; (\inv{\po} \cap \fr) ; [\tagcwrite] = \emptyset$ and
  $\so = \iso \cup \rfe \cup \pf \cup \fr \cup \mo$. Note that here \mo is
  necessarily included in \ppo and is not relevant by itself.

  Let us define $g$.
  \begin{itemize}
  \item For $\action' = (\threadt, \_, (\rblsub,(x,\vect{v}), \text{true}))$,
    from the definition of the implementation and the abstraction $f$, there is
    some events $\action_w = (\threadt, \_, (\brlwrite,(h^x,v), ()))$ and
    \linebreak
    $\action_b = (\threadt, \_, (\brlbr,(h^x,\wid_x,s_n), ()))$ in
    $\inv{f}(\action')$, where
    $s_n = \setpred{\nodefun{\threadt_i}}{\threadt_i \in \rblthdr(x)} \setminus
    \set{\nodefun{\threadt}}$. We define
    $g(\action', \tagcwrite) = (\action_w, \tagcwrite)$, and for every
    $\node \in s_n$ we define $g(\action', \tagnrw) = (\action_b, \tagnrw)$.
  \item For $\action' = (\threadt, \_, (\rblsub,(x,\vect{v}), \text{false}))$,
    the first event of the implementation is \linebreak
    $\action_r = (\threadt, \_, (\brlread,(h^x), v))$ and we define
    $g(\action', \tagwait) = (\action_r, \tagcread)$.
  \item For $\action' = (\threadt, \_, (\rblrec,(x), r))$, the second
    event of the implementation is of the form
    $\action_r = (\threadt, \_, (\brlread,(h^x), v))$. If the \rblrec succeeds
    (\ie $r = \vect{v}$) we define
    $g(\action', \tagcread) = (\action_r, \tagcread)$. If the \rblrec fails (\ie
    $r = \bot$) we define $g(\action', \tagwait) = (\action_r, \tagcread)$.
  \end{itemize}
  Since the stamps \tagcread and \tagwait have the same relations to other stamps
  (see \Cref{fig:to}), the first property of $g$ holds.

  For every event in $E'$, we note \rblin and \rblout the values of the
  corresponding counter ($h^x$ for a \rblsub, $h^x_{\threadt_i}$ for a \rblrec)
  before and after the function call.
  \begin{itemize}
  \item For $\action' = (\threadt, \_, (\rblsub,(x,\vect{v}), r)) \in E'$,
    there is some event
    $\action_r = (\threadt, \_, (\brlread,(h^x), v)) \in \inv{f}(\action')$. We
    define $\rblin(\action') = v$. If this function fails (\ie
    $r = \texttt{false}$), we define $\rblout(\action') = v$ as well. Otherwise,
    from the implementation there is
    $\action_w = (\threadt, \_, (\brlwrite,(h^x,v'), ())) \in \inv{f}(\action')$
    and we define $\rblout(\action') = v'$.
  \item Similarly for
    $\action' = (\threadt, \_, (\rblrec,(x), r)) \in E'$. There is
    $\action_r = (\threadt, \_, (\brlread,(h^x_\threadt), v)) \in
    \inv{f}(\action')$, and in case of success there is
    $\action_w = (\threadt, \_, (\brlwrite,(h^x_\threadt,v'), \_)) \in
    \inv{f}(\action')$. For a failure we have
    $\rblin(\action') = \rblout(\action') = v$, and for a success
    $\rblin(\action') = v$ and $\rblout(\action') = v'$.
  \end{itemize}

  We extend the notation to subevents:
  $\rblin((\action', \tagt)) \defeq \rblin(\action')$, and similarly for
  \rblout. We have some basic properties about \rblin and \rblout.
  \begin{itemize}
  \item The first event of each thread has an \rblin value of $0$, and we always
    have $0 \leq \rblin(\action') \leq \rblout(\action')$
  \item For an event $\action'$ with label
    $(\rblsub, (x, (v_1,\ldots,v_V)), \texttt{true})$, we have
    $\rblout(\action') = \rblin(\action') + V + 1$
  \item For an event $\action'$ with label
    $(\rblrec, (x), (v_1,\ldots,v_V))$, we have
    $\rblout(\action') = \rblin(\action') + V + 1$
  \item Let $E'\rst{\rblsub,x}$ be the subset of $E'$ for calls to \rblsub on
    $x$, and $\po'\rst{\rblsub,x}$ the corresponding subset of $\po'$. If
    $(\action'_1, \action'_2) \in \imm{(\po'\rst{\rblsub,x})}$, then
    $\rblout(\action'_1) = \rblin(\action'_2)$. \Ie, if we have two consecutive
    \rblsub calls ($\action'_1$ and $\action'_2$) on $x$, the value of $h^x$ at
    the end of the execution of $\action'_1$ is equal to the value at the
    beginning of the execution of $\action'_2$.

    This comes from the semantics of \brl. A \brlread is required to read the
    last value written in program order. It cannot read from another thread or a
    broadcast as there is no other writing on $h^x$ by definition of the
    implementation. It cannot read from a later write, since
    $[\tagcwrite];(\rf \cap \inv{\po});[\tagcread] \subseteq (\rfe \cap
    \inv{\ppo}) \subseteq (\hb \cap \inv{\hb}) = \emptyset$. It cannot read from
    an earlier write than the last, since
    $[\tagcread] ; (\inv{\po} \cap \fr) ; [\tagcwrite] = \emptyset$.
  \item Similarly for \rblrec, we can define $E'\rst{\rblrec,x}$ and
    $\po'\rst{\rblrec,x}$. If
    $(\action'_1, \action'_2) \in \imm{(\po'\rst{\rblrec,x})}$, then
    $\rblout(\action'_1) = \rblin(\action'_2)$ by the same reasoning.
  \end{itemize}

  We then choose the following relation $\rf'$.
  $$\rf' \defeq \bigcup_{\node,x} \rf^\node_x \qquad
  \rf^\node_x \defeq (\Write^\node_x \times \Read^\node_x) \cap \setpred{(\saction'_1,
    \saction'_2)}{\rblin(\saction'_1) = \rblin(\saction'_2)}$$

  We take $\so' = \rf' \cup \rblfr'$, where
  $\rblfr' \defeq \bigcup_{\node,x} \left( \exec'.\rblfailread^\node_x \times
    \exec'.\Write^\node_x \setminus (\inv{\po'} ; \inv{\rf'}) \right)$. We need to
  prove that $g(\so') \subseteq \hb$ and that
  $\exec' = \tup{E', \po', \gettags', \so', \_}$ is \rbl-consistent. Since $E'$
  respects the functions \rblthdr and \rblthdw, we only need to check that
  $\rf'$ is well-formed for the latter.

  \bigskip

  As an intermediary result, let us show that if
  $(\saction'_1, \saction'_2) \in \rf'$, with
  $\saction'_1 = (\action'_1,\tagnrw)$ and
  $\saction'_2 = (\action'_2,\tagcread)$ (\ie they are on a location $x$ with
  $\rblin(\action'_1) = \rblin(\action'_2)$), then the two events write/read the
  same tuple $\vect{v}$ and we have $\rblout(\action'_1) = \rblout(\action'_2)$.
  We have
  $\action'_1 = (\threadt_1, \_, (\rblsub, (x, \vect{v}), \texttt{true}))$
  and $\action'_2 = (\threadt_2, \_, (\rblrec, (x), \vect{v}'))$. Let us
  name $H = \rblin(\action'_1)$, $V = \texttt{len}(\vect{v})$, and
  $V' = \texttt{len}(\vect{v}')$. we aim to show that $\vect{v} = \vect{v}'$,
  which would also imply
  $\rblout(\action'_1) = \rblin(\action'_1) + V + 1 = \rblin(\action'_2) + V' +
  1 = \rblout(\action'_2)$.

  From the implementation of $\action'_1$, we have
  $\action_{w_1} = (\threadt_1, \_, (\brlwrite, (h^x, H+V+1), ())) \in
  \inv{f}(\action'_1)$ and
  $\action_{b_1} = (\threadt_1, \_, (\brlbr, (h^x, \wid_x, \_), ())) \in
  \inv{f}(\action'_1)$. We necessarily have
  $((\action_{w_1}, \tagcwrite), (\action_{b_1}, \tagnlr)) \in \rf$, and thus
  $\vw((\action_{b_1}, \tagnrw)) = \vr((\action_{b_1}, \tagnlr)) = H+V+1 =
  \rblout(\action'_1)$. This is because $\action_{b_1}$ cannot read from an
  earlier read, as that would be ignoring $\action_{w_1}$ which is forbidden
  (from $\fr \in \so$), and cannot read from a later read because of the
  $\brlwait(\wid_x)$ operation (from $\pf \in \so$) placed within each
  successful execution of $\rblsub(x,\_)$, making sure $h^x$ is read by the
  broadcast before we can modify it again. This also holds for any broadcast on
  $h^x$ of other events.

  From the implementation of $\action'_2$ we have
  $\action_r = (\threadt_2, \_, (\brlread, (h^x), H')) \in \inv{f}(\action'_2)$.
  From the inequality in the implementation we have
  $H' > H = \rblin(\action'_2)$, so $H' \neq 0$ and the value is read from a
  broadcast from thread $\threadt_1$. There is $\action'_3 \in E'$ such that
  $\action_{b_3} = (\threadt_1, \_, (\brlbr, (h^x, \wid_x, \_), ())) \in
  \inv{f}(\action'_3)$ with
  $\vw((\action_{b_3}, \tagnrw)) = H' = \rblout(\action'_3)$ and
  $((\action_{b_3}, \tagnrw), (\action_r, \tagcread)) \in \rfe$. We might have
  $\action'_1 = \action'_3$ and $\action_{b_1} = \action_{b_3}$, but not
  necessarily. Since $\rblout(\action'_3) = H' > H = \rblin(\action'_1)$, we
  necessarily have $(\action'_1, \action'_3) \in (\po')^*$ and thus by transitivity
  $((\action_{b_1}, \tagnrw), (\action_r, \tagcread)) \in \hb$. Note that this
  can be written $(g(\saction'_1), g(\saction'_2)) \in \hb$, which proves
  $g(\rf') \subseteq \hb$.

  The implementation of $\action'_1$ makes several write and broadcast events of
  the form \linebreak
  $\action_{w_i} = (\threadt_1, \_, (\brlwrite, (x_i, v_i), ())) \in
  \inv{f}(\action'_1)$ and
  $\action_{b_i} = (\threadt_1, \_, (\brlbr, (x_i, \_, \_), ())) \in
  \inv{f}(\action'_1)$, with $i=(H+k)\%S$ for $0 \leq k \leq V$. Note: no
  location $x_i$ is written twice, since $V+1 \leq S$ from the condition in the
  implementation. Similarly, the implementation of $\action'_2$ makes several
  read events of the form
  $\action_{r_i} = (\threadt_2, \_, (\brlread, (x_i), v_i')) \in
  \inv{f}(\action'_2)$. It would be enough to check that each of these read
  event reads the value written by the corresponding write (\ie $v_i = v_i'$).

  Firstly, the value written is available. We have
  $(\action_{w_i}, \tagcwrite) \arr{\ppo} (\action_{b_i}, \tagnlr) \arr{\iso}
  (\action_{b_i}, \tagnrw) \arr{\ppo} (\action_{b_1}, \tagnrw) \arr{\hb}
  (\action_{r}, \tagcread) \arr{\ppo} (\action_{r_i}, \tagcread)$, so since \hb
  is transitive and irreflexive this implies
  $((\action_{b_i}, \tagnlr), (\action_{w_i}, \tagcwrite)) \not\in \fr$ and
  $((\action_{r_i}, \tagcread), (\action_{b_i}, \tagnrw)) \not\in \fr$, and we
  cannot read from earlier values.

  Secondly, we need to check that $\action_{b_i}$ and $\action_{r_i}$ cannot
  read from later values. Let us take
  $\action_{w'_i} = (\threadt_1, \_, (\brlwrite, (x_i, \_), ())) \in
  \inv{f}(\action'_3)$ and
  $\action_{b'_i} = (\threadt_1, \_, (\brlbr, (x_i, \_, \_), ())) \in
  \inv{f}(\action'_3)$ from some later (in $\po'$) successful \rblsub
  $\action'_3$ on $x$. We use the index $\__3$ to indicate values of the
  execution of $\action'_3$. We have $i$ of the form $(H_3 + k_3)\%S$, for some
  $0 \leq k_3 \leq V_3$. Since $(\action'_1, \action'_3) \in \po'$ we have
  $H_3 = \rblin(\action'_3) \geq \rblout(\action'_1) > (H+k)$. Thus from
  $(H+k)\%S = i = (H_3+k_3)\%S$ we have $H+k + S \leq H_3+k_3 \leq H_3 + V_3$,
  \ie the indices before modulo differ by at least the size $S$ of the buffer.
  From the condition in the implementation of $\rblsub$, we have
  $(H_3 - M_3) + (V_3 +1) \leq S$, and so
  $M_3 \geq H_3 + V_3 -S + 1 \geq H+k+1 > H = \rblin(\action'_2)$. Intuitively,
  this large value of $M_3$ indicates that $\action'_2$ is already finished. The
  implementation of $\action'_3$ makes a read
  $\action_{r_3} = (\threadt_1, \_, (\brlread, (h^x_{\threadt_2}), v_3))$ with
  $v_3 \geq M_3 > \rblin(\action'_2)$. The implementation of $\action'_2$ makes
  a write
  $\action_{w_2} = (\threadt_2, \_, (\brlwrite, (h^x_{\threadt_2},
  \rblout(\action'_2)), ()))$. By our properties of \rblin and \rblout, this is
  the first write on $h^x_{\threadt_2}$ with value greater than $H$. Thus we
  have $((\action_{w_2}, \tagcwrite), (\action_{r_3}, \tagcread)) \in \hb$ by
  \ppo transitivity to the write being read, and via the intermediary of some
  broadcast. Since
  $((\action_{r_i}, \tagcread), (\action_{w_2}, \tagcwrite)) \in \ppo$,
  $((\action_{r_3}, \tagcread), (\action_{w'_i}, \tagcwrite))) \in \ppo$ (thus
  $((\action_{b_i}, \tagnlr), (\action_{w'_i}, \tagcwrite))) \in \hb$), and
  $((\action_{r_3}, \tagcread), (\action_{b'_i}, \tagnrw))) \in \ppo$, we have
  that $\action_{r_i}$ cannot read from a later broadcast and $\action_{b_i}$
  cannot read from a later write.

  Thus $\vect{v} = \vect{v}'$ and $\rblout(\action'_1) = \rblout(\action'_2)$,
  which concludes our intermediary result.
  The same property holds for
  $((\action'_1,\tagcwrite), (\action'_2,\tagcread)) \in \rf'$ for similar
  reasons. Except that if both threads are on the same node the reader can
  directly read the data without the help of broadcasts.

  \bigskip

  From this intermediary result, it is easy to check that $\rf'$ is well-formed.
  \begin{itemize}
  \item $\rf'$ is total and functional on its range. It is functional as two
    different successful \rblsub events on $x$ necessarily have different \rblin
    values. We can check $\inv{\rf'}$ is total by contradiction, by taking the
    first (lowest \rblin value) event $\saction'_r \in \exec'.\Read^\node_x$ that is
    not related in $\rf'$. If there is $\saction'_2 \in \exec'.\Read^\node_x$ with
    $(\saction'_2, \saction'_r) \in \imm{(\po'\rst{\rblrec,x})}$, then by
    hypothesis there is $\saction'_1$ such that
    $(\saction'_1, \saction'_2) \in \rf'$. From our intermediary result we have
    $\rblin(\saction'_r) = \rblout(\saction'_2) = \rblout(\saction'_1)$. If
    there is a next successful \rblsub event $\saction'_w$, then it would
    necessarily have
    $\rblin(\saction'_w) = \rblout(\saction'_1) = \rblin(\saction'_r)$, and thus
    we would have $(\saction'_w, \saction'_r) \in \rf'$, a contradiction. Such
    an event must exist because $\saction'_r$ is successful: from the
    implementation, $\saction'_r$ reads $h^x$ and finds a value strictly higher
    than $\rblout(\saction'_1)$, which requires the existence of later \rblsub
    events.

  \item Events related in $\rf'$ write and read the same tuple of values, from
    our intermediary result.

  \item Each thread can read each value once. This is because two successful
    \rblrec calls on $x$ from the same thread will have different \rblin values
    and cannot read from the same \rblsub.

  \item Threads cannot jump a value. This is easily checked by induction. The
    first successful \rblrec (with \rblin value $0$) must read the first
    successful \rblsub (with \rblin value $0$). Whenever a successful \rblrec
    occurs reading a specific \rblsub, the following successful \rblsub/\rblrec
    events will have the same \rblin value, and thus have to be related by
    $\rf'$.
  \end{itemize}

  Finally, we are left to prove that $g(\so') \subseteq \hb$. During the proof
  of the intermediary result, we already checked $g(\rf') \subseteq \hb$. Let
  $(\saction'_f, \saction'_w) \in \rblfr'$, where
  $\saction'_f \in \exec'.\rblfailread^\node_x$ is a failed \rblrec on $x$ and
  $\saction'_w \in \exec'.\Write^\node_x$ a successful \rblsub. We will assume they
  are on different nodes, $\saction'_f = (\action'_f, \tagwait)$ and
  $\saction'_w = (\action'_w, \tagnrw)$, but the same reasoning can be adapted
  (without the broadcasts) to threads on the same node. Note that we necessarily
  have $\rblin(\action'_f) \geq \rblin(\action'_w)$. By contradiction, if we had
  $\rblin(\action'_w) < \rblin(\action'_f)$ then by induction and using our
  intermediary result we can see there is $\saction'_r \in \exec'.\Read^\node_x$
  such that $(\saction'_r, \saction'_f) \in \po'$ and
  $\rblin(\saction'_r) = \rblin(\saction'_w)$, thus
  $(\saction'_w, \saction'_r) \in \rf'$ contradicting
  $(\saction'_f, \saction'_w) \in \rblfr'$.
  The implementation of $\action'_f$ comprises two events:
  $\action_1 = (\threadt_f, \_, (\brlread, (h^x_{\threadt_f}), H))$ and
  $\action_2 = (\threadt_f, \_, (\brlread, (h^x), H'))$ with
  $H' \leq H = \rblin(\action'_f)$. The implementation of $\action'_w$ ends
  with two events:
  $\action_3 = (\threadt_w, \_, (\brlwrite, (h^x, \rblout(\action'_w)), ()))$
  and $\action_4 = (\threadt_w, \_, (\brlbr, (h^x, \wid_x, \_), ()))$. We have
  $g(\saction'_f) = (\action_2, \tagcread)$ and
  $g(\saction'_w) = (\action_4, \tagnrw)$, both events accessing the location
  $h^x$. As seen previously, we have
  $\vw((\action_4, \tagnrw)) = \rblout(\action'_w)$ as the broadcast can only
  read from $\action_3$. We have
  $\vr(g(\saction'_f)) = H' \leq \rblin(\action'_f) \leq \rblin(\action'_w) <
  \rblout(\action'_w) = \vw(g(\saction'_w))$. Since the value of $h^x$
  increases, we necessarily have
  $(g(\saction'_f), g(\saction'_w)) \in \fr \subseteq \so \subseteq \hb$.
\end{proof}

\begin{corollary}
  The implementation \implrbl is sound.
\end{corollary}



\subsection{\rdmawait to \rdmatso}

\begin{theorem}
  \label{thm:proof-wait-to-tso}
  Let $\progps$ be a program using only the $\rdmawait$ library. Then we have \\
  $\outcome_{\rdmatso}(\impliw{\progps}) \suq \outcome_{\{\rdmawait\}}(\progps)$.
\end{theorem}
\begin{proof}
  By definition, we are given $\exec = \tup{E, \po, \gettags, \so}$
  $\rdmatso$-consistent (\Cref{def:rtso-consistency}) such that
  $\tup{\vect{v},\tup{E,\po}} \in \interp{\impliw{\progps}}$. Among others, it
  means $\tup{E, \po}$ respects nodes and there exists well-formed $\vr$, $\vw$,
  $\rf$, $\mo$, $\ro$, and $\pf$ such that $\ib$ is irreflexive,
  $\gettags = \gettagsrtso$,
  $\so = \iso \cup \rfe \cup [\tagnlw[]];\pf \cup \ro \cup \fr \cup
  \mo \cup ([\Inst];\ib)$, and $\hb \defeq (\ppo \cup \so)^+$ is irreflexive.

  From \Cref{lem:find-abstraction}, since $\progps$ uses only $\rdmawait$, there
  is $E',\po',f$ such that $\tup{\vect{v}, \tup{E', \po'}} \in \interp{\progps}$
  and $\absf{\rdmawait}{\implwait}{\tup{E, \po}}{\tup{E', \po'}}$. Note that
  this clearly implies $\tup{E, \po}$ also respects nodes, as the implementation
  $\implwait$ keeps the same locations. Our objective is to find $\gettags'$,
  $\so'$, and $\hb'$ such that $\exec' = \tup{E', \po', \gettags', \so', \hb'}$
  is $\{\rdmawait\}$-consistent (\Cref{def:lambda-cons,def:rl-consistency}). Of
  course, we pick $\gettags' \defeq \gettagsrl$ as it is the only choice for
  consistency. We will also pick $\hb' \defeq (\ppo' \cup \so')^+$ since there
  is no external constraints. Thus, we only need to carefully pick $\so'$ and
  show it works.

  While our objective is not exactly local soundness (\Cref{def:locally-sound}),
  we still use a concretisation function
  $g : \tup{E',\po',\gettags'}.\SEvents \rightarrow \exec.\SEvents$ to then
  define $\so'$.

  \begin{itemize}
  \item For $\action' = (\threadt, \_, (\rlwrite,(x,v), ()))$, from the
    definition of the implementation $\implwait$ and the abstraction $f$, there
    is some event
    $\action = (\threadt, \_, (\rtsowrite, (x,v), ())) \in \inv{f}(\action')$.
    We define $g(\action', \tagcwrite) = (\action, \tagcwrite)$. For events
    calling $\rlread$, $\rlcas$, $\rlmf$, and $\rlrf$, we proceed similarly and
    let $g$ map each subevent to their counterpart in the implementation.
  \item For $\action' = (\threadt, \_, (\rlget,(x,y,\wid), ()))$, there
    is some event
    $\action = (\threadt, \_, (\rtsoget, (x,y), (v))) \in \inv{f}(\action')$.
    We define $g(\action', \tagnrr[\node(y)]) = (\action, \tagnrr[\node(y)])$
    and $g(\action', \tagnlw[\node(y)]) = (\action, \tagnlw[\node(y)])$. We
    proceed similarly for \rlput events.
  \item Finally for $\action' = (\threadt, \_, (\rlwait,(\wid), ()))$, there is
    in $\inv{f}(\action')$ some last event (in $\po$ order) of the form
    $\action = (\threadt, \_, (\rtsoempty, (\wid^N), \true))$ confirming the set
    $\wid^N$ tracking operations towards the last node $N$ is empty. We define
    $g(\action', \tagwait) = (\action, \tagmf)$.
  \end{itemize}

  We can see that $g(\tup{\action', \tagt'}) = \tup{\action, \tagt}$ implies
  that $f(\action) = \action'$ and that $\tagt$ is more restrictive than
  $\tagt'$.

  Each subevent in $\exec'.\Read$ (resp. $\exec'.\Write$) is mapped through $g$
  to a subevent in $\exec.\Read$ (resp. $\exec.\Write$) using the same stamp and
  location. Thus it is straightforward to define $\vr'$, $\vw'$, $\rf'$, $\mo'$,
  and $\ro'$ by relying on their counterparts in $\exec$. \Eg
  $\vr'(\saction') \defeq \vr(g(\saction'))$ and
  $\rf' \defeq \setpred{(\saction_1',\saction_2')}{(g(\saction_1'),
    g(\saction_2')) \in \rf}$. The well-formedness of $\vr$, $\vw$, $\rf$,
  $\mo$, and $\ro$ trivially implies that of $\vr'$, $\vw'$, $\rf'$, $\mo'$, and
  $\ro'$. From this, we have all the expected derived relations, including
  $\pfget'$, $\pfput'$, and
  $\ib' \defeq (\ippo' \cup \iso' \cup \rf' \cup \pfget' \cup \pfput' \cup \ro'
  \cup \fri')^+$. We then define
  $\so' \defeq \iso' \cup \rfe' \cup \pfget' \cup \ro' \cup \fr' \cup \mo' \cup
  ([\Inst];\ib')$, and as previously mentioned
  $\hb' \defeq (\ppo' \cup \so')^+$.

  To show $\{\rdmawait\}$-consistency, we are left to prove that $\ib'$ and
  $\hb'$ are irreflexive. For this, it is enough to show that $g(\ib') \suq \ib$
  and $g(\hb') \suq \hb \defeq (\ppo \cup \so)^+$ since we know both $\ib$ and
  $\hb$ to be irreflexive.

  For all subevent $\saction'$, $g(\saction')$ has a more restrictive stamp than
  $\saction'$ (in most cases it is the same stamp, but for \code{Wait} the stamp
  \tagmf is more restrictive than \tagwait); this implies that
  $g(\ppo') \subseteq \ppo$. Then, by definition, it is trivial to check that
  $g(\rf') \subseteq \rf$, $g(\mo') \subseteq \mo$, $g(\ro') \subseteq \ro$,
  $g(\ippo') \subseteq \ippo$, $g(\rfe') \subseteq \rfe$,
  $g(\iso') \subseteq \iso$, $g(\fr') \subseteq \fr$, and
  $g(\fri') \subseteq \fri$.

  To finish the proof, we need the following crucial pieces:
  $g(\pfput') \suq \ib$, $g(\pfget') \suq \ib$, and $g(\pfget') \suq \hb$. In
  fact, it is enough to show that $g(\pfput')$ and $g(\pfget')$ are both
  included in $\pf;\ppo^+$. This is because $\pf;\ppo^+ \suq \ib$,
  $[\tagnlw[]];\pf;\ppo^+ \suq \hb$, and the domain of $g(\pfget')$
  is included in $\cup_\node \ \exec.\tagnlw$ by definition.

  Let $((\action'_1, \tagnrw),(\action'_2, \tagwait)) \in \pfput'$. By
  definition they are of the form
  $\action'_1 = (\threadt, \_, (\rlput, (x, y, \wid), ()))$ and
  $\action'_2 = (\threadt, \_, (\rlwait, (\wid), ()))$, for some $\threadt$,
  $x$, $y$, and $\wid$, with $(\action'_1, \action'_2) \in \po'$ and
  $\node = \node(x)$ the remote node of this operation.
  By definition of the implementation and the abstraction, $\inv{f}(\action'_1)$
  contains two events $\action_1 = (\threadt, \_, (\rtsoput, (x, y), (v)))$ and
  $\action_a = (\threadt, \_, (\rtsoadd, (\wid^\node, v), ()))$, with
  $\action_1 \arr{\po} \action_a$. Meanwhile $\inv{f}(\action'_2)$ contains a
  last event $\action_2 = (\threadt, \_, (\rtsoempty, (\wid^N), \true))$ and an
  earlier event \linebreak
  $\action_3 = (\threadt, \_, (\rtsoempty, (\wid^\node), \true))$, with
  $\action_3 \arr{\po^*} \action_2$, confirming operations towards $\node$ are
  done (if $\node = N$ then $\action_2 = \action_3$).

  Since $f(\action_a) = \action'_1 \arr{\po'} \action'_2 = f(\action_3)$ and $f$
  is an abstraction, we have $\action_a \arr{\po} \action_3$, \ie the value $v$
  is added to $\wid^\node$ before the moment $\wid^\node$ is confirmed empty. By
  consistency (\Cref{def:rtso-consistency}), there is an in-between event
  $\action_4 = (\threadt, \_, (\rtsorm, (\wid^\node, v), ()))$ that removes this
  value, with $\action_a \arr{\po} \action_4 \arr{\po} \action_3$. From the
  definition of the implementation, such an event $\action_4$ is immediately
  preceded (with maybe other \rtsorm in-between) by an event
  $\action_p = (\threadt, \_, (\rtsopoll, (\node), (v)))$.
  Now we argue that we necessarily have
  $((\action_1, \tagnrw), (\action_p, \tagwait)) \in \pf$. From the
  well-formedness of \pf, we know that $(\action_p, \tagwait)$ has a preimage
  (\pf is total and functional on its range) and that this preimage outputs the
  value $v$. By consistency (\Cref{def:rtso-consistency}), $\action_1$ is the
  only \rtsoget or \rtsoput with output $v$. Thus $(\action_1, \tagnrw)$ is the
  preimage of $(\action_p, \tagwait)$ by $\pf$.

  Finally we have
  $g(\action'_1, \tagnrw) = (\action_1, \tagnrw) \arr{\pf} (\action_p, \tagwait)
  \arr{\ppo} (\action_4, \tagmf) \arr{\ppo} (\action_3, \tagmf) \arr{\ppo^*}
  (\action_2, \tagmf) = g(\action'_2, \tagwait)$, which shows
  $g(\pfput') \suq \pf;\ppo^+$.

  We similarly have $g(\pfget') \suq \pf;\ppo^+$ via the same reasoning. Thus
  $\ib'$ and $\hb'$ are irreflexive, and $\exec'$ is $\{\rdmawait\}$-consistent.
\end{proof}

\input{msw}

%% file: waittotso.tex
\section{\rdmawait implementation into \rdmatso}
\label{sec:rdmawait-to-rdmatso}



\subsection{Background: \rdmatso}

Our definition of \rdmatso is closer to an independent language than a library.
Unlike the definition of an execution in \Cref{def:execution}, we do not need a
relation \hb to represent the potential rest of the program, as \rdmatso is not
a library in the sense of \Cref{def:library}. A program \emph{cannot} combine
instructions from \rdmatso and other libraries presented in this paper, as
polling would interfere with RDMA operations of other libraries.

We use the following 11 methods:
\begin{align*}
m(\vect{v}) & ::=
\rtsowrite(x,v)
\mid \rtsoread(x)
\mid \rtsocas(x,v_1,v_2)
\mid \rtsomf() \\
& \quad
\mid \rtsoget(x,y)
\mid \rtsoput(x,y)
\mid \rtsopoll(\node)
\mid \rtsorf(\node) \\
& \quad
\mid \rtsoadd(x,v)
\mid \rtsorm(x,v)
\mid \rtsoempty(x)
\end{align*}

\begin{multicols}{2}
\begin{itemize}
\item $\rtsowrite : \Loc \times \Val \rightarrow ()$
\item $\rtsoread : \Loc \rightarrow \Val$
\item $\rtsocas : \Loc \times \Val \times \Val \rightarrow \Val$
\item $\rtsomf : () \rightarrow ()$
\item $\rtsoget : \Loc \times \Loc \rightarrow \Val$
\item $\rtsoput : \Loc \times \Loc \rightarrow \Val$
\item $\rtsopoll : \Nodes \rightarrow \Val$
\item $\rtsorf : \Nodes \rightarrow ()$
\item $\rtsoadd : \Loc \times \Val \rightarrow ()$
\item $\rtsorm : \Loc \times \Val \rightarrow ()$
\item $\rtsoempty : \Loc \rightarrow \mathbb{B}$
\end{itemize}
\end{multicols}

As expected, the \code{Wait} operation is replaced with a \code{Poll} operation.
Compared to \rdmatso from \cite{OOPSLA-24}, we slightly extend the language
so that \textput/\textget operations return an arbitrary unique identifier, and
polling also returns the same identifier of the operation being
polled\footnote{In practice, the identifier is not random and can be chosen by
  the program}. In addition, we also assume basic set operations $\rtsoadd$,
$\rtsorm$, and $\rtsoempty$ to store these new identifiers, where the locations
used for sets do not overlap with locations used for other operations.

\paragraph{Consistency predicate} An execution of an \rdmatso program is of the
form $\exec = \tup{E, \po, \gettags, \so}$, similarly to \cref{def:library} but
$\hb = (\ppo \cup \so)^+$ does not have the flexibility of containing additional
external constraints.

We define the only valid stamping function $\gettagsrtso$ as follows:
\begin{itemize}
\item A poll has stamp \tagwait:
  $\gettagsrtso((\_, \_, (\rtsopoll, \_, \_))) = \set{\tagwait}$.
\item Auxiliary set operations have stamp \tagmf:
  $\gettagsrtso((\_, \_, (\rtsoadd, \_, \_))) =$ \\
  $\gettagsrtso((\_, \_, (\rtsorm, \_, \_))) = \gettagsrtso((\_, \_,
  (\rtsoempty, \_, \_))) = \set{\tagmf}$.
\item Other events follow \gettagsrl (\cf \Cref{sec:rdmawait-app}). E.g., events
  calling \rtsowrite have stamp \tagcwrite, while events calling \rtsoget
  towards node $\node$ have stamps \tagnrr and \tagnlw. We also define \loc on
  subevents similarly to \rdmawait.
\end{itemize}

We mark set operations with \tagmf to simplify the consistency conditions, as we
do not want to explicitly integrate them in the read ($\Read$) and write
($\Write$) subevents.

Given $\exec = \tup{E, \po, \gettagsrtso, \so}$, we say that \vr, \vw, \rf, \mo, \ro, and \pf are well-formed if:

\begin{itemize}
\item \vr, \vw, \rf, \mo, and \ro are well-formed, as in \rdmawait;
\item Let
  $P_\node \defeq \setpred{(\action, \tagwait)}{\action = (\_, \_, (\rtsopoll,
    (\node), \_)) \in E}$ be the set of poll (sub)events towards node
  $\node$. Then
  $\pf \suq \bigcup_{\node \in \Nodes} (\exec.\tagnlw \cup \exec.\tagnrw) \times P_\node$ is
  the \emph{polls-from} relation, relating earlier NIC writes to later polls.
  Moreover:
  \begin{itemize}
  \item $\pf \suq \po$ (we can only poll previous operations of the same
    thread);
  \item $\pf$ is functional on its domain (every NIC write can be polled at
    most once);
  \item $\pf$ is total and functional on its range (every $\rtsopoll$ polls from
    exactly one NIC write);
  \item $\rtsopoll$ events poll-from the oldest non-polled remote operation towards the given node:\\
    for each node $\node$, if $w_1,w_2 \in (\exec.\tagnlw \cup \exec.\tagnrw)$ and
    $w_1 \arr{\po} w_2 \arr{\pf} p_2$, then there exists $p_1$ such that
    $w_1 \arr{\pf} p_1 \arr{\po} p_2$;
  \item and a $\rtsopoll$ returns the unique identifier of the polled
    operation:\\ if
    $((\_, \_, (\_, \_, v_1)), \_) \arr{\pf} ((\_, \_, (\rtsopoll, \_, v_2)),
    \tagwait)$ then $v_1 = v_2$.
  \end{itemize}
\end{itemize}

We use the derived relations $\fr$, $\fri$, $\rfe$, $\rfi$, $\ippo$, and $\iso$
as defined for \rdmawait. We can then define $\ib$ as follows:
$$\ib \defeq (\ippo \ \cup \ \iso \ \cup \ \rf \ \cup \ \pf \ \cup \ \ro
\ \cup \ \fri)^+$$

\begin{definition}[\rdmatso-consistency]
  \label{def:rtso-consistency}
  $\exec = \tup{E, \po, \gettags, \so}$ is \rdmatso-consistent if:
  \begin{itemize}
  \item $(\ppo \cup \so)^+$ is irreflexive (similarly to \Cref{def:lambda-cons});
  \item $\tup{E, \po}$ respects nodes (as in \rdmawait);
  \item $\gettags = \gettagsrtso$;
  \item there exists well-formed \vr, \vw, \rf, \mo, \ro, and \pf such that
    $\ib$ is irreflexive and \\
    $\so = \iso \cup \rfe \cup [\tagnlw[]];\pf \cup \ro \cup \fr \cup
    \mo \cup ([\Inst];\ib)$;
  \item identifiers for \textget/\textput operations are unique:\\ if
    $\action_1$ and $\action_2$ are both of the form
    $(\_, \_, (\rtsoget, \_, v))$ or $(\_, \_, (\rtsoput, \_, v))$, then
    $\action_1 = \action_2$;
  \item and the set operations are (per-thread) sound: if \rtsoempty returns
    $\true$, then every value added to the set was subsequently removed. I.e.,
    if $\action_1 = (\threadt, \_, (\rtsoadd, (x,v), \_)$, \linebreak
    $\action_3 = (\threadt, \_, (\rtsoempty, (x), \true))$, and
    $\action_1 \arr{\po} \action_3$, then there exists
    $\action_2 = (\threadt, \_, (\rtsorm, (x,v), \_)$ such that
    $\action_1 \arr{\po} \action_2 \arr{\po} \action_3$.
  \end{itemize}
\end{definition}



\input{rdmaloco}

\subsection{Implementation Function}

In \Cref{fig:implwait} we define the implementation $\implwait$ from a full
program using only the \rdmawait library into a program using only \rdmatso.
We assume threads use disjoint work identifiers $\wid \in \Wid$, otherwise it is
straightforward to rename them.

For each location $x$ of \rdmawait, we also use a location $x$ for \rdmatso. For
each work identifier $\wid$ of \rdmawait, we use new \rdmatso locations
$\set{\wid^1,\ldots,\wid^N}$ where $N \defeq \cardinal{\Nodes}$ is the number of
nodes. Each location $\wid^\node$ is used as a set containing the identifiers of
ongoing operations towards node $\node$.

Most \rdmawait operations ($\rlwrite$, $\rlread$, $\rlcas$, $\rlmf$, and
$\rlrf$) are directly translated into their \rdmatso counterparts. An operation
$\rlget(x,y,\wid)$ towards node $\node$ is translated into a similar
$\rtsoget(x,y)$ whose output is added to the set $\wid^\node$; We proceed
similarly for \textputs. Finally, a $\rlwait(\wid)$ operation needs to poll
until all relevant operations are finished, \ie the sets
$\set{\wid^1,\ldots,\wid^N}$ are all empty. Whenever we poll, we obtain the
identifier of a finished operation, and we remove it from \emph{all} sets where
it might be held. We remove it from $\wid^\node$ but also from any other set
$\wid_k^\node$ tracking a different group of operations, as otherwise a later
call to $\rlwait(\wid_k)$ would hang and never return.

To simplify the notation of the implementation, we use the intuitive for-loops
and while-loops. As no information is carried between the loops, these for-loops
can be inlined, and the while-loops can easily be turned into loop-break
similarly to \Cref{fig:loco-bar}.

\begin{figure}
  $$\texttt{For a thread } \threadt \texttt{ using work identifiers } \set{\wid_1,\ldots,\wid_K} \texttt{:}$$
\begin{minipage}{.48\textwidth}
\begin{align*}
  & \implwait(\threadt, \rlwrite, (x,v)) \defeq \rtsowrite(x,v) \\
  & \implwait(\threadt, \rlread, (x)) \defeq \rtsoread(x) \\
  & \implwait(\threadt, \rlcas, (x,v_1,v_2)) \defeq \rtsocas(x,v_1,v_2) \\
  & \implwait(\threadt, \rlmf, ()) \defeq \rtsomf() \\
  & \implwait(\threadt, \rlrf, (\node)) \defeq \rtsorf(\node) \\
  & \\
  & \implwait(\threadt, \rlget, (x,y,\wid)) \defeq \\
  & \LetC{v}{\rtsoget(x,y)}{\rtsoadd(\wid^{\nodefun{y}}, v)}
\end{align*}
\end{minipage}
\qquad
\begin{minipage}{.45\textwidth}
\begin{align*}
  & \implwait(\threadt, \rlwait, (\wid)) \defeq \\
  & \texttt{For } \node \texttt{ in } 1,\ldots,N \texttt{ do \{} \\
  & \quad \texttt{While } (\rtsoempty(\wid^\node) \neq \true) \texttt { do \{} \\
  & \qquad \LetC{v}{\rtsopoll(\node)}{} \\
  & \qquad \texttt{For } k \texttt{ in } 1,\ldots,K \texttt{ do \{} \\
  & \qquad \quad \rtsorm(\wid_k^\node, v) \quad \} \ \} \ \} \\
  & \\
  & \implwait(\threadt, \rlput, (x,y,\wid)) \defeq \\
  & \LetC{v}{\rtsoput(x,y)}{\rtsoadd(\wid^{\nodefun{x}}, v)}
\end{align*}
\end{minipage}
\caption{Implementation \implwait of \rdmawait into \rdmatso}
\label{fig:implwait}
\end{figure}



\subsection{Proof}

We do not prove that the implementation above is locally sound
(\Cref{def:locally-sound}), as \Cref{thm:locally-sound-impl} does \emph{not}
apply in this case. It is not possible to combine a program following \rdmatso
with programs of the other libraries presented in this paper. Instead, we assume
a full program using only the \rdmawait library and compile it into \rdmatso.

\begin{theorem}
  \label{thm:wait-to-tso}
  Let $\progps$ be a program using only the $\rdmawait$ library. Then we have \\
  $\outcome_{\rdmatso}(\impliw{\progps}) \suq \outcome_{\{\rdmawait\}}(\progps)$, where:
  \begin{align*}
   \outcome_{\{\rdmawait\}}(\progps) &= \setpred{\vect{v}}{ \exists \tup{E, \po,
      \gettags, \so, \hb} \ \{\rdmawait\}\text{-consistent.\ }
    \tup{\vect{v},\tup{E,\po}} \in \interp{\progps}} \\
   \outcome_{\rdmatso}(\impliw{\progps}) &= \setpred{\vect{v}}{ \exists \tup{E,
      \po, \gettags, \so} \ \rdmatso\text{-consistent.\ }
    \tup{\vect{v},\tup{E,\po}} \in \interp{\impliw{\progps}} }
  \end{align*}

\end{theorem}
\begin{proof}
  See \Cref{thm:proof-wait-to-tso}.
\end{proof}

%% file: rdmaloco.tex
\subsection{\rdmawait Library}
\label{sec:rdmawait-app}

This appendix completes \Cref{sec:rdmawait} on the definition of \rdmawait. As
mentioned, we have the 8 methods:
\begin{align*}
m(\vect{v}) & ::=
\rlwrite(x,v)
\mid \rlread(x)
\mid \rlcas(x,v_1,v_2)
\mid \rlmf() \\
& \quad
\mid \rlget(x,y,\wid)
\mid \rlput(x,y,\wid)
\mid \rlwait(\wid)
\mid \rlrf(\node)
\end{align*}

\begin{multicols}{2}
\begin{itemize}
\item $\rlwrite : \Loc \times \Val \rightarrow ()$
\item $\rlread : \Loc \rightarrow \Val$
\item $\rlcas : \Loc \times \Val \times \Val \rightarrow \Val$
\item $\rlmf : () \rightarrow ()$
\item $\rlget : \Loc \times \Loc \times \Wid \rightarrow ()$
\item $\rlput : \Loc \times \Loc \times \Wid \rightarrow ()$
\item $\rlwait : \Wid \rightarrow ()$
\item $\rlrf : \Nodes \rightarrow ()$
\end{itemize}
\end{multicols}

We also define \loc as expected:
$\loc(\rlwrite(x,v)) = \loc(\rlread(x)) = \loc(\rlcas(x,v_1,v_2)) =~\set{x}$;
$\loc(\rlget(x,y,d)) = \loc(\rlput(x,y,d)) = \set{x;y}$; and
$\loc(\action) = \emptyset$ otherwise.

We assume that each location $x$ is associated with a specific node
$\nodefun{x}$. We say that $\tup{E, \po}$ respects nodes if for all event on
thread $\threadt$ with label of the form $(\rlwrite,(x,\_),\_)$,
$(\rlread,(x),\_)$, $(\rlcas,(x,\_,\_),\_)$, $(\rlget,(x,\_, \_),\_)$, or
$(\rlput,(\_,x,\_),\_)$, we have $\nodefun{x} = \nodefun{\threadt}$. \Ie
arguments corresponding to local locations should be locations of the current
node.
Given $\tup{E, \po}$, we now define the only valid stamping function
$\gettagsrl$. Since the thread is not relevant, we note
$\gettagsrl(m(\vect{v}),v')$ for
$\gettagsrl(\tup{\_, \_, \tup{m, \vect{v}, v'}})$.

\vspace{0.5em}

\begin{minipage}{0.55\linewidth}
\begin{itemize}[leftmargin=2.05em]
\item $\gettagsrl(\rlwrite(x,v),()) = \set{\tagcwrite}$
\item $\gettagsrl(\rlread(x),v) = \set{\tagcread}$
\item $\gettagsrl(\rlmf(),()) = \set{\tagmf}$
\item $\gettagsrl(\rlcas(x,v_1,v_2),v_1) = \set{\tagcas}$
\item $\gettagsrl(\rlcas(x,v_1,v_2),v_3) = \set{\tagmf ; \tagcread}$ if
  $v_1\neq v_3$
\end{itemize}
\end{minipage}
\hspace{-0.12\linewidth}
\begin{minipage}{0.55\linewidth}
\begin{itemize}[leftmargin=2em]
\item $\gettagsrl(\rlwait(\wid),()) = \set{\tagwait}$
\item
  $\gettagsrl(\rlget(x,y,d),()) = \set{\tagnrr[\nodefun{y}] ; \tagnlw[\nodefun{y}]}$
\item
  $\gettagsrl(\rlput(x,y,d),()) = \set{\tagnlr[\nodefun{x}] ; \tagnrw[\nodefun{x}]}$
\item $\gettagsrl(\rlrf(\node),()) = \set{\tagnf}$
\item[]
\end{itemize}
\end{minipage}

\vspace{0.5em}

Put and get operations perform both a NIC read and a NIC write, and as such
are associated to two stamps. A succeeding \rlcas can be represented as a single
stamp \tagcas, while a failing \rlcas behaves as both a memory fence (\tagmf)
and a CPU read (\tagcread).

We extend \loc to subevents. For events with zero or one locations, the
subevents have the same set of locations. For \rlget/\rlput, each of the two
subevent is associated to the relevant location. \Eg if
$\action = (\_, \_, (\rlget, (x, y, d), \_))$, then
$\loc(\tup{\action, \tagnrr[\nodefun{y}]}) = \set{y}$ and
$\loc(\tup{\action, \tagnlw[\nodefun{y}]}) = \set{x}$.

Given an execution $\exec = \tup{E, \po, \gettagsrl, \_, \_}$, recall we define
the set of \emph{reads} as
$\exec.\Read \defeq \exec.\tagcread \cup \exec.\tagcas \cup \exec.\tagnlr[] \cup
\exec.\tagnrr[]$ and \emph{writes} as
$\exec.\Write \defeq \exec.\tagcwrite \cup \exec.\tagcas \cup \exec.\tagnlw[] \cup
\exec.\tagnrw[]$. We say that \vr, \vw, \rf, \mo, and \ro are well-formed if:
\begin{itemize}
\item $\vr : \exec.\Read \rightarrow \Val$ associates each read subevent with a value,
  matching the value returned if available: if $\action$ has a label of the form
  $(\rlread, \_, v)$ or $(\rlcas, \_, v)$, then $\vr(\action) = v$.
\item $\vw : \exec.\Write \rightarrow \Val$ associates each write subevent with a value,
  matching the value written if known in $\exec$: if $\action$ has a label of
  the form $(\rlwrite, (\_, v), \_)$ or $(\rlcas, (\_, v', v), v')$, then
  $\vw(\action) = v$.
\item RDMA operations write the value read: if
  $\saction_1 = \tup{\action, \tagnlr} \in E$ and
  $\saction_2 = \tup{\action, \tagnrw} \in E$, then
  $\vr(\saction_1) = \vw(\saction_2)$; and similarly for $\tagnrr$ and
  $\tagnlw$.
\item $\rf \suq \exec.\Write \times \exec.\Read$ is the `\emph{reads-from}'
  relation on events of the same location with matching values; \ie
  $(\saction_1, \saction_2) \in \rf \Rightarrow
  \loc(\saction_1) = \loc(\saction_2) \land \vw(\saction_1) = \vr(\saction_2)$.
  $\rf$ is functional on its range: every read in $\exec.\Read$ is related to at
  most one write in $\exec.\Write$. If a read is not related to a write, it
  reads the initial value of zero:
  $\saction_2 \in \exec.\Read \land (\_, \saction_2) \not\in \rf \Rightarrow
  \vr(\saction_2) = 0$.
\item $\mo \defeq \bigcup_{x \in \Loc} \mo_x$ is the
  `\emph{modification-order}', where each $\mo_x$ is a strict total order on
  $\exec.\Write_x$ describing the order in which writes on $x$ reach the memory.
\item $\ro$ is the `\emph{NIC flush order}', such that for all $\node$ and
  $(\saction_1, \saction_2) \in \exec.\SEvents$ with
  $\threadfun{\saction_1} = \threadfun{\saction_2}$, if
  $(\saction_1, \saction_2) \in \exec.\tagnlr \times \exec.\tagnlw$ then
  $(\saction_1, \saction_2) \in \ro \cup \inv\ro$, and if
  $(\saction_1, \saction_2) \in \exec.\tagnrr \times \exec.\tagnrw$ then
  $(\saction_1, \saction_2) \in \ro \cup \inv\ro$.
\end{itemize}
The definitions above are similar to the relations defined for \brl (see
\cref{sec:brlib}), with the addition of $\ro$ representing the PCIe guarantees
that NIC reads flush previous NIC writes.

For each subevent, we distinguish the moment the subevent \emph{starts}
executing and the moment it \emph{finishes} executing. The relation $\so$
represents dependency between the end of executions of subevents. To express the
semantics of \rdmawait, we also need to consider the \emph{issued-before}
relation $\ib$ representing dependency between the start of executions of
subevents. Note that neither $\ib$ or $\so$ is a subset of the other. The
starting (when sent to the store buffer of \pcie fabric) and finishing (reaching
memory) points of some write subevents might differ. We define the set of
\emph{instantaneous subevents} as
$\exec.\Inst \defeq \exec.\SEvents \setminus (\exec.\tagcwrite \cup
\exec.\tagnlw[] \cup \exec.\tagnrw[])$, regrouping the subevents that start and
finish at the same time.

Given $\exec$ and well-formed \vr, \vw, \rf, \mo, and \ro, we derive additional
relations.

\begin{mathpar}
\fr \defeq \setpred{(r,w)}
{\begin{matrix}
  r \in \exec.\Read \land w \in \exec.\Write \land \loc(r) = \loc(w) \\
  \land \ ((r,w) \in (\inv\rf; \mo) \lor r \not\in \img{\rf})
\end{matrix}} \setminus [\exec.\SEvents]
\and
\rfe \defeq \rf \setminus \rfi
\and
\pfget \defeq \setpred{((\action_1, \tagnlw),(\action_2, \tagwait))}
{\begin{matrix}
  \exists d. \ (\action_1,\action_2) \in \po \\
  \land \ \action_1 = (\_, \_, (\rlget, (\_, \_, d), \_)) \\
  \land \ \action_2 = (\_, \_, (\rlwait, (d), \_))
\end{matrix}}
\and
\rfi \defeq [\tagcwrite] ; (\po \cap \rf) ; [\tagcread]
\and
\pfput \defeq \setpred{((\action_1, \tagnrw),(\action_2, \tagwait))}
{\begin{matrix}
  \exists d. \ (\action_1,\action_2) \in \po \\
  \land \ \action_1 = (\_, \_, (\rlput, (\_, \_, d), \_)) \\
  \land \ \action_2 = (\_, \_, (\rlwait, (d), \_))
\end{matrix}}\!
\and
\fri \defeq [\tagcread] ; ((\po \cup \inv{\po}) \cap \fr) ; [\tagcwrite]
\end{mathpar}

\begin{align*}
  \iso \defeq & \phantom{\cup~} \setpred{((\action, \tagmf),(\action, \tagcread))} {\action = (\_, \_, (\rlcas, \_, \_)) \in E \land \gettagsrl(\action) = \set{\tagmf; \tagcread}} \\
              & \cup \setpred{((\action, \tagnrr),(\action, \tagnlw))} {\action = (\_, \_, (\rlget, \_, \_)) \in E \land \gettagsrl(\action) = \set{\tagnrr; \tagnlw}} \\
              & \cup \setpred{((\action, \tagnlr),(\action, \tagnrw))} {\action = (\_, \_, (\rlput, \_, \_)) \in E \land \gettagsrl(\action) = \set{\tagnlr; \tagnrw}}
\end{align*}

\pfget (resp \pfput) represent the synchronisation between the write part of a
get (resp put) and a later \rlwait on the same work identifier. While both are
included in $\ib$, only \pfget is included in $\so$ as waiting for a put does
not guarantee the NIC remote write has finished. We define \rfe, \rfi, and \fr
similarly to the semantics of \brl, and we also define \fri as expected. The
internal synchronisation order $\iso$ represents ordering between subevents of
the same event. We ask that puts and gets read before writing, and that a
failing \rlcas performs a memory fence before reading.

Finally we can define $\ib$ as follows. $\ib$ includes a larger subset of $\po$
than $\ppo$, as we guarantee the starting order of the cases corresponding to
cells B1, B5, G10, and I10 of \cref{fig:to}. \Ie, while a later CPU read might
finish before an earlier CPU write, they have to start in order; and while a
remote fence does not guarantee previous NIC writes have finished, it guarantees
they have at least started.
$$\ippo \defeq \ppo \ \cup \ [\exec.\tagcwrite] ; \po ; [\exec.\tagcread \cup \exec.\tagwait]
\ \cup \ \bigcup_{\node \in \Nodes} ([\exec.\tagnrw \cup \exec.\tagnlw] ; \po ;
[\exec.\tagnf])$$
$$\ib \defeq (\ippo \ \cup \ \iso \ \cup \ \rf \ \cup \ \pfget \ \cup \ \pfput
\ \cup \ \ro \ \cup \ \fri)^+$$

And from this we define the consistency predicate for \rdmawait, similarly to
the semantics of \rdmatso. We ask that \ib and \so be irreflexive, the second
being implied by \cref{def:lambda-cons}. The inclusion of $([\Inst];\ib)$ in
$\so$ indicates that, if an instantaneous subevent starts before another
subevent, then they also finish in the same order.

\begin{definition}[\rdmawait-consistency]
  \label{def:rl-consistency}
  $\exec = \tup{E, \po, \gettags, \so, \hb}$ is \rdmawait-consistent if:
  \begin{itemize}
  \item $\tup{E, \po}$ respects nodes;
  \item $\gettags = \gettagsrl$;
  \item there exists well-formed \vr, \vw, \rf, \mo, and \ro such that $\ib$ is
    irreflexive and \\
    $\so = \iso \cup \rfe \cup \pfget \cup \ro \cup \fr \cup \mo \cup
    ([\Inst];\ib)$.
  \end{itemize}
\end{definition}

We can easily check that this predicate satisfies monotonicity and
decomposability.

%% file: msw.tex
\subsection{Mixed-size writes Library}
\label{sec:msw-app}


\subsubsection{The \msw Library}
\label{sec:msw}

A limitation of the \rdmawait library is that each location corresponds to a
specific memory location, and thus can only contain a fixed amount of data. LOCO
wants to provide abstractions simulating shared memory with distributed objects.
As such, we want to hide away the atomicity constraints of the underlying RDMA
technology and provide methods to manipulate large objects without the risk of
wrong manipulations and corrupted data.
A first step for this is the mixed-size write library (\msw) that can manipulate
data of any size with the same semantics as \rdmawait. The library uses
similar methods, with a syntax defined as follows.
\begin{align*}
  m(\vect{v}) & ::=
  \mswwrite(x,\tup{v_1,\ldots,v_k})
  \mid \mswread(x) \\
  & \quad
  \mid \mswget(x,y,d)
  \mid \mswput(x,y,d)
  \mid \mswwait(d)
\end{align*}

There is two differences with the methods of the $\rdmawait$ library. Firstly,
the read function $\mswread : \Loc \rightarrow \Val^* \uplus \set{\bot}$ can
fail if the underlying data is not in a stable state (\ie corrupted or being
modified). Secondly, the reads and writes
$\mswwrite : \Loc \times \Val^* \rightarrow ()$ methods manipulate tuples of
values, whereas \rdmawait locations can only hold a single value. If
necessary, a more usual read method can be derived by simply looping calls to
$\mswread$ until it succeeds.

The consistency predicate is then a copy of the one from \rdmawait, except
failing reads are ignored. This semantics guarantees there is no
out-of-thin-air: if a \mswread operation succeeds, then it reads a value that
was explicitly written by some \mswwrite operation.

This library can then be used to implement an MSW-Broadcast library where each
shared variable contains a tuple of values, similarly to how \brl is built on
top of \rdmawait.

\paragraph{Implementation}

We assume given a function $\mswsize : \Loc \rightarrow \mathbb{N}$ associating
locations to the amount of data they hold. From this, we define the
implementation $\implmsw$ of the \msw library into \rdmawait. We assume some
function $\mswhash$, such that $\mswhash(\vect{v}) = \mswhash(\vect{v'})$
implies $\vect{v} = \vect{v'}$. For each location $x$ of the \msw library, we
create $\mswsize(x)+1$ locations $\set{x_0, x_1, \ldots, x_{\mswsize(x)}}$ of
the \rdmawait library. The location $x_0$ holds the hash of the data, which is
written to $x_1, \ldots, x_{\mswsize(x)}$.

For events that do not respect \mswsize or the nodes, the implementation is
simply an infinite loop, similarly to the previous implementations. Otherwise,
as shown in \cref{fig:implmsw}, we apply the \rdmawait methods to each location,
and a read succeeds if the hash corresponds to the accompanying data.

\begin{figure}
\scalebox{0.9}{
\begin{minipage}[t]{.6\textwidth}\addtolength{\jot}{-0.5mm}
\begin{align*}
  \implmsw&(\threadt, \mswwrite, (x, \tup{v_1, \ldots, v_{\mswsize(x)}})) \defeq \\
  & \rlwrite(x_0, \mswhash((v_1,\ldots,v_{\mswsize(x)})))  \\
  & \rlwrite(x_1, v_1) ; \\
  & \ldots ; \\
  & \rlwrite(x_{\mswsize(x)}, v_{\mswsize(x)}) ; \\
  \implmsw&(\threadt, \mswread, (x)) \defeq \\
  & \texttt{let } v_0 = \rlread(x_0) \texttt{ in} \\
  & \texttt{let } v_1 = \rlread(x_1) \texttt{ in} \\
  & \ldots \\
  & \texttt{let } v_{\mswsize(x)} = \rlread(x_{\mswsize(x)}) \texttt{ in} \\
  & \texttt{if } v_0 = \mswhash(\tup{v_1, \ldots, v_{\mswsize(x)}}) \texttt{ then } \tup{v_1, \ldots, v_{\mswsize(x)}} \texttt{ else } \bot
\end{align*}
\end{minipage}}
\hspace{-0.2\linewidth}
\scalebox{0.9}{
\begin{minipage}[t]{.5\textwidth}\addtolength{\jot}{-0.5mm}
\begin{align*}
  \implmsw&(\threadt, \mswput, (x, y, \wid)) \defeq \\
  & \rlput(x_0, y_0, \wid) ; \\
  & \ldots ; \\
  & \rlput(x_{\mswsize(x)}, y_{\mswsize(x)}, \wid)) ; \\
  \implmsw&(\threadt, \mswget, (x, y, \wid)) \defeq \\
  & \rlget(x_0, y_0, \wid) ; \\
  & \ldots ; \\
  & \rlget(x_{\mswsize(x)}, y_{\mswsize(x)}, \wid)) ; \\
  \implmsw& (\threadt, \mswwait, (\wid)) \defeq \rlwait(\wid)
\end{align*}
\end{minipage}}
\vspace{-10pt}
\caption{Implementation \implmsw of the \msw library into \rdmawait}
\vspace{-10pt}
\label{fig:implmsw}
\end{figure}



\begin{theorem}
  The implementation \implmsw is locally sound.
\end{theorem}
\begin{proof}
  See \Cref{thm:msw-sound}.
\end{proof}

\subsubsection{Correctness}
This appendix completes \Cref{sec:msw} on the definition of \rdmawait. Our model
assumes a size function $\mswsize : \Loc \rightarrow \mathbb{N}$ associating
each location to the amount of data it stores. As mentioned, we have the 5
methods:
$$m(\vect{v}) ::=
\mswwrite(x,(v_1,\ldots,v_k))
\mid \mswread(x)
\mid \mswget(x,y,d)
\mid \mswput(x,y,d)
\mid \mswwait(d)
$$

\begin{multicols}{2}
\begin{itemize}
\item $\mswwrite : \Loc \times \Val^* \rightarrow ()$
\item $\mswread : \Loc \rightarrow \Val^* \uplus \set{\bot}$
\item $\mswget : \Loc \times \Loc \times \Wid \rightarrow ()$
\item $\mswput : \Loc \times \Loc \times \Wid \rightarrow ()$
\item $\mswwait : \Wid \rightarrow ()$
\end{itemize}
\end{multicols}

While this syntax does not include a TSO memory fence (similarly to \bal in
\ref{sec:brlib}), a program can use both this library and the memory fence from
\rdmawait.

We also define \loc as expected:
$\loc(\mswwrite(x,v)) = \loc(\mswread(x)) = \set{x}$; \linebreak
$\loc(\mswget(x,y,d)) = \loc(\mswput(x,y,d)) = \set{x;y}$; and
$\loc(\action) = \emptyset$ otherwise.



\paragraph{Consistency predicate} Given an execution
$\exec = \tup{E, \po, \gettags, \so, \hb}$, we define consistency similarly to
\rdmawait. The main difference is that the $\mswread$ function reading a
location can fail without justification.

We define the only valid stamping function $\gettagsmsw$ as follows:
\begin{itemize}
\item A succeeding \mswread has stamp \tagcread:
  $\gettagsmsw((\_, \_, (\mswread, \_, \vect{v}))) = \set{\tagcread}$.
\item A failing \mswread has stamp \tagwait: $\gettagsmsw((\_, \_, (\mswread, \_, \bot))) = \set{\tagwait}$.
\item Other events follow \gettagsrl (\cf \cref{sec:rdmawait-app}): events calling
  \mswwrite, \mswput, \mswget, and \mswwait have respectively stamps \tagcwrite,
  \tagnrr and \tagnlw, \tagnlr and \tagnrw, and \tagwait.
\end{itemize}
We mark failed read events with the stamp \tagwait to simplify the definition.
This stamp has the same \tagppo relation as \tagcread (\cf \ref{fig:to}), and is
thus equivalent, but we do not need to change our definition of $\exec.\Read$
covering all events stamped \tagcread.

\begin{definition}[\msw-consistency]
  $\exec = \tup{E, \po, \gettags, \so, \hb}$ is \msw-consistent if:
  \begin{itemize}
  \item $\tup{E, \po}$ is well-formed (as in \rdmawait);
  \item $E$ respects the function $\mswsize$. \Ie, for all event with label
    $(\mswwrite,(x,(v_1,\ldots,v_k)),())$ or $(\mswread,(x),(v_1,\ldots,v_k))$
    we have $k = \mswsize(x)$, and for all event with label
    $(\mswget,(x,y,\wid),())$ or $(\mswput,(x,y,\wid),())$ we have
    $\mswsize(x) = \mswsize(y)$.
  \item $\gettags = \gettagsmsw$;
  \item there exists well-formed \vr, \vw, \rf, \mo, and \ro (defined as in
    \rdmawait) such that $\ib$ is irreflexive and
    $\so = \iso \cup \rfe \cup \pfget \cup \ro \cup \fr \cup \mo \cup
    ([\Inst];\ib)$.
  \end{itemize}
\end{definition}



\paragraph{Note} The components $\vr$, $\rf$, and $\fr$ do not cover failed read
events. This weak semantics does not guarantee any read will eventually succeed,
as they are allowed to fail for any reason. It means synchronisation (\eg
barrier) do not force written data to be available.

This semantics only guarantees that there is no out-of-thin-air; \ie if a read
succeeds then it returns a value that was explicitly written.

A more complex semantics ensuring that properly written data is not corrupted
would be interesting. It would require a proper notion of data races, and lead
to a semantics much more complex than that of \rdmawait.



\paragraph{Implementation} The implementation $\implmsw$ of the \msw library
into \rdmawait is discussed in \Cref{sec:msw}.

%% file: proof.tex
\section{Correctness Proof of \code{kvstore}}
\label{sec:proof}

The \code{kvstore} object described in \cref{sec:applications} is
linearisable~\cite{herlihy-toplas-1990}, and we here provide an proof
of safety.  Note that our proof leverages both the composition of
linearisability and the mutual exclusion property of our locks, their
use are simplified by the composable nature of LOCO channels.

Updates to the indices are protected by an array of
\pcode{ticket\_lock}.  When a node tries to insert or delete a key,
it first acquires the lock with index \pcode{key \% NUM\_LOCKS}. It
then looks the key up in its local index. In the case of an insertion,
if the key does not yet exist, the node first writes the value to a
free slot in its local data array with the valid bit unset, increments
the counter corresponding to that slot, updates the checksum, and then
broadcasts the value's location and counter to other nodes on a
\code{ringbuffer} called the \emph{tracker}. Each node monitors the
set of other nodes' trackers with a dedicated thread, which applies
requested updates to the local index and then acknowledges the
message.  The inserter waits until all nodes have acknowledged its
message, meaning the location of the key is present in all indices,
and then marks the entry valid and releases the lock.  Deletion is the
reverse under the lock; marking the entry invalid, then broadcasting
the deletion, and removing the entry once all nodes have acknowledged
it.

To update the value mapped to a key, a node takes the lock
corresponding to that key and looks up its location in the local
index.  If it exists, it writes the new value to that location
(retaining the counter and valid bit), updates the checksum, then
releases the lock. This write is fenced, %
to ensure it is ordered with the
subsequent lock release. 

To retrieve the value mapped to a key, a node need not take a lock,
but simply looks up the key in the local index, failing if it is not
found, and reads the value and accompanying metadata from their
location on the corresponding node.  If the checksum is incorrect due
to a torn update, it retries.  If the valid bit is not set (indicating
an incomplete insertion/deletion), or the counter mismatches
(indicating a stale local index), the reader can safely return
\pcode{EMPTY}.

\subsection{Preliminaries}

We choose linearisation points~\cite{herlihy-toplas-1990} 
for each modification operation as follows.
A \code{write} linearises when the key, value, and checksum
are fully placed on the host node.  A \code{delete} linearises
when the \code{valid} bit is unset (before all nodes have
modified their local index and acknowledged the deletion).
An \code{insert} linearises when the \code{valid} bit is set
(after all nodes have
modified their local index and acknowledged the insertion).

The linearisation points of reads are determined retrospectively
depending on the read value.

Investigation of the algorithm determines every read consists of two steps (possibly repeated).
(1) A fetch from the local index to determine the node and address of the key's associated value.
(2) A remote read to this location.
The remote read can result in one of three possible scenarios.
\begin{enumerate}
\item If the read contents match the associated counter and checksum and the valid bit is set, the read linearises
at the point of the remote read's execution and returns the read value.  
\item If the read contents and the associated checksum do not match, 
the read overlaps with an ongoing (torn) update, and the read is retried in its entirety.
\item If the read contents match the requested counter but the valid bit is unset, this implies
either that an in-progress insert has not yet linearised, or an in-progress
deletion has already linearised but not yet updated the local index. The read linearises
at the point of the remote read's execution and returns \code{EMPTY}.
\item If the read contents do not match the requested counter, this implies an in-progress
\code{delete} has completed but had not yet updated the local index when the read was initiated,
and later operations have reused the slot.  In this case, the
read linearises immediately after the delete and returns \code{EMPTY}.
\end{enumerate}

\subsection{Proof of Safety}

\begin{lemma}
\label{lm:tmo}
All \code{write}s, \code{delete}s, and \code{insert}s
for a given key form a total modification order which respects the real-time
ordering of the operations.
\end{lemma}

\begin{proof}
By mutual exclusion on the per-key lock, each operation's
effects are completed before any subsequent operation.
\end{proof}

\begin{lemma}
\label{lm:read}
Every \code{read} returns a value consistent with the total modification order
and which respects real-time ordering of the operations.
\end{lemma}

\begin{proof}
We break our proof into three cases contingent on the
result of the remote read.  In the first, the local index counter
matches the result of the remote read, in the second, the local
index does not match, in the third, the checksum does not match
and the read cannot determine the case.  We validate the
linearisation of the read for each case in reverse order.

In the case where the checksum does not match, this is an
atomicity violation, and the operation retries without linearising.

In the case where the local index does not match,
the counter value read by the remote read indicates that the local
index is out of date.  This case 
implies an in-progress
\code{delete} has linearised but not yet updated the local index, 
and later operations have reused the slot.  As the remote \code{delete}
cannot complete until the local index is updated, the read must have
overlapped in real-time with the delete, and thus can return \code{EMPTY}.

In the case where the local index matches, the remote read may discover
a either a valid or invalid value.  If the value is valid,
the read can return the read value, as this value respects
the most recent linearisation of a modification to the location.
If the read discovers an invalid flag --- this indicates
that its local index is out-of-date with respect to an ongoing
\code{delete} or \code{insert}.  Returning \code{EMPTY} respects
the linearisation point of both operations (note the asymmetry of the
modifying operations to enable this possibility).
\end{proof}

By lemma~\ref{lm:tmo} and~\ref{lm:read}, and by composition of linearisable
objects~\cite{herlihy-toplas-1990},
\begin{theorem}
The presented hashmap is linearisable.
\end{theorem}